\documentclass[twocolumn]{aastex63}
\usepackage{amssymb,amsmath}
\usepackage{graphicx}
\usepackage{xcolor,natbib}
\usepackage{multirow}
\usepackage[T1]{fontenc}
\usepackage{ae,aecompl}
\usepackage{newtxtext, newtxmath}
\usepackage{float}
\usepackage{subfigure}
\usepackage{longtable}   
\usepackage{enumitem}
\usepackage{longtable,listings}
\usepackage{longtable}
\usepackage{array} 
\usepackage{booktabs}
\usepackage{subfigure}
\usepackage{multirow}
\usepackage[flushleft]{threeparttable}
\usepackage{graphicx} 
\usepackage{parcolumns}
\def\oiii{[O~{\sc~iii}]\ }
\def\nii{[N~{\sc ii}]\ }

\def\oiii{[O~{\sc iii}]\ }
\def\feii{Fe~{\sc ii}\ }

\submitjournal{ApJS}

\shorttitle{type 1 AGNs with double-peaked \oiii}
\shortauthors{Zheng et al.}

\begin{document}

\title{On type 1 active galactic nuclei with double-peaked [O~{\sc iii}]. I. data sample and basic results}

\correspondingauthor{Xueguang Zhang; Qirong Yuan}
\email{xgzhang@gxu.edu.cn; yuanqirong@njnu.edu.cn}

\author{Qi Zheng}
\affiliation{School of Physics and Technology, Nanjing Normal University, No. 1,	Wenyuan Road, Nanjing, 210023, P. R. China}

\author{Yansong Ma}
\affiliation{School of Physics and Technology, Nanjing Normal University, No. 1,	Wenyuan Road, Nanjing, 210023, P. R. China}

\author{Xueguang Zhang$^{*}$}
\affiliation{Guangxi Key Laboratory for Relativistic Astrophysics, School of Physical Science and Technology,
	GuangXi University, No. 100, Daxue Road, Nanning, 530004, P. R. China}

\author{Qirong Yuan$^{*}$}
\affiliation{School of Physics and Technology, Nanjing Normal University, No. 1,	Wenyuan Road, Nanjing, 210023, P. R. China}

\author{Weihao Bian}
\affiliation{School of Physics and Technology, Nanjing Normal University, No. 1,	Wenyuan Road, Nanjing, 210023, P. R. China}

\begin{abstract}
Double-peaked narrow emission lines (DPNELs) might be evidence for the existence of kpc-scale dual AGNs.
There are so far large samples of objects with DPNELs in narrow emission line galaxies. Here, a systematic search is made to build a sample of type 1 AGNs with double-peaked \oiii from Data Release 16 of the Sloan Digital Sky Survey (SDSS).
Through visually inspecting and fitting \oiii, fitting broad H$\alpha$ emission lines, performing F-test for \oiii~profiles, and checking broad H$\beta$ and \oiii~emission lines, we select  62 type 1 AGNs with reliable double-peaked \oiii~from 11557 QSOs with z < 0.3.
After visually checking the 62 SDSS multi-color images, we find only seven objects with signs of merging. 
Four possible models for the double-peaked \oiii~observed in our sample are discussed: the superposition model, AGN outflow model, dual AGN model, and rotating disk model. However, the current results can not provide any one explanation conclusively, and additional observational data are needed to provide the details of narrow line regions.
But at least 22 objects with different velocity offsets between double-peaked \oiii~and narrow H$\alpha$ emission lines could be excluded as dual AGN candidates.
The relative velocity offsets of the \oiii~blue-shifted/red-shifted components are negative to their line flux ratios, which is consistent with dual AGN model.
This work provides a new sample of 62 type 1 AGNs with double-peaked \oiii~for further study.
\end{abstract}


\section{Introduction}
It is widely accepted that central supermassive black holes (SMBHs) are present in most bulge-dominated galaxies \citep{Ko95,Ko13,He14}.
According to the hierarchical formation models, galaxy mergers frequently occur \citep{Si98,Ma19,Zu22}, and then facilitate the transportation of gas towards their central regions, initiate episodes of nuclear star formation, possibly trigger the activation of central supermassive black holes \citep{Sp05,Pe23,Li23}.
As a result, the force of dynamical friction comes into  guiding each SMBH towards the center of the emerging mass distribution. 
This process can lead to the creation of a dual AGN with separations from kpc scale to pc scale \citep{De19}. Subsequently, it progresses into a binary black hole (BBH) \citep{La20,Ko21} with sub-pc separation driven by gravitational wave emission, ultimately merging to form a larger SMBH.
The studies on merging galaxies would provide valuable insights into the drivers of galaxy evolution, the estimation of the galaxy merger rate and the production of gravitational waves \citep{Me05,Ko06,Ar20}.
In this manuscript, we mainly focus on dual AGN systems around kpc-scale, and hereafter `dual AGN' mentioned in the manuscript means system around kpc-scale.

While it is expected that dual AGN systems would be common in merger scenario, there are actually rare identified dual AGN systems as discussed in \citet{Bh23}. Considerable work has been put into seeking observational support for the existence of dual AGN systems through spatially resolved images in these years, such as in \citet{Ko03,Li11,Sh19a,Go19,Zh21,Sa21,Bh23,Zh23}.
For spatially resolved systems, dual AGN systems can be directly identified by two bright cores as morphological characteristics.

However, due to the lack of spatial resolution, it is in most cases impossible to resolve the two nuclei.
Therefore, some other spectroscopic features are applied to search for dual AGN systems, such as DPNELs \citep{Xu09} or shifted single-peaked narrow emission lines relative to absorption lines \citep{Zh23}.
In this scenario, the orbital radial velocity of two nuclei is
non-negligible, and it is spectroscopically observed as AGN emission lines produced by ionized gas. These emission lines exhibit velocity offsets relative to the stellar absorption lines. The latter are assumed to trace the systemic velocity of the host galaxy \citep{Ba16}.
The main focus of the paper is the study of DPNELs related to kpc dual AGN systems.

The double-peaked narrow emission profiles are detected in both type 1 and type 2 AGNs and they are generally most prominent in \oiii $\lambda$$\lambda$ 4959, 5007.
\citet{Zh04} found a type 2 quasar SDSS J1048+0055 with double-peaked profiles of \oiii $\lambda$$\lambda$ 4959, 5007 and two radio cores in 8.4 GHz, and first suggested that DPNELs might be an effective way to find dual AGN systems. Then there are several objects reported as dual AGN systems with both double-peaked profiles of \oiii and two cores in images as discussed in \citet{Co11,Mc11,Li13,Wo14}, and several dual AGN candidates with DPNELs but no spatially resolved images as reported in \citet{Ge07,Ba12,An13,Zh24}.
Moreover, besides the double-peaked profiles can be found in \oiii, this features can also be found in the higher-ionization lines.
\citet{Ba13} firstly reported the candidates of dual AGN systems at higher redshift with double-peaked [Ne~{\sc v}]$\lambda$3426 or [Ne~{\sc iii}]$\lambda$3869 from the Sloan Digital Sky Survey (SDSS) Data Release 7 (DR7) to support dual AGN systems.
Furthermore, as suggested by \citet{Xu09}, the presence of double-peaked profiles in all of its narrow emission lines, rather than just one, can be considered as a better proof
to support exceptionally strong contenders for kpc-scale dual AGN systems. Up to the present, there are only SDSS J131642.90+175332.5 \citep{Xu09}, SDSS J143132.84+435807.20 \citep{Se21}, and SDSS J222428.53+261423.2 \citep{Zh24} reported such unambiguous double-peaked profiles in all optical narrow emission lines.

Unfortunately, it is difficult to confirm these objects with DPNELs as dual AGN systems because of lacking two spatially resolved cores.
In addition to dual AGN systems, there are several alternative explanations for the double-peaked narrow emission profiles.
The first alternative explanation for the DPNELs is the superposition of two extra-galactic emission line objects (superposition model). 
\citet{Xu09} discussed this scenario in SDSS J131642.90+175332.5, and \citet{Do101} investigated the peak separation between blue-shifted and red-shifted components ($\Delta\upsilon$) of a chance superposition.
The second alternative explanation for the DPNELs is AGN outflow model.
Several studies \citep{Ml15,Ne16,Co18,Ru19} suggested that the prevailing influence on objects with DPNELs is, indeed, outflows, and \citet{Ne16} claimed that this percentage can be as high as 86\% from 71 type 2 AGNs at z < 0.1.
The third alternative explanation for the DPNELs is rotating disk model.
\citet{Sm12} showed that it is likely that equal-peaked narrow emission profiles represent rotating disk.

Thus detecting double-peaked profiles in narrow emission lines
alone is not sufficient to confirm objects as kpc-scale dual AGN systems.
All these cases simply indicate the diverse nature of narrow line region (NLR) gas dynamics, so it is necessary for follow-up observations to 
confirm or refute kpc-scale dual AGN systems from the double-peaked \oiii sample.
These follow-up observations include high-resolution imaging to detect spatially resolved cores.
\citet{Fu11} published images of 50 AGNs (z < 0.6) with double-peaked \oiii from the Keck II laser guide star adaptive optics system (LGSAO;\citealt{Wi06}) and found that $\sim$30\% of the objects in their sample show discernible companions with a spatial resolution of $0.1^{\prime \prime}$.
With the same instrument, \citet{Ro11} showed six objects with double galaxy structures and four objects in galaxy mergers from 12 QSOs (0.2 < z < 0.6) with double-peaked \oiii.
\citet{Li13} examined and confirmed two dual AGNs from four candidates with double-peaked \oiii through the Hubble Space Telescope (HST) and Chandra imaging.
By the same method, \citet{Co15} confirmed a dual AGN with a 2.2 kpc separation in an extreme minor merger, which resulted in the double-peaked \oiii emission lines.
While high-resolution imaging can detect two components with small separations, it can not confirm whether the double-peaked profiles are related to only one component.

In addition, spatially resolved spectroscopy is another method of follow-up observations besides high-resolution imaging.
Long-slit spectroscopy is used to spectrally separate the DPNELs to determine whether the emission lines are from the regions along the slit. 
As discussed in \citet{Co12}, the double AGN emission components aligned with the host galaxy major axis are expected in case of dual AGN system orbiting in the host galaxy potential.
Integral field spectroscopy offers high-quality and delicate spectra with smaller diameter apertures to support robust evidence of the origin of the double emission lines.
As shown in \citet{Fu12}, integral-field spectroscopy resolves the kinematic components spatially, linking each to its respective core.


It is worth noting that many studies have combined imaging with spatially resolved spectroscopy to gain a more comprehensive understanding of the sources with DPNELs.
\citet{Co09b} showed evidence for a dual AGN COSMOS J100043.15+020637.2, showing double-peaked \oiii, with HST image ($1.75\pm0.03~h^{-1} \rm kpc$ projected spatial offset) and slit spectroscopy.
\citet{Liu10b} reported four dual AGNs, whose locations of two cores in deep near-infrared images are coincident with those of two components of double-peaked \oiii in the slit spectra.
\citet{Mc11} detected that SDSS J095207.62+255257.2 with double-peaked \oiii is confirmed as a dual AGN separated by 4.8 kpc, using LGSAO imaging and near-infrared integral-field spectroscopy.
\citet{Fu11b} confirmed SDSS J150243.09+111557.3 with double-peaked \oiii as a dual AGN separated by 7.4 kpc through the optical integral-field spectroscopy and high-resolution radio images.
\citet{Sh11} reported that roughly 10\% of 31 type 2 AGNs with double-peaked \oiii are dual AGNs and about 50\% of them exhibit the kinematic signatures of a single NLR determined by both near infrared image and optical slit spectroscopy.
\citet{Fu12} discovered that only 2\% of the double-peaked \oiii are produced by the orbital motion of the merging nuclei through combining integral-field spectroscopy with LGSAO imaging.
A few years later, \citet{Mc15} claimed that 5\% of double-peaked
objects are dual AGNs through high-resolution imaging and spatially resolved integral field and long-slit spectroscopy.

Unfortunately, \citet{Vi15} showed that the combination of deep multi-band imaging and long-slit spectroscopy would sometimes fail to unambiguously identify the objects with double-peaked \oiii as dual AGNs.
Moreover, the techniques mentioned above are suitable for identifying individual dual AGN systems with DPNELs, but quite expensive for discovering large samples of dual AGN system candidates by DPNELs. 
However, a larger sample of dual AGN system candidates with DPNELs is needed to understand the physical origins of DPNELs through statistical results.
There are a growing number of objects with DPNELs
through systematic searches in the SDSS \citep{Wa09,Sm10,Liu10,Ly16,Co18,Ma20,Ki202}, the Large Sky Area Multi-Object Fiber Spectroscopic Telescope \citep{Sh14,Wa19}, and the DEEP2 Galaxy Redshift Survey \citep{Co09}.
\citet{Ma20} reported the largest sample to date of 5,663 galaxies displaying DPNELs, based on the Reference Catalogue of Spectral Energy Distribution \citep{Ch17}, to analyze host-galaxy properties and kinematics; however, this dataset remains non-public. In comparison, \citet{Ge12} reported a publicly available sample of 3,030 DPNEL galaxies from SDSS DR7.

However, the majority of the reported objects with DPNELs are narrow emission line galaxies (type 2 AGNs, star-forming galaxies, etc.).
Combining with the narrow emission line galaxies with DPNELs,
broad emission line galaxies with DPNELs will provide more clues to test and/or identify the physical origin of DPNELs.
Therefore, we carry out a systematic search for type 1 AGNs with double-peaked \oiii from the SDSS DR16 \citep{Ah20}.

The paper is organized as follows.
In Section 2, we show our main procedure to collect the type 1 AGNs with double-peaked \oiii from the parent sample of quasars in SDSS DR16.
In Section 3, basic physical parameters are measured for the collected type 1 AGNs with DPNELs. 
In Section 4, the possible explanations for the DPNELs in this paper are given.
The main conclusions are given in Section 5.
In this paper, we adopt the cosmological parameters of $H_{0}=70{\rm km\cdot s}^{-1}{\rm Mpc}^{-1}$,
$\Omega_{\Lambda}=0.7$ and $\Omega_{\rm m}=0.3$.

\section{SAMPLE SELECTION}
\begin{figure*}
	\centering\includegraphics[width=14cm,height=17cm]{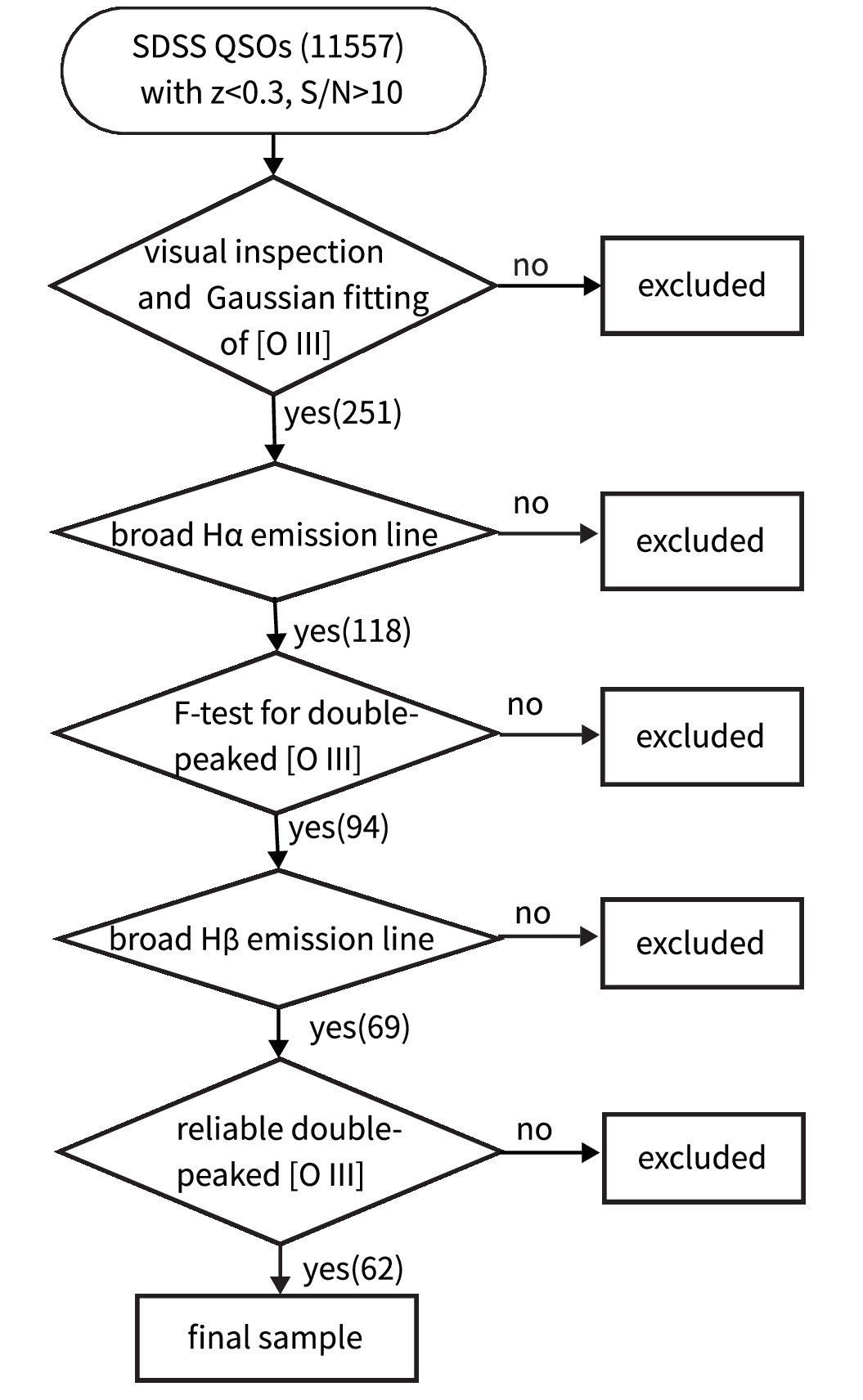}
	\caption{The flow chart in this work.}
	\label{mg}
\end{figure*}

For the goal of selecting type 1 AGNs with double-peaked \oiii, the objects classified as QSOs in \citet{Ro12,Pa18,Ly20} (\url{https://www.sdss4.org/dr16/algorithms/qso_catalog/}) in SDSS DR16 \citep{Ah20} are mainly considered. 
The details are described below in six steps, and a corresponding
brief flow chart of the whole procedure is shown in Figure \ref{mg}.

In the first step, a parent sample of SDSS QSOs are created through the SDSS SQL tool by the following two criteria of spectroscopic results of $z < 0.3$ and signal-to-noise ratio (S/N) larger than 10. 
The criterion of $z < 0.3$ is applied to ensure H$\alpha$ totally covered in SDSS spectra.
The criterion of S/N is applied for the reliable measurements of emission line parameters.
Then the applied SQL query (\url{https://skyserver.sdss.org/dr16/en/tools/search/sql.aspx}) in detail is as follows:
\begin{lstlisting}
SELECT s.ra, s.dec, s.z, s.snmedian
FROM specobjall as s
WHERE 
s.class=`qso' and s.z<0.3 and 
s.zwarning=0 and s.snmedian>10
\end{lstlisting}
The database of SpecObjAll (\url{https://skyserver.sdss.org/dr16/en/help/browser/browser.aspx?cmd=description+galSpecLine+U#&&history=description+SpecObjAll+U}) contains all the basic spectroscopic information.
As a result, the parent sample, containing 11557 QSOs from SDSS DR16, is obtained. 

\begin{figure*} \centering\includegraphics[width = 15cm,height=9cm]{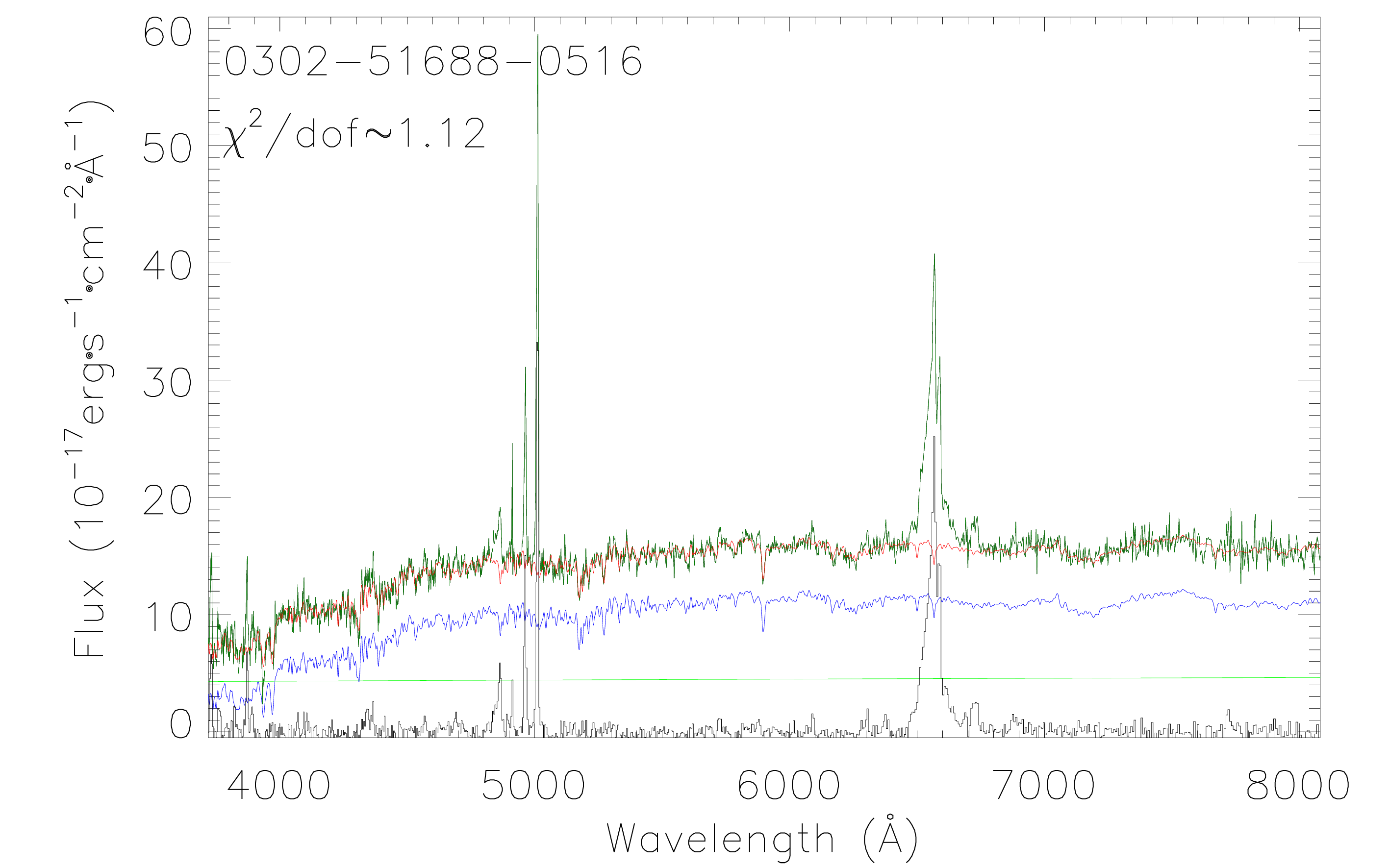} \caption{An example of SDSS spectrum in the final sample. 
		The solid dark green line represents the spectrum with Plate-Mjd-Fiberid shown in the top-left of corner, the solid red line shows the best fitting results with $\chi^2/\rm dof$ shown in the top-left of corner, the solid blue line shows starlight, the solid green line represents the AGN continuum emission, and the solid black line represents the line spectrum calculated by the SDSS spectrum minus the best fitting results.
	} \label{fig01} \end{figure*}

\begin{figure*} \centering\includegraphics[width = 15cm,height=9cm]{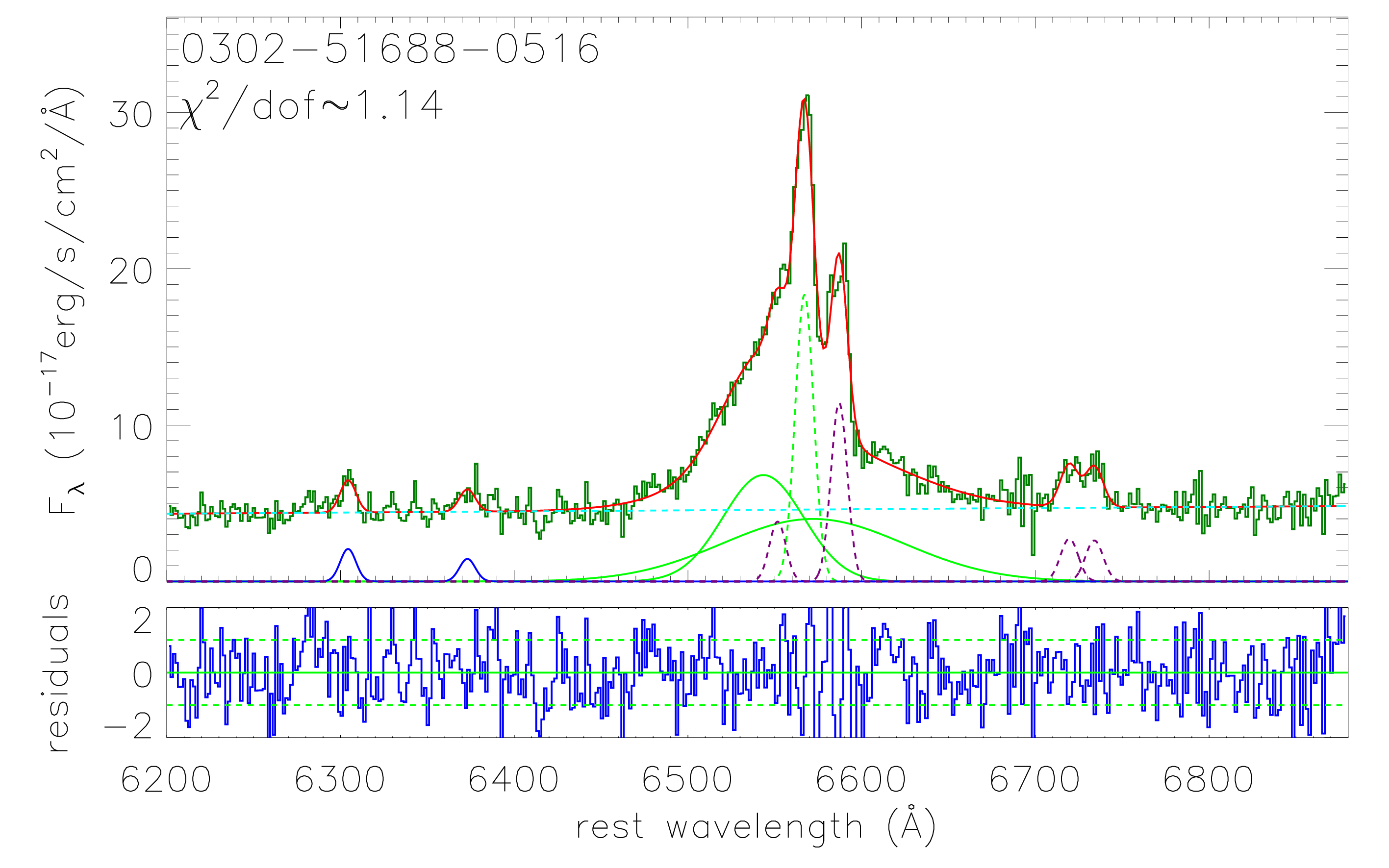} \caption{An example of the best fitting results of the emission lines around H$\alpha$.
		In the top panel, the solid dark green line shows the line spectrum after subtracting starlight from host galaxy with Plate-Mjd-Fiberid shown in the top-left of corner, the solid red line represents the best fitting results with $\chi^2/\rm dof$ shown in the top-left of corner, the dashed cyan line represents continuum emission,
		the solid green lines represent the determined components of broad H$\alpha$ emission line, the dashed green line represents narrow H$\alpha$ emission line, the solid blue lines represent [O~{\sc i}] emission lines, and the dashed purple lines represent [N~{\sc ii}] and [S~{\sc ii}] doublets. 
		In the bottom panel, the solid blue line represents the residuals calculated by the line spectrum minus the best fitting results and then divided by uncertainties of SDSS spectrum, the horizontal solid and dashed green lines show residuals=0, ±1, respectively.} \label{image2} \end{figure*}

\begin{figure*}
	\centering\includegraphics[width=18cm,height=9.5cm]{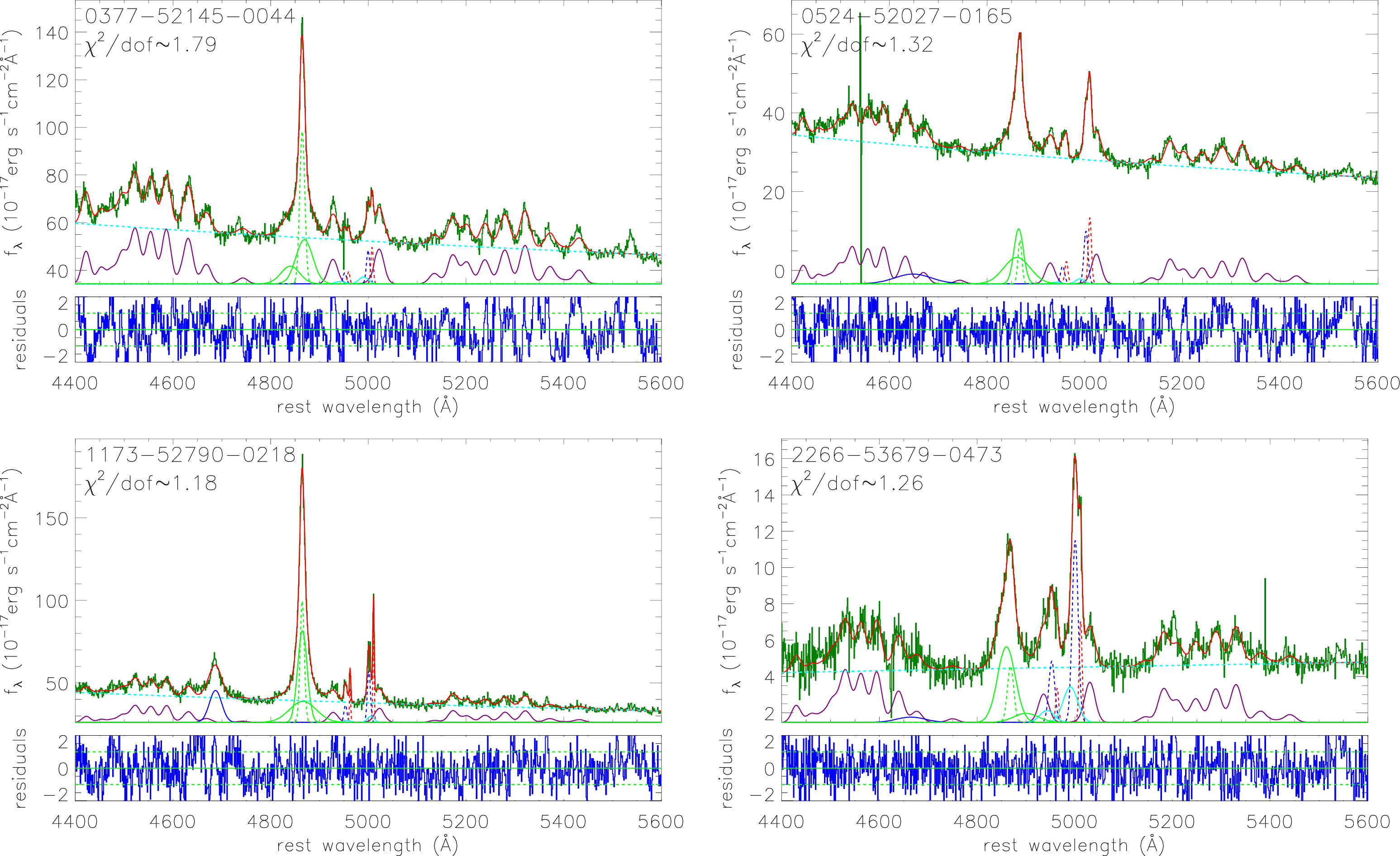}
	\caption{Four examples on the best fitting results of Fe{\sc ii} emission lines. 
		In the top of each panel, the solid dark green line shows the line spectrum after subtracting starlight (if present) determined from host galaxy with Plate-Mjd-Fiberid shown in the top-left of corner, the solid red line represents the best fitting results with $\chi^2/\rm dof$ shown in the top-left of corner, the solid blue line represents He~{\sc ii} emission line (if present), the solid purple lines represent Fe~{\sc ii} emission lines, and the dashed cyan line represents continuum emission, the solid green lines represent broad H$\beta$ emission line, the dashed green line represents narrow H$\beta$ emission line, the solid cyan lines represent extended components of \oiii, and the dashed blue and red lines represent the blue-shifted and red-shifted components of \oiii, respectively. 
		In the bottom of each panel, the solid blue line represents the residuals calculated by the line spectrum minus the best fitting results and then divided by uncertainties of SDSS spectrum, the horizontal solid and dashed green lines show residuals=0, ±1, respectively.
	}
	\label{fig03}
\end{figure*}

\begin{figure*} \centering\includegraphics[width = 15cm,height=9cm]{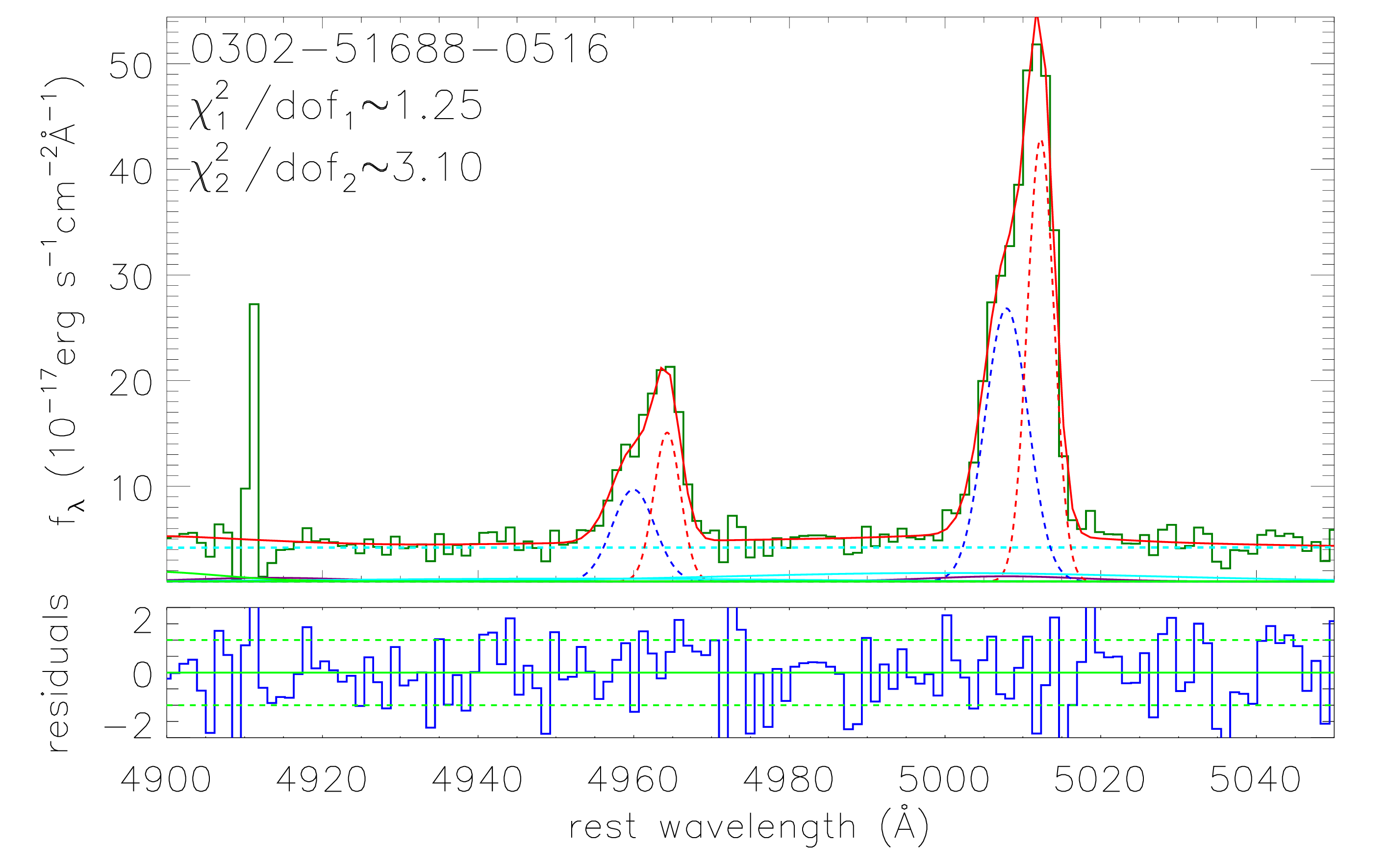} \caption{An example of the best fitting results of the emission lines around \oiii.
		In the top panel, the solid dark green line shows the line spectrum after subtracting starlight determined from host galaxy with Plate-Mjd-Fiberid shown in the top-left of corner, the solid red line represents the best fitting results with two narrow Gaussian functions describing each core component of \oiii~doublet, and the corresponding $\chi_1^2/\rm dof_1$ is shown in the top-left of corner, the $\chi_2^2/\rm dof_2$ determined by one narrow Gaussian function for each core component of \oiii~doublet is shown in the top-left of corner, the dashed cyan line represents continuum emission, the solid purple lines represent Fe~{\sc ii} emission lines, the solid green lines represent broad H$\beta$ emission line, the solid cyan lines represent extended component of \oiii, and the dashed blue and red lines represent the blue-shifted and the red-shifted components of \oiii, respectively. 
		In the bottom panel, the solid blue line represents the residuals calculated by the line spectrum minus the best fitting results and then divided by the uncertainties of the SDSS spectrum, the horizontal solid and dashed green lines show residuals=0,±1, respectively.} \label{f04} \end{figure*}

\begin{figure*}
	\centering\includegraphics[width=8cm,height=5cm]{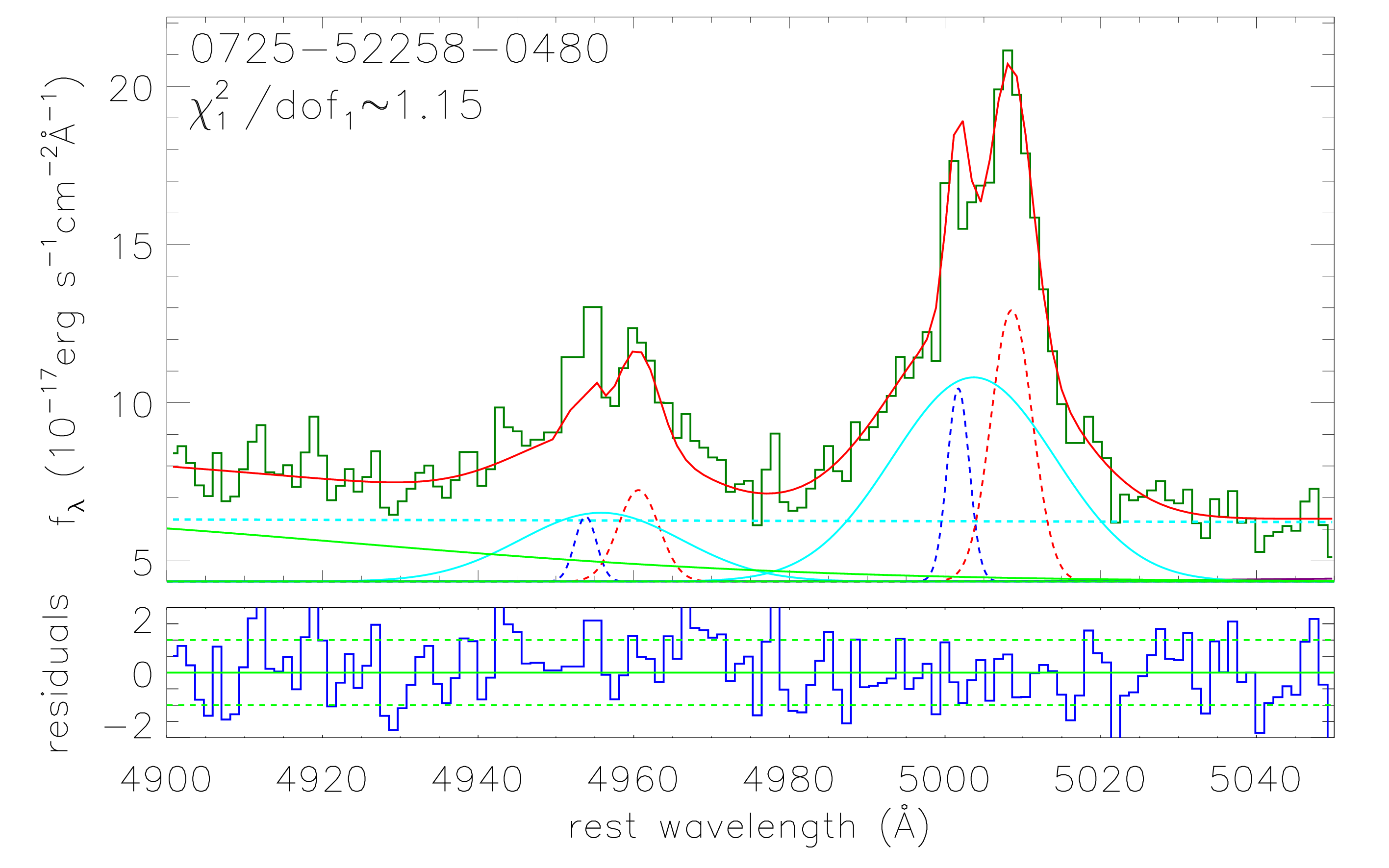}
	\centering\includegraphics[width=8cm,height=5cm]{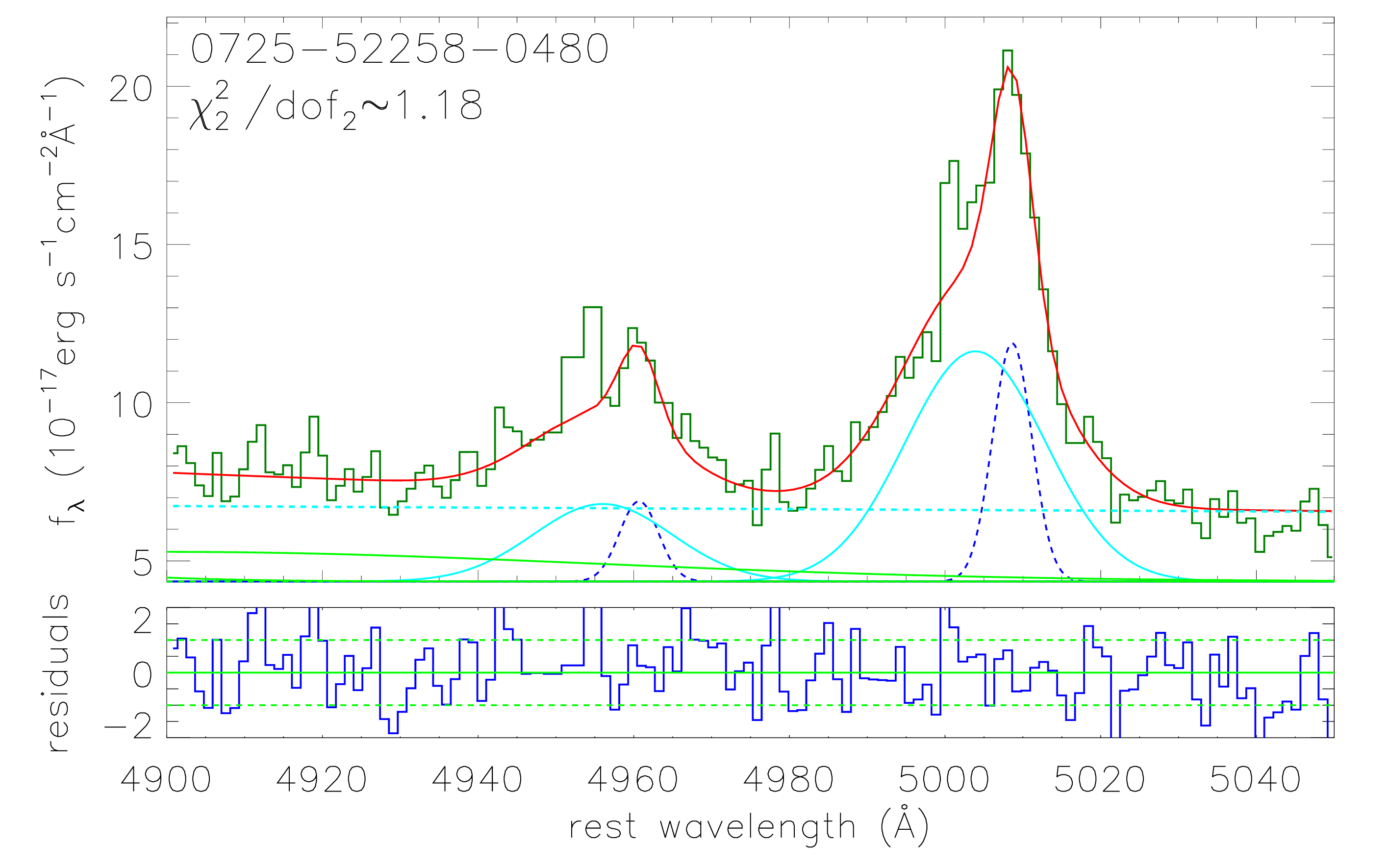}
	\caption{An example of type 1 AGN excluded by F-test.
		In the top of each panel, the solid dark green line shows the line spectrum after subtracting starlight determined from host galaxy with Plate-Mjd-Fiberid shown in the top-left of corner, the solid red line represents the best fitting results with $\chi_1^2/\rm dof_1$ ($\chi_2^2/\rm dof_2$) of two (one) narrow Gaussian functions for each core component of \oiii~doublet, the dashed cyan line represents continuum emission, the solid green lines represent broad H$\beta$ emission line, the solid cyan lines represent extended \oiii components. 
		The dashed blue and red lines (in the top-left panel) represent the blue-shifted and red-shifted components of \oiii, respectively.
		The dashed blue line (in the top-right panel) represents the single peak of \oiii. 
		In the bottom of each panel, the solid blue line represents the residuals calculated by the line spectrum minus the best fitting results and then divided by uncertainties of SDSS spectrum, the horizontal solid and dashed green lines show residuals=0,±1, respectively.
	}
	\label{fig08}
\end{figure*}

In the second step, the spectra around \oiii~are visually inspected and simply fitted by Gaussian functions. 
The spectra of all the quasars in the parent sample are displayed on the screen with rest wavelength coverage from 4900 to 5050 \AA, and five people independently inspect the spectra {\bf by eyes}.
Not only the objects with apparent double-peaked \oiii emission features but also the objects with plateau and strange emission features around \oiii are chosen as candidates with double-peaked \oiii.
Here, the `double-peak' means the presence of two distinct peaks near [O~{\sc~iii}]$\lambda$5007 separated by a dip, while `plateau' indicates a single peak near [O~{\sc~iii}]$\lambda$5007, with adjacent points maintaining values comparable to the maximum.
One object would be eventually collected as a candidate, if it is thought to have double-peaked emission features of \oiii by at least three people.
Besides, \oiii~doublet is simply described by six narrow Gaussian functions (four Gaussian functions for the possible double-peaked profiles and two Gaussian functions for the extended components). 
If \oiii~doublet of one object can be fitted by six Gaussian functions or by four Gaussian functions (with very weak extended components), it is left as a candidate with DPNELs.
All the objects selected according to the two methods mentioned above are saved.
The undeniable fact is that those methods might mistakenly identify Fe~{\sc ii} lines as peaks of \oiii~lines, but this is just a preliminary screening, and these objects will be further processed in detail afterwards.
Among all the 11557 QSOs in parent sample, 251 objects are firstly selected as candidates of type 1 AGNs with probable double-peaked \oiii. 

In the third step, the emission lines around H$\alpha$ are modeled, in order to reconfirm the collected objects as type 1 AGNs with broad H$\alpha$.
Due to the apparent host galaxy contributions in some of the SDSS spectra of the collected 251 objects, the simple stellar population (SSP) method \citep{Br93,Br03,Ka03,Ci05,ca17} is firstly applied to describe the contributions of starlight.
The same method can be found in \citet{zxg14,zxg21,zxg24,zxg242}.
Host galaxy contribution can be modeled by combination of the 39 broadened, strengthened and shifted SSPs with 13 population ages ranging from 5 Myr to 12 Gyr and 3 metallicities (Z=0.008, 0.05, 0.02) as described and discussed in \citet{Ka03,Br03}.
Meanwhile, a power law function is applied to describe AGN continuum emission.
Then through the Levenberg-Marquardt least-squares minimization technique (the MPFIT package; \citealp{Ma09}), the observed spectrum of the object with apparent host galaxy contribution can be described by SSPs plus a power law function.
In addition, when the SSP method is applied, the emission lines are masked out with FWZI (Full Width at Zero Intensity) about 400 km/s, and broad Blamer lines and possible \feii~emission lines are masked out with rest wavelength from 4400 to 5600~\AA~and from 6400 to 6700~\AA.
Here, we do not show the best fitting results of all the firstly collected 251 objects, but Figure \ref{fig01} shows an example, SDSS J141003.66+001250.2 (Plate-Mjd-Fiberid: 0302-51688-0516), as one of the targets in our final sample of 62 type 1 AGNs with double-peaked \oiii. And the best fitting results to all the 62 targets are shown in the Figure \ref{figure11} in the Appendix A.
As shown in Figure \ref{fig01} and Figure \ref{figure11}, 26 objects do not show host galaxy contributions due to the very stronger AGN emission features.
It is worth noting that there is an object (Plate-Mjd-Fiberid: 2652-54508-0025) with very weak AGN continuum emission but confirmed with both broad H$\alpha$ and H$\beta$ emission lines in the later.

After subtracting the host galaxy contributions, the emission lines around H$\alpha$ (rest wavelength from 6200 to 6880 \AA) can be modeled as follows.
A narrow plus two broad Gaussian functions (second moment $\sigma$ smaller or larger than 600 ${\rm km/s}$) are applied to describe the H$\alpha$ emission line.
Each of [S{\sc ii}] ([O{\sc i}], [N{\sc ii}]) doublet is described by a Gaussian function, and the Gaussian components of doublet have the same redshift and line width in velocity space.
Meanwhile, the flux ratio of [N{\sc ii}] doublet is set as the theoretical flux ratio 3.
A power law component is applied to describe possible
continuum emission underneath the emission lines around H$\alpha$.
Based on the Levenberg-Marquardt least-squares minimization
technique, the emission lines around H$\alpha$ can be determined.
Then, among the firstly collected 251 objects, 118 objects with reliable broad H$\alpha$ emission lines are selected through the following criteria:
\begin{equation}
\sigma >13\textsc{\AA}~(\sim 600 {\rm~km/s}),
\frac{\sigma}{\sigma_{\rm error}}>3,
\frac{f}{f_{\rm error}}>3,
\end{equation}
where $\sigma$ and $\sigma_{\rm error}$ (in units of \AA) are the second moment and the corresponding uncertainty determined by MPFIT of one of the Gaussian components applied to describe broad H$\alpha$, and $f$ and $f_{\rm error}$ (in units of $10^{-17}{\rm erg/s/cm^2}$) are the flux and the corresponding uncertainty of emission components of broad H$\alpha$.
The other 133 objects are rejected since there are no reliable broad H$\alpha$ emission lines.
According to the fitting results, the redshift relative to the central wavelength of narrow H$\alpha$ emission line is consistent to that of \nii, which supports the reasonable measurements of narrow lines, even there are weak \nii.
Figure \ref{image2} (complete results for all 62 objects shown in Figure \ref{figure12} in Appendix A) shows the best fitting results of the emission lines around H$\alpha$ of the final 62 type 1 AGNs with double-peaked \oiii. 
Among Figure \ref{image2} and Figure \ref{figure12} in Appendix A, one point should be noted.
There are very broad components around 6730\AA~in some objects (Plate-Mjd-Fiberid: 11040-58456-0394, 11347-58440-0066, 1403-53227-0485, 1944-53385-0120 and 7283-57063-0660) and it is hard to accurately describe the [S~{\sc ii}] doublet.

In the fourth step, the emission lines around H$\beta$ (rest wavelength from 4400 to 5600 \AA), including probable optical Fe~{\sc ii} emission features, are modeled and the F-test technique \citep{Ma97, Ge12} is applied to check whether there are reliable double-peaked profiles of \oiii.
Considering probable double-peaked emission features, two narrow Gaussian functions for the core component, and a broad Gaussian function ($\sigma$ > 400 km/s)
for the extended component \citep{gh05a, sh112, zxg212} are applied to describe each of \oiii~doublet.
Meanwhile, corresponding Gaussian components of \oiii~doublet have the same redshift and line width in velocity space, and the flux ratio set to 3.
And a power law component is applied to describe AGN continuum emission.
A broad Gaussian function is applied to describe He~{\sc ii} emission line. 
The optical Fe~{\sc ii} emission templates in the four groups discussed in \citet{kp10} are applied to describe the optical Fe~{\sc ii} features. 
Two broad Gaussian functions ($\sigma$ > 600 km/s) are applied to describe the profile of broad H$\beta$ emission line.
A narrow Gaussian function is applied to describe narrow H$\beta$ emission line since the following two reasons. First, we mainly focus on the double-peaked profiles of \oiii. Second, the H$\beta$ emission line is rather weak, which will not affect the reliability of broad H$\beta$ discussed in the next step.
Moreover, we also try to fix the central wavelengths of narrow H$\beta$ and H$\alpha$ with these of double-peaked \oiii~to describe possible double-peaked narrow H$\beta$ and H$\alpha$ emission lines, but could not find appropriate results. Only nine objects show clear double-peaked narrow H$\alpha$ emission lines as shown afterwards.

\begin{table*}[htbp] 
	\centering 
	\caption{sample information} 
	\label{tab1} 
	\begin{tabular}{c|c|c|c|c|c|c|c|c} 
		\toprule
		Plate-Mjd-Fiberid & RA & DEC & z& $\rm mag_{r}$ & S/N & FIRST &$R$& Reference\\
		(1)&(2)&(3)&(4)&(5)&(6)&(7)&(8)&(9)\\
		\midrule
		0302-51688-0516	&	212.51527	&	0.21395	&	0.14	&	18.4$\pm$0.01	&	15.19	&	0	&	0	&	-		\\ \hline
		0307-51663-0219	&	219.25504	&	-1.07168	&	0.29	&	19.03$\pm$0.08	&	11.15	&	0	&	0	&	-		\\ \hline
		0332-52367-0639	&	184.03066	&	-2.23828	&	0.1	&	17.28$\pm$0.01	&	29.36	&	0	&	0	&	 (1),(2)		\\ \hline
		0377-52145-0044	&	340.12026	&	-1.11381	&	0.13	&	17.25$\pm$0.02	&	27.98	&	0	&	0	&	-		\\ \hline
		0394-51913-0111	&	13.6107	&	-0.339	&	0.17	&	19.03$\pm$0.01	&		25.54	&	10.53	&	237.98	&	 -	\\ \hline
		0448-51900-0084	&	133.73845	&	54.80571	&	0.26	&	18.2$\pm$0.02	&	15.73	&	1.31	& 11.78	&	-		\\  \hline
		0452-51911-0080	&	145.43677	&	57.85658	&	0.16	&	17.97$\pm$0.01	&	15.12	&	0	&	0	&	(1),(2),(6),(7)		\\  \hline
		0524-52027-0165	&	196.07083	&	2.09365	&	0.23	&	17.33$\pm$0.02	&	25.71	&	0	&	0	&	-		\\ 
		0555-52266-0033	&	144.88269	&	54.81924	&	0.29	&	18.31$\pm$0.01	&	15.56	&	0	&	0	&	-		\\  \hline
		0609-52339-0435	&	221.95334	&	62.74577	&	0.23	&	18.49$\pm$0.02	&	11.75	&	0	&	0	&	(1),(2),(8)		\\  
		\bottomrule
	\end{tabular} \\
	Notes. Column 1: Plate, MJD, Fiberid of spectroscopic observation; Column 2: RA; Column 3: DEC; Column 4:  Redshift of spectroscopic observation; Column 5: Apparent psf magnitude in r band; Column 6: Median signal-to-noise over all good pixels; Column 7: Integrated FIRST radio flux(mJy); Column 8: Radio-loudness; Column 9: Reference:(1) \citet{Zh16}; (2) \citet{Sm10}; (3) \citet{Ge12}; (4)\citet{Sm12}; (5) \citet{Zh24}; (6) \citet{Fu12}; (7) \citet{Co18}; (8) \citet{Ki202}; (9) \citet{Co12}; (10) \citet{Mc15}; (11) \citet{Ro11}; (12) \citet{Fu11}.
\end{table*}

It is known that there are apparent emission features of the optical Fe~{\sc ii} around \oiii, leading to effects on detecting double-peaked \oiii. Therefore, the emission lines around H$\beta$ and \oiii~are described by model functions, with the rest wavelength from 4400 to 5600 \AA~to totally cover optical Fe~{\sc ii}.
Figure \ref{fig03} shows several objects with strong Fe~{\sc ii}. It seems that the apparent peaks around \oiii are actually produced by optical Fe~{\sc ii}, which could potentially be misidentified as double-peaked \oiii,  
especially in the top-left panel of Figure \ref{fig03}.

Due to the main focus on double-peaked \oiii, the best fitting results around \oiii doublet (rest wavelength from 4900 to 5050 \AA) with two narrow plus one broad Gaussian functions for each of the \oiii doublet, referred as `model 1', are shown in Figure \ref{f04} (complete results for all 62 objects shown in Figure \ref{figure13} in Appendix A). The corresponding $\chi^2_1$ in Figure \ref{f04} is re-calculated related to the best fitting results with the rest wavelength from 4900 to 5050 \AA, and the $\rm dof_1$ (degrees of freedom) is determined by numbers of data points minus eleven (nine in the three Gaussian functions for [O~{\sc iii}]$\lambda5007$ and two in the power law component underneath the \oiii~doublet).
Here, the value of $\rm dof_1$ for the model 1 does not consider the number of model parameters for the optical Fe~{\sc ii}, because the same optical Fe~{\sc ii}, treated as a constant component, determined by the best fitting results with rest wavelength from 4400 to 5600 \AA~
are considered both in the model 1 and the following model 2.
So there is a little difference of shown $\chi^2/\rm dof$ between Figure \ref{fig03} and \ref{f04}.

Large wavelength range is not appropriate to do the 
following F-test, therefore, the range between 4900 to 5050 \AA~is 
considered. 
In order to test reliability of the double-peaked \oiii through the F-test technique, 
another model, referred as `model 2', is applied to describe the [O~{\sc iii}] emission lines.
In the model 2, there is only a narrow plus a broad Gaussian functions applied to describe each of the [O~{\sc iii}] doublet.
There is also a power law component applied to describe possible continuum emission.
Then, the same restrictions used in model 1 are applied in model 2 in \oiii~doublet.
Through the same MPFIT package, the best fitting results with $\chi^2_2/\rm dof_2$ can be determined to the \oiii~doublet.
Based on the different $\chi^2/\rm dof$
values for the model 1 and the model 2 for the [O~{\sc iii}]
emissions line fittings, the calculated $F_p$ \citep{Ma97,Ge12} value can be described as 
\begin{equation}
	F_p=\frac{\frac{\chi^2_2-\chi^2_1}{\rm dof_2-\rm dof_1}}{\chi^2_1/\rm dof_1}.
\end{equation}
According to $\rm dof_2-\rm dof_1=3$ and $\rm dof_1$ as number of dofs of the F distribution numerator and denominator, the expected value from the statistical F-test with confidence level about 3$\sigma$ will be near to $F_p$ value of 4.99.
For the spectra with wavelength ranging from 4900 to 5050 \AA, the numbers of data are all about 131 ($\rm dof_1=120$ and $\rm dof_2=123$).
Here, the 94 AGNs with confidence level higher than 3$\sigma$ are preferred to have double-peaked \oiii, rather than single-peaked \oiii. For example, the AGN (Plate-Mjd-Fiberid: 7875-56980-0296) is preferred to have DPNELs due to $\chi^2_1=109$, $\chi^2_2=1004$ with confidence level higher than 5$\sigma$ ($F_p$=328).
Figure \ref{fig08} shows an example excluded by F-test with confidence level less than 3$\sigma$ ($F_p$=2) with  $\chi^2_1= 138$ and $\chi^2_2= 145$.



In the fifth step, the modeled parameters of H$\beta$ emission lines are checked to identify true type 1 AGNs with broad H$\beta$ emission lines.
Similar criteria as described in equation (1) are applied here.
Then, 69 type 1 AGNs in our work are retained.

In the sixth step, we refine our selection process by identifying more reliable candidates for the final sample based on the following criterion. We select 62 type 1 AGNs into the final sample with second moment and line flux of double-peaked \oiii~three times larger than their uncertainties.
The 62 type 1 AGNs in our work with reliable double-peaked \oiii are confirmed as true type 1 AGNs with both broad H$\beta$ and H$\alpha$ emission lines. 
Basic information of the 62 AGNs in the final sample is listed in Table \ref{tab1} including Plate-Mjd-Fiberid, RA, DEC, redshift, photometric magnitude in $r$-band, S/N.
Since the broad H$\beta$ components are applied only to verify true type 1 AGNs, the parameters of the central wavelength, line width and flux of broad H$\beta$ emission lines 
determined from line profiles of two broad Gaussian functions are listed in Table \ref{tab5}. 
Similarly, parameters of broad H$\alpha$ emission lines are listed in Table \ref{tab5}.
The features of [N~{\sc ii}] and narrow H$\alpha$ emission lines are listed in Table \ref{tab4}.
Meanwhile, the modeled parameters of double-peaked \oiii and the results of F-test are shown in Table \ref{tab2}.
For clarity, here, we present the information of several AGNs of the final sample in the main text of the manuscript; the corresponding full information of all 62 type 1 AGNs is provided in Appendix B.



\clearpage
\begin{table*}[htbp] 
	\centering 
	\caption{Line parameters of broad H$\alpha$ and broad H$\beta$} 
	\label{tab5} 
	\begin{tabular}{c|c|c|c|c|c|c|c} 
		\toprule
		Plate-Mjd-Fiberid & $\lambda_{H\alpha}$  & $\rm FWHM_{H\alpha}$  & $F_{H\alpha}$ & $\lambda_{H\beta}$  & $\rm FWHM_{H\beta}$  & $F_{H\beta}$ & mass\\ 
		(1)&(2)&(3)&(4)&(5)&(6)&(7)&(8)\\ 
		\midrule
		0302-51688-0516	&	6559.7 	$\pm$	5.0 	&	72.3 	$\pm$	5.7 	&	918.7 	$\pm$	116.2 	&	4850.9 	$\pm$	3.1 	&	35.9 	$\pm$	5.4 	&	96.4 	$\pm$	17.7 	&	7.81	$\pm$	0.10 	\\ \hline
		0307-51663-0219	&	6569.2 	$\pm$	0.6 	&	47.7 	$\pm$	5.0 	&	881.3 	$\pm$	55.5 	&	4864.9 	$\pm$	0.8 	&	43.7 	$\pm$	2.6 	&	188.6 	$\pm$	9.3 	&	7.50	$\pm$	0.11 	\\ \hline
		0332-52367-0639	&	6565.3 	$\pm$	1.6 	&	99.4 	$\pm$	3.9 	&	3770.8 	$\pm$	226.9 	&	4870.1 	$\pm$	2.1 	&	102.5 	$\pm$	5.9 	&	768.8 	$\pm$	50.6 	&	7.98	$\pm$	0.05 	\\ \hline
		0377-52145-0044	&	6546.1 	$\pm$	0.5 	&	73.3 	$\pm$	5.4 	&	4751.0 	$\pm$	153.0 	&	4859.9 	$\pm$	1.6 	&	50.8 	$\pm$	2.8 	&	1154.8 	$\pm$	66.7 	&	7.80	$\pm$	0.07 	\\ \hline
		0394-51913-0111	&	6566.3 	$\pm$	1.7 	&	35.2 	$\pm$	4.0 	&	1141.8 	$\pm$	94.9 	&	4863.3 	$\pm$	1.9 	&	23.2 	$\pm$	4.6 	&	46.0 	$\pm$	10.5 	&	7.77	$\pm$	0.12 	\\ \hline
		0448-51900-0084	&	6567.9 	$\pm$	0.3 	&	40.8 	$\pm$	4.5 	&	1700.1 	$\pm$	58.8 	&	4865.3 	$\pm$	1.7 	&	35.2 	$\pm$	5.9 	&	544.7 	$\pm$	28.4 	&	7.28	$\pm$	0.11 	\\ \hline
		0452-51911-0080	&	6555.0 	$\pm$	0.8 	&	48.8 	$\pm$	2.1 	&	2213.3 	$\pm$	33.9 	&	4857.4 	$\pm$	4.3 	&	130.1 	$\pm$	12.4 	&	369.4 	$\pm$	31.6 	&	7.59	$\pm$	0.04 	\\ \hline
		0524-52027-0165	&	6557.9 	$\pm$	0.6 	&	48.8 	$\pm$	3.0 	&	2387.4 	$\pm$	110.1 	&	4857.4 	$\pm$	4.3 	&	29.3 	$\pm$	12.4 	&	369.4 	$\pm$	31.6 	&	7.79	$\pm$	0.07 	\\ \hline
		0555-52266-0033	&	6566.7 	$\pm$	0.7 	&	113.3 	$\pm$	1.3 	&	2854.9 	$\pm$	66.8 	&	4869.4 	$\pm$	16.5 	&	87.2 	$\pm$	33.1 	&	824.1 	$\pm$	64.6 	&	8.44	$\pm$	0.02 	\\
		\hline
		0609-52339-0435	&	6561.2 	$\pm$	0.8 	&	76.5 	$\pm$	1.7 	&	1051.7 	$\pm$	30.6 	&	4878.4 	$\pm$	15.8 	&	51.3 	$\pm$	35.6 	&	541.9 	$\pm$	72.0 	&	7.69	$\pm$	0.03 	\\ 
		\bottomrule
	\end{tabular} \\
	Notes.
	Column 1: Plate, MJD, Fiberid of spectroscopic observation; Column 2 and Column 3: the central wavelength (the first moment) of board H$\alpha$ in units of \AA~and corresponding FWHM of broad H$\alpha$ in units of \AA~ determined through the line profiles described by two Gaussian functions, respectively; Column 4: corresponding flux of broad H$\alpha$ in units of $10^{-17}{\rm erg/s/cm^{2}}$ determined by the line flux described by two Gaussian functions;
	Column 5 and Column 6: the central wavelength (the first moment) of board H$\beta$ in units of \AA~and corresponding FWHM of broad H$\beta$ in units of \AA~determined through the line profiles described by two Gaussian functions, respectively;
	Column 7: corresponding flux of broad H$\beta$ in units of $10^{-17}{\rm erg/s/cm^{2}}$ determined by the line flux described by two Gaussian functions; Column 8: the virial black hole mass log($\rm M_{\rm BH}$/$\rm M_{\odot}$).
\end{table*}

\begin{table*}[htbp] 
	\centering 
	\caption{Features of narrow H$\alpha$ and \nii emission lines} 
	\label{tab4} 
	\begin{tabular}{c|c|c|c|c|c|c} 
		\toprule
		Plate-Mjd-Fiberid & $\lambda_{H\alpha}$  & $\sigma_{H\alpha}$   & $F_{H\alpha}$  & $\lambda_{[N~{\textsc{ii}}]}$   & $\sigma_{[N~{\textsc{ii}}]}$  & $F_{[N~{\textsc{ii}}]}$   \\ 
		(1)&(2)&(3)&(4)&(5)&(6)&(7)\\ 
		\midrule
		0302-51688-0516&6567.1$\pm$0.2&5.3$\pm$0.3&244.6$\pm$23.8&6587.0$\pm$0.2&4.7$\pm$0.3&135.8$\pm$14.0\\ \hline
		0307-51663-0219&6569.1$\pm$0.2&2.3$\pm$0.2&77.4$\pm$7.8&6589.3$\pm$0.3&3.2$\pm$0.3&70.7$\pm$8.8\\ \hline
		0332-52367-0639&6563.3$\pm$0.2&4.3$\pm$0.3&254.6$\pm$25.8&6583.9$\pm$0.2&4.2$\pm$0.3&289.2$\pm$26.5\\ \hline
		0377-52145-0044&6567.8$\pm$0.1&7.2$\pm$0.1&3502.1$\pm$55.5&6586.1$\pm$0.3&4.9$\pm$0.3&409.6$\pm$38.0\\ \hline
		0394-51913-0111&6559.6$\pm$0.1&2.3$\pm$0.1 &125.7$\pm$5.2&	6585.2$\pm$0.1&	6.9$\pm$0.2&471.1$\pm$28.3\\ \hline
		0448-51900-0084&6566.7$\pm$0.1&2.2$\pm$0.2&113.0$\pm$8.9&6587.7$\pm$0.2&4.2$\pm$0.2&235.3$\pm$13.6\\ \hline
		0452-51911-0080&6564.0$\pm$0.2&4.7$\pm$0.2&339.4$\pm$13.2&6585.0$\pm$0.2&4.6$\pm$0.2&281.5$\pm$11.4\\ \hline
		0524-52027-0165&6569.8$\pm$0.1&7.1$\pm$0.2&1176.8$\pm$54.4&6592.7$\pm$0.2&2.9$\pm$0.2&107.9$\pm$11.4\\ \hline
		0555-52266-0033&6563.8$\pm$0.4&6.5$\pm$0.2&401.0$\pm$19.8&6580.6$\pm$1.0&8.0$\pm$0.9&190.7$\pm$35.3\\ \hline
		0609-52339-0435&6567.5$\pm$0.1&5.0$\pm$0.1&600.3$\pm$13.8&6586.4$\pm$0.1&4.7$\pm$0.2&276.5$\pm$12.0\\ 
		\bottomrule
	\end{tabular} \\
	Notes. Column 1: Plate, MJD, Fiberid of spectroscopic observation; Column 2: the central wavelength of narrow H$\alpha$ in units of~\AA; Column 3: corresponding second moment of narrow H$\alpha$ in units of \AA; Column 4: corresponding flux of narrow H$\alpha$ in units of $10^{-17}{\rm erg/s/cm^{2}}$; Column 5: the central wavelength of narrow \nii~in units of~\AA; Column 6: corresponding second moment of~\nii~in units of~\AA; Column 7:  corresponding flux of \nii~in units of $10^{-17}{\rm erg/s/cm^{2}}$.
\end{table*}

{\raggedright
	\begin{table*}[htbp] 
		\centering 
		\caption{Features of \oiii} 
		\label{tab2} 
		\begin{tabular}{c|c|c|c|c|c|c|c|c|c} 
			\toprule
			Plate-Mjd-Fiberid & $\lambda_b$ & $\sigma_b$  & $F_b$& $\lambda_r$ & $\sigma_r$ & $F_r$&F-test&$V_b$&$V_r$\\
			(1)&(2)&(3)&(4)&(5)&(6)&(7)&(8)&(9)&(10)\\ 
			0302-51688-0516	&	5007.9 	$\pm$	0.5 	&	2.8 	$\pm$	0.3 	&	178.9 	$\pm$	29.6 	&	5012.3 	$\pm$	0.1 	&	1.7 	$\pm$	0.1 	&	180.2 	$\pm$	28.4 	&	5 	&	173.2 	$\pm$	41.9 	&	435.6 	$\pm$	19.4 	\\ \hline
			0307-51663-0219	&	5001.0 	$\pm$	0.9 	&	4.6 	$\pm$	0.6 	&	78.2 	$\pm$	13.8 	&	5009.4 	$\pm$	0.3 	&	3.1 	$\pm$	0.2 	&	87.3 	$\pm$	13.0 	&	3 	&	-			&	-			\\ \hline
			0332-52367-0639	&	5005.8 	$\pm$	0.1 	&	1.9 	$\pm$	0.1 	&	360.5 	$\pm$	16.0 	&	5011.2 	$\pm$	0.2 	&	1.8 	$\pm$	0.1 	&	181.2 	$\pm$	15.6 	&	5 	&	112.5 	$\pm$	16.2 	&	214.0 	$\pm$	19.8 	\\ \hline
			0377-52145-0044	&	4999.5 	$\pm$	0.7 	&	3.6 	$\pm$	0.7 	&	121.2 	$\pm$	24.1 	&	5007.5 	$\pm$	0.5 	&	2.5 	$\pm$	0.4 	&	94.5 	$\pm$	21.8 	&	3 	&	-			&	-			\\ \hline
			0394-51913-0111	&	5003.7 	$\pm$	0.1 	&	1.8 	$\pm$	0.1 	&	69.5 	$\pm$	7.0 	&	5008.5 	$\pm$	0.2 	&	6.0 	$\pm$	0.2 	&	409.3 	$\pm$	32.4 	&	5 	&	203.3 	$\pm$	15.6 	&	83.8 	$\pm$	22.1 	\\ \hline
			0448-51900-0084	&	5004.9 	$\pm$	0.6 	&	3.9 	$\pm$	0.5 	&	239.1 	$\pm$	47.1 	&	5009.9 	$\pm$	0.1 	&	1.9 	$\pm$	0.1 	&	164.6 	$\pm$	30.4 	&	5 	&	-		&	-			\\ \hline
			0452-51911-0080	&	5006.1 	$\pm$	0.1 	&	2.6 	$\pm$	0.1 	&	776.8 	$\pm$	19.7 	&	5011.8 	$\pm$	0.1 	&	1.7 	$\pm$	0.1 	&	390.0 	$\pm$	17.9 	&	5 	&	21.2 	$\pm$	20.4 	&	319.3 	$\pm$	20.0 	\\ \hline
			0524-52027-0165	&	5002.4 	$\pm$	1.2 	&	4.3 	$\pm$	0.7 	&	143.0 	$\pm$	36.4 	&	5010.3 	$\pm$	0.6 	&	3.3 	$\pm$	0.3 	&	136.9 	$\pm$	35.8 	&	3 	&	-			&	-			\\ \hline
			0555-52266-0033	&	5004.4 	$\pm$	0.3 	&	2.8 	$\pm$	0.2 	&	227.4 	$\pm$	25.4 	&	5010.2 	$\pm$	0.3 	&	2.8 	$\pm$	0.2 	&	210.8 	$\pm$	26.4 	&	5 	&	-			&	-			\\ \hline
			0609-52339-0435	&	4997.7 	$\pm$	0.7 	&	5.8 	$\pm$	1.0 	&	93.6 	$\pm$	18.8 	&	5009.3 	$\pm$	0.2 	&	2.8 	$\pm$	0.2 	&	74.7 	$\pm$	9.7 	&	5 	&	387.7 	$\pm$	69.4 	&	311.2 	$\pm$	40.5 	\\ 
			\bottomrule
		\end{tabular} \\
		Notes. Column 1: Plate, MJD, Fiberid of spectroscopic observation; Column 2: the central wavelength of blue-shifted component of \oiii in units of \AA; Column 3: corresponding second moment of blue-shifted component of \oiii in units of \AA; Column 4: corresponding flux of blue-shifted component of \oiii~in units of $10^{-17}{\rm erg/s/cm^{2}}$; Column 5: the central wavelength of red-shifted component of \oiii in units of \AA; Column 6: corresponding second moment of red-shifted component of \oiii in units of \AA; Column 7: corresponding flux of red-shifted component of \oiii~in units of $10^{-17}{\rm erg/s/cm^{2}}$; Column 8: the  confidence level ($\sigma$) of F-test; Column 9 and Column 10: the velocity offset of blue-shifted component and red-shifted component of \oiii determined by shifted velocity from SSP method in units of km/s, respectively, and `-' represents an object with no obvious absorption line.
	\end{table*}
}

\begin{table*}[htbp] 
	\centering 
	\caption{Features of double-peaked narrow H$\alpha$ emission lines} 
	\label{tab3} 
	\begin{tabular}{c|c|c|c|c|c|c} 
		\toprule
		Plate-Mjd-Fiberid & $\lambda_{b}$  & $\sigma_{b}$ & $F_{b}$& $\lambda_{r}$  & $\sigma_{r}$ & $F_{r}$ \\ 
		(1)&(2)&(3)&(4)&(5)&(6)&(7)\\ 
		\midrule
		0907-52373-0295&6559.33$\pm$0.79&5.27$\pm$0.96&136.39$\pm$35.45&6568.32$\pm$0.31&2.37$\pm$0.35&60.61$\pm$15.78\\ \hline
		1048-52736-0416&6561.37$\pm$0.26&2.57$\pm$0.18&277.44$\pm$25.33&6567.45$\pm$0.26&2.34$\pm$0.18&231.5$\pm$25.08\\ \hline
		1679-53149-0532&6560.57$\pm$0.44&3.88$\pm$0.28&422.8$\pm$42.6&6568.22$\pm$0.36&2.94$\pm$0.25&264.44$\pm$44.67\\ \hline
		2022-53827-0553&6561.93$\pm$0.5&2.98$\pm$0.44&97.4$\pm$13.99&6568.42$\pm$0.3&1.94$\pm$0.24&71.24$\pm$12.52\\ \hline
		2365-53739-0359&6561.19$\pm$0.24&1.85$\pm$0.17&144.58$\pm$18.89&6566.42$\pm$0.26&2.25$\pm$0.2&195.86$\pm$20.08\\ \hline
		2791-54556-0005&6560.62$\pm$0.44&3.5$\pm$0.45&122.18$\pm$15.51&6567.76$\pm$0.34&1.75$\pm$0.33&45.76$\pm$11.71\\ \hline
		3830-55574-0154&6564.2$\pm$0.47&7.2$\pm$0.31&596.14$\pm$31.58&6572.06$\pm$0.25&2.45$\pm$0.4&63.97$\pm$20.11\\ \hline
		7723-58430-0620-&6563.42$\pm$0.34&5.97$\pm$0.23&616.97$\pm$29.47&6571.56$\pm$0.16&2.56$\pm$0.23&131.63$\pm$24.96\\ \hline
		7875-56980-0296&6561.96$\pm$0.13&1.87$\pm$0.15&35.65$\pm$3.05&6569.17$\pm$0.14&1.87$\pm$0.17&33.24$\pm$3.66\\ 
		\bottomrule
	\end{tabular} \\
	Notes. Column 1: Plate, MJD, Fiberid of spectroscopic observation; Column 2: the central wavelength of blue-shifted component of narrow H$\alpha$ in units of \AA; Column 3: corresponding second moment of blue-shifted component of narrow H$\alpha$ in units of \AA; Column 4: corresponding flux of blue-shifted component of narrow H$\alpha$ in units of $10^{-17}{\rm erg/s/cm^{2}}$; Column 5: the central wavelength of red-shifted component of narrow H$\alpha$ in units of \AA; Column 6: corresponding second moment of red-shifted component of narrow H$\alpha$ in units of \AA; Column 7: corresponding flux of red-shifted component of narrow H$\alpha$ in units of $10^{-17}{\rm erg/s/cm^{2}}$.
\end{table*}

\section{Basic properties of type 1 AGNs with DPNELs}
In this section, basic properties of the 62 type 1 AGNs with DPNELs in this work are measured and discussed.

\subsection{Black hole mass}
Based on the virialization assumption to emission clouds in broad line regions (BLRs) \citep{Pe04,Gr05,sh112,Ra11}, the virial black hole mass \citep{Gr05,An22} in this work can be determined in the 62 type 1 AGNs with double-peaked \oiii~by incorporating the established correlation between broad line region size (estimated via reverberation mapping) and continuum luminosity \citep{Be13}.
The equation \citep{Pe04,Gr05,Me22} can be described as:
\begin{equation}
\begin{split}	
\rm M_{BH}=2.67\times10^6(\frac{\rm L_{H\alpha}}{10^{42}\rm~erg /s})^{0.55}\\
\times(\frac{\rm FWHM_{H\alpha}}{10^3\rm~km/s})^{2.06}\rm M_\odot,
\end{split}
\end{equation}
Here, the equation is used due to consideration of the best match H$\beta$-related (RM) measurements.
In this equation, $\rm FWHM_{H\alpha}$ is the full width at half maximum,
and $\rm L_{H\alpha}$ is the luminosity of broad H$\alpha$ emission line after considering interstellar extinction given by \citet{Fi99} with assumption of the intrinsic Balmer decrement 3.1 for broad H$\alpha$ to broad H$\beta$.
The mean value of the estimated black hole masses for the 62 type 1 AGNs is about log$\rm M_{\rm BH}$/$\rm M_{\odot}$ $\sim$7.96$\pm$0.05, and the uncertainty is determined by the uncertainties of $\rm FWHM_{H\alpha}$ and $\rm L_{H\alpha}$. Meanwhile, the estimated black hole masses are listed in Table \ref{tab5} (complete results for all 62 objects shown in Table \ref{tabb5} in Appendix B).

\begin{figure*}
\centering\includegraphics[width=18cm,height=9cm]{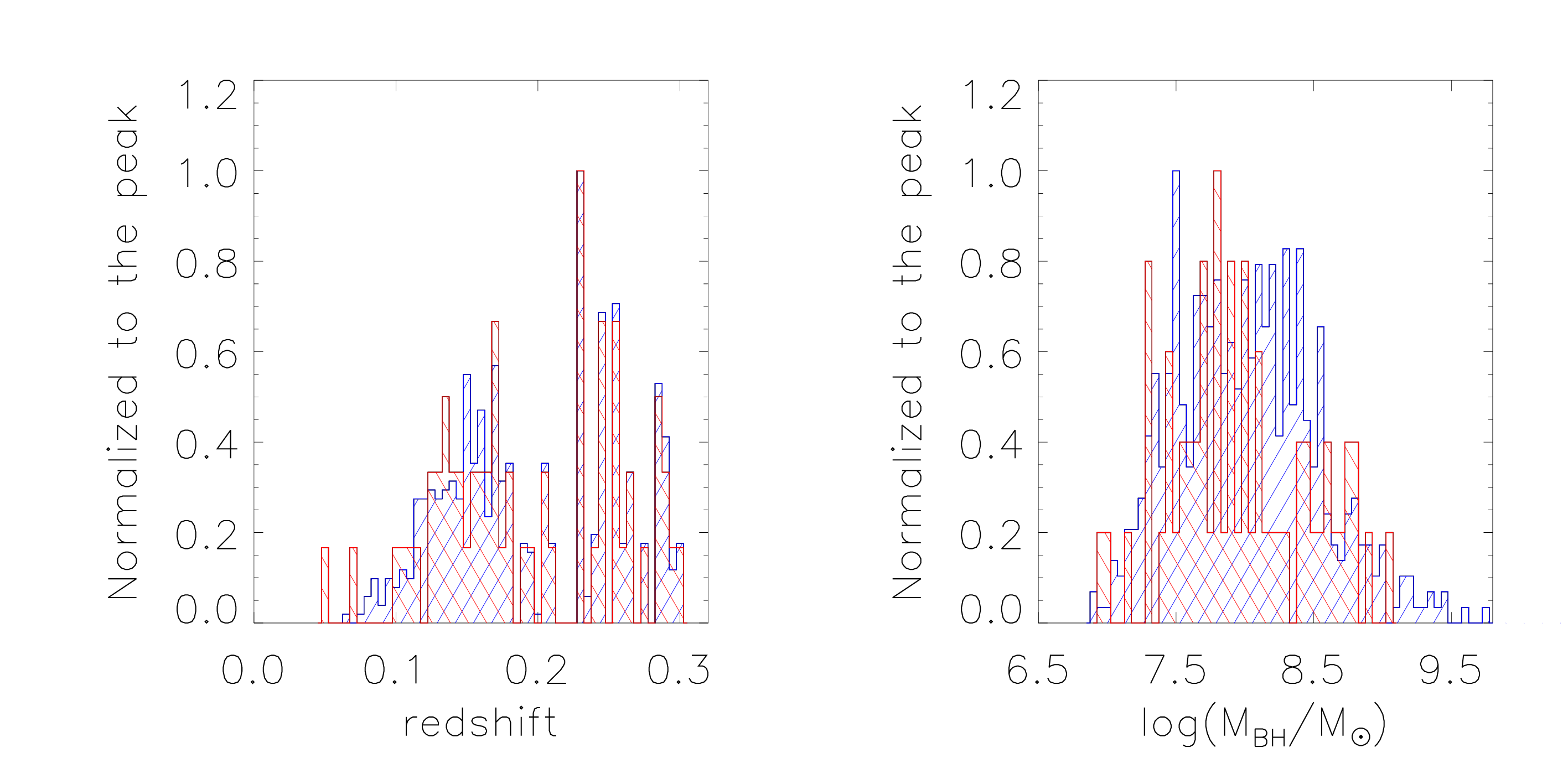}
\caption{Distributions of redshift (left panel), black hole masses $M_{\rm BH}$ (right panel) for the 558 typical type 1 AGNs from \citet{sh112} (histogram filled with blue line) and the 62 type 1 AGNs with double-peaked \oiii~in our sample (histogram filled with red line).
}
\label{new}
\end{figure*}

Then, it is possible to examine whether the black hole masses of our sample are different from black hole masses of normal type 1 AGNs with single-peaked narrow emission lines.
Here, a sample of normal type 1 AGNs is constructed from the \citet{sh112} dataset, which includes 105,783 QSOs from SDSS DR7. 
To minimize the influence of differing evolutionary histories, a new sample of 558 normal type 1 AGNs with single-peaked \oiii$\lambda$5007\AA~is randomly selected from the \citet{sh112} dataset, ensuring a redshift distribution similar to that of our sample.
According to a Kolmogorov–Smirnov (K-S) test \citep{Ko33,Sm48}, the redshift distributions of our sample and the newly constructed sample of 558 AGNs are statistically similar, with a probability of 99.99\%. The redshift distributions are displayed in the left panel of Figure \ref{new}.

Here, in order to ignore effects of the different equations applied on virial BH masses, the $M_{\rm BH}$ of the 558 objects is calculated by the same way based on equation (3) with the parameters of broad H$\alpha$ emission lines from Table 1 in \citet{sh112}, and the mean value is log$\rm M_{\rm BH}$/$\rm M_{\odot}$ $\sim$8.01$\pm$0.06.
Meanwhile, the distributions of the black hole masses $M_{\rm BH}$ of the 62 type 1 AGNs with double-peaked \oiii~in our sample and the 558 typical type 1 AGNs are shown in the right panel of Figure \ref{new}.
Then, the Student's t-test technique \citep{St08} is applied to check whether the 62 type 1 AGNs with DPNELs and the normal type 1 AGNs have significantly different mean values of virial black hole masses.
The 62 type 1 AGNs show the same mean value of $M_{\rm BH}$ with the 558 typical type 1 AGNs with the probability of 47.86\%.

\subsection{Radio-loudness}
The radio-loudness $R$ is a classical tool to indicate the power of AGN outflows related to the accretion flow \citep{Ke89, Gi23, ky23},
and the $R$ can be defined as  $R$=$\nu_{5GHz}L_{5GHz}/\nu_{opt}L_{opt}$ \citep{Si07},
where $L_{opt}$ represents the luminosity at 5100\AA~after
subtracting starlight (if present) and $L_{5GHz}$ is the luminosity at 5GHz determined by the integrated flux from the Faint Images of the Radio Sky at Twenty cm (FIRST) \citep{Be95,Wh97,He15}, and $\nu_{5GHz}$ and $\nu_{opt}$ are their corresponding frequencies. The radio fluxes can be obtained through SQL query as follows:
\lstset{
	breaklines=true,        
	breakatwhitespace=true
}
\begin{lstlisting}
SELECT s.ra, s.dec, s.z, s.snmedian,
F.integr
FROM specobjall  AS s
JOIN FIRST AS F
ON F.objID=s.bestobjid
WHERE 
class=`qso' and s.z<0.3 and s.zwarning=0 and s.snmedian>10
\end{lstlisting}
The SDSS provided public database of FIRST  (\url{https://skyserver.sdss.org/dr16/en/help/browser/browser.aspx?cmd=description+FIRST+U}) contains matched parameters of SDSS objects that match to FIRST objects.
The collected integrated radio fluxes are listed in Table \ref{tab1} (complete results for all 62 objects shown in Table \ref{tabb1} in Appendix B). 

There are 9 type 1 AGNs outside the footprint of the FIRST survey, 34 type 1 AGNs below the detection limit (radio flux as zero) and 19 type 1 AGNs with radio fluxes beyond zero in our final sample.	
For our final sample, the fraction of type 1 AGNs with radio fluxes in the FIRST footprint is 35\% (19/53), whereas this fraction is 16\% in our parent sample.
Typically, an object is classified as radio-loud if $R > 10$ \citep{Ke89}.
In total, there are 11 radio loud AGNs in our sample and the results of $R$ are listed in Table \ref{tab1} (complete results for all 62 objects shown in Table \ref{tabb1} in Appendix B).
However, the 34 type 1 AGNs in our sample with radio fluxes reported as zero do not necessarily lack radio fluxes. Some of these AGNs may have radio fluxes that fall below the FIRST detection limit, potentially due to the orientation of their outflows \citep{Sm10}.
In our final sample, the fraction of radio-loud type 1 AGNs in the FIRST footprint is 20\% (11/53).
For comparison, the fraction of radio-loud AGNs is 23\% in \citet{Ba13} sample, which shows double-peaked profiles in [Ne~{\sc v}] or [Ne~{\sc iii}], and this fraction in their parent sample is 10\%.

Both our sample and the \citet{Ba13} sample clearly indicate a preferential selection of radio objects among AGNs with DPNELs compared to the parent sample.
This suggests that the origin of the double-peaked profiles in narrow emission lines may be associated with the presence of radio jets in some objects.


\subsection{Photometric image properties}
\begin{figure*}
	\centering\includegraphics[width=15cm,height=16cm]{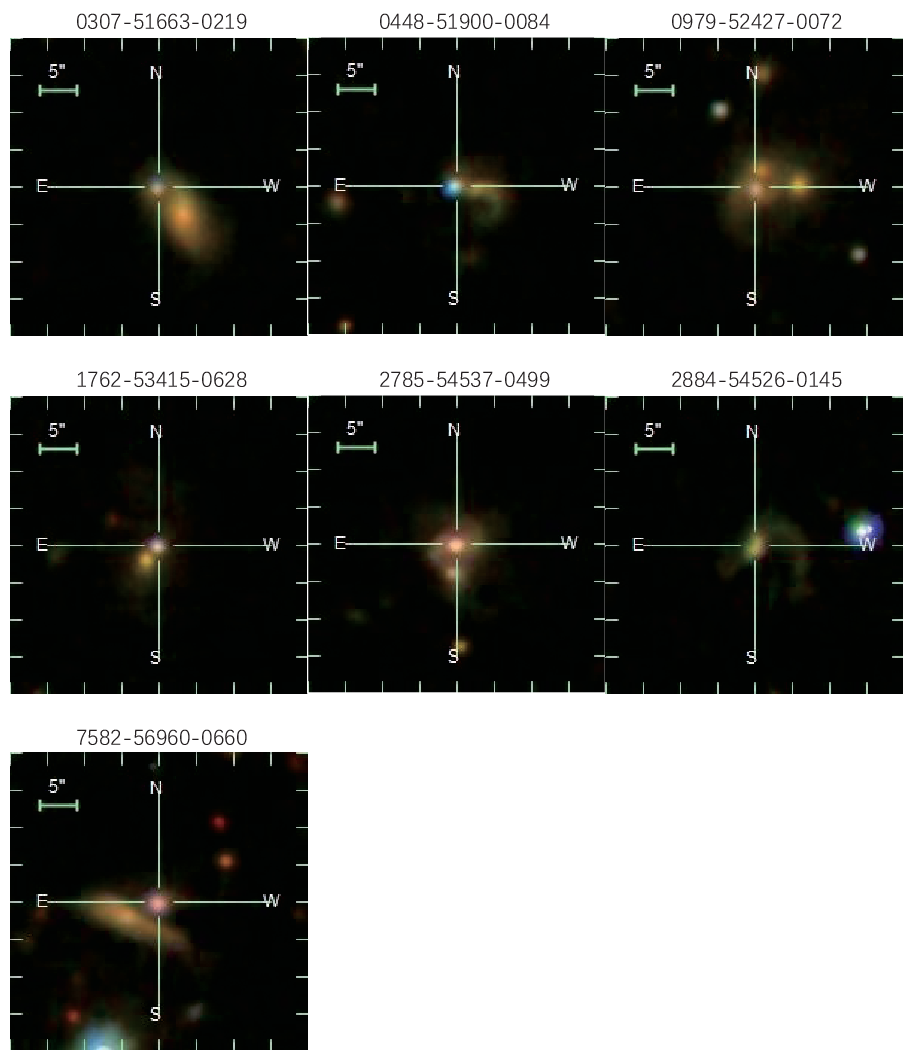}
	\caption{Photometric image properties of the seven type 1 AGNs with double-peaked \oiii.}
	\label{phot}
\end{figure*}

If the double-peaked \oiii emission lines were the results of dual AGN systems, it is expected for dual-cored galaxies or merging galaxies in the images. We visually inspect the SDSS images of the 62 type 1 AGNs in our sample and the SDSS multi-color images of seven visually interesting type 1 AGNs are shown in Figure \ref{phot}.

SDSS J143701.20-010417.9 (Plate-Mjd-Fiberid: 0307-51663-0219) may be a major merger between two galaxies, and the projected distance between two cores is about $4.7^{\prime \prime}$.
The galaxies are clearly undergoing interaction, while the scope of fiber position is not large enough to encompass the NLRs of both galaxies, after considering the $3^{\prime \prime}$ in diameter of SDSS fibers.

SDSS J085457.23+544820.5 (Plate-Mjd-Fiberid: 0448-51900-0084) and SDSS J161950.67+500535.3 (Plate-Mjd-Fiberid: 2884-54526-0145) do not appear two cores, but large tidal tails in the southwest in the image.

SDSS J172120.48+263658.4 (Plate-Mjd-Fiberid: 0979-52427-0072) exhibits three cores in the image and the primary galaxy shows a large tidal tail. The NLR of the companion galaxy in the west of the primary galaxy should be out of the scope of fiber position (the projected distance between the two cores about $6.6^{\prime \prime}$), while the NLR of the companion galaxy in the north of the primary galaxy may be covered by the scope of fiber position (the projected distance between the two cores about $3.0^{\prime \prime}$). 

SDSS J115638.15+150500.9 (Plate-Mjd-Fiberid: 1762-53415-0628) exhibits two cores with projected distance about $2.3^{\prime \prime}$ in the image. The companion galaxy in the southeast is close to the primary galaxy, and its NLR may be encompassed by the scope of fiber position.

SDSS J141041.50+223337.0 (Plate-Mjd-Fiberid: 2785-54537-0499) shows two cores with projected distance about $4.0^{\prime \prime}$ in the image, but the scope of fiber position is not large enough to encompass the NLR of the companion galaxy.

SDSS J222518.67+210203.6 (Plate-Mjd-Fiberid: 7582-56960-0660) shows a long and narrow companion galaxy in the southeast of the primary galaxy, and the projected distance between two cores is about $4.8^{\prime \prime}$.

\subsection{Properties of~ \rm\oiii}
\begin{figure*}
\centering\includegraphics[width=18cm,height=12cm]{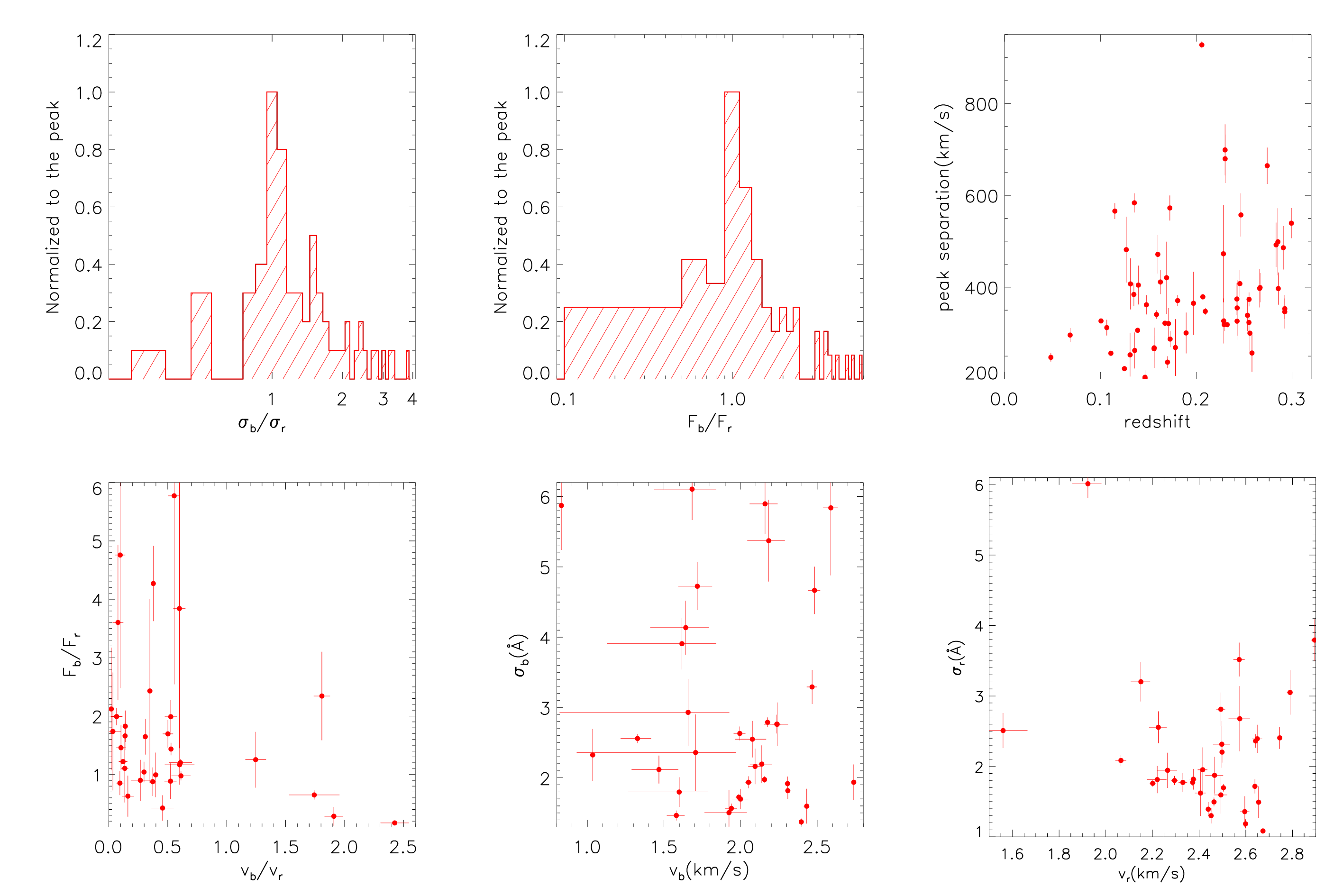}
\caption{Top-left panel: the distribution of line width of the blue-shifted/red-shifted \oiii~components ($\sigma_{b}/\sigma_{r}$); Top-middle panel: the distribution of line flux ratio ($F_{b}/F_{r}$);
Top-right panel: the relationship between peak separations and redshifts;	
Bottom-left panel: the relationship between $V_{b}/V_{r}$ and $F_{b}/F_{r}$;
Bottom-middle panel: the relationship between $V_{b}$ and $\sigma_{b}$;
Bottom-right panel: the relationship between $V_{r}$ and $\sigma_{r}$.
}
\label{fig12}
\end{figure*}

The top-left panel of Figure \ref{fig12} shows the distribution of the line width ratios of blue-shifted components to red-shifted components of \oiii with the mean value about 1.47$\pm$0.27. 
The top-middle panel of Figure \ref{fig12} shows the distribution of the line flux ratios of blue-shifted components to red-shifted components of \oiii with the mean value about 1.63$\pm$0.47.

The top-right panel of Figure \ref{fig12} shows the relationship between the redshifts and the peak separations of our double-peaked AGNs, and the Spearman Rank correlation coefficient is 0.36 with null-hypothesis probability of 0.44\%.
Objects at greater distance generally exhibit higher luminosities, and more luminous sources tend to have broader \oiii~lines \citep{Sa07}. Consequently, larger peak separations are required to resolve double-peaked profiles. Therefore, it is reasonable to expect an increase in observed peak separation with redshift. Our sample aligns with this trend, showing that at higher redshift, the typical peak separation tends to increase.


According to Kepler's law and the Magorrian relation \citep{Ma98,Ma03}, the velocity offsets of \oiii~blue-shifted/red-shifted components ($V_{b}/V_{r}$) have a negative dependence on their relative fluxes ($F_{b}/F_{r}$) for a dual AGN system, and it has been confirmed in \citet{Wa09} through a study of type 2 AGNs with double-peaked \oiii.
In the dual AGN scenario, the brighter shifted component, which is assuming typically from the more massive black hole, tends to be closer to the rotating center and has the lower orbital velocity.
Since there are 26 type 1 AGNs with double-peaked \oiii~in our sample showing no host galaxy contributions, it is hard to calibrate the precise redshift and then calculate the velocity offset of each component with respect to the absorption lines.
The remaining 36 type 1 AGNs in our sample is considered to test the relationship between $V_{b}/V_{r}$ and $F_{b}/F_{r}$. 
Here, the $V_{b}$=$\mid$($\lambda_b$-5008.24)$\times$3$\times10^5$/5008.24+$V_s$$\mid$ and $V_{r}$=$\mid$($\lambda_r$-5008.24)$\times$3$\times10^5$/5008.24+$V_s$$\mid$
are measured, where $V_s$ is the shifted velocity determined by SSP method, and $\lambda_b$ ($\lambda_r$) is the central wavelength of blue-shifted (red-shifted) component of \oiii.
The relationship between $V_{b}/V_{r}$ and $F_{b}/F_{r}$ is shown in the bottom-left panel of Figure \ref{fig12}, and the Spearman Rank correlation coefficient is -0.34 with null-hypothesis probability of 4.23\% after accepting the redshift calibrated by the shifted velocity from the SSP method is accurate.

The bottom-middle panel of Figure \ref{fig12} shows the relationship between the velocity offsets and line widths of blue-shifted components of double-peaked \oiii~in our sample, and the Spearman Rank correlation coefficient is -0.03 with null-hypothesis probability of 86.93\%.
The bottom-right panel of Figure \ref{fig12} shows the relationship between the velocity offsets and line widths of red-shifted components, and the Spearman Rank correlation coefficient is -0.08 with null-hypothesis probability of 64.73\%.
The velocity offset is almost independent of line width, which will provide clues on discussions on AGN outflow model to explain the double-peaked profiles.

\begin{figure*}
\centering\includegraphics[width=18cm,height=10.5cm]{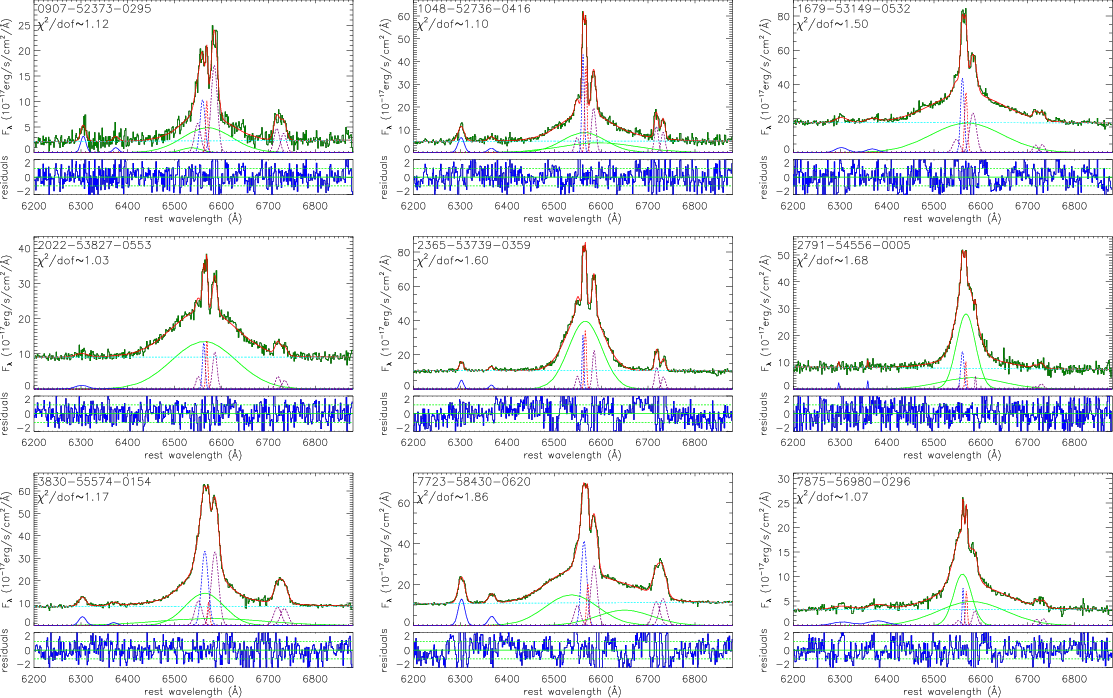}
\caption{The best fitting results of the emission lines around double-peaked narrow H$\alpha$ emission lines. 
In the top of each panel, the solid dark green line shows the line spectrum after subtracting starlight (if present) determined from host galaxy with Plate-Mjd-Fiberid shown in the top-left of corner, the solid red line represents the best fitting results with $\chi^2/\rm dof$ shown in the top-left of corner, the dashed cyan line represents continuum emission, the dashed purple lines represent [N~{\sc ii}] and [S~{\sc ii}] doublets, the solid green lines represent broad H$\alpha$ emission line, and the dashed blue line and red line represent the blue-shifted and red-shifted components of narrow H$\alpha$, respectively. 
In the bottom of each panel, the solid blue line represents the residuals calculated by the line spectrum minus the best fitting results and then divided by uncertainties of SDSS spectrum. The horizontal solid and dashed green lines show residuals=0,±1, respectively.
}
\label{fig06}
\end{figure*}

\section{DISCUSSION}
In this section, the basic properties of double-peaked \oiii~in the previous sections are synthesized to analyze the probable physical scenarios that produce double-peaked \oiii, such as superposition, AGN outflows, dual AGN and rotating disk.
\subsection{Superposition}
In this scenario, along the same line-of-sight, the blue-shifted emission lines originate in the foreground QSO, while the red-shifted emission lines are associated with another background object \citep{De14}.
However, it is hard for two systems with small separation in spatial and velocity to be unrelated.
\citet{Ko08} thought the possibility of a chance superposition is considerably improbable.
\citet{Sh09} estimated that the probability of superposition within 1 arcsec (observed in the same SDSS fiber) of two systems with velocity offset less than 2650 $\rm km/s$ is about $10^{-8}$ for a given QSO.
In order to increase the almost negligible probability of superposition, the QSO is assumed to superimposed within rich galaxy clusters \citep{He09,Sh09}.
Based on this assumption, \citet{Sh09} proposed that the probability of a chance superposition within 1 arcsec (observed in the same SDSS fiber) is $10^{-4.3}$, and then this prediction is confirmed by \citet{Do10}.

Considering the 11557 AGNs in our parent sample, it is hard to find even one object with a chance superposition due to the low probability.
But it is hard to totally rule out the superposition scenario, in the near future, detecting SDSS spectrum with potential two sets of absorption line systems should be an interesting objective to support probable superposition.

\subsection{AGN outflows}
AGN outflows \citep{Du88,Wh04,Ya21} are able to produce double-peaked profiles. 
Powerful AGNs are accepted to be capable of driving large-scale outflows.
NGC 1068 (Seyfert 2) \citep{Ax98,Ma17} is a prototypical case which has double-peaked emission lines caused by outflows.
In the scenario of bi-conical outflows, the blue-shifted and red-shifted components of double-peaked \oiii~profiles are determined by the projection of outflows moving toward and away from the observer. 
As shown in Table \ref{tab1} and Table \ref{tabb1} in Appendix B, there are 11 radio loud AGNs in our sample based on the calculated $R$, and the fraction of the type 1 AGNs with radio fluxes in our final sample is two times larger than the parent sample.

If generally accepted that peak separation in NLR is produced by outflows, the NLR around the AGNs should be stratified.
If the double-peaked profiles of narrow emission lines originate from distinct regions, it would naturally result in variations in peak separations and line widths.
In another words, in the radially decelerating outflow of NLR, the emission lines produced closer to the AGNs would exhibit higher velocity offsets, and more broadened line widths, which is dominated by the bulge gravitational potential \citep{Ba13}.
The correlation between peak separation and line width can be found in \citet{Ko08,Liu10}.
By contrast, there is a very weak correlation between line widths and velocity offsets in both blue-shifted and red-shifted components of \oiii, as shown in Figure \ref{fig12}.
So the physical explanations for the velocity offsets in our sample are likely different from the samples in \citet{Ko08,Liu10}.

When the outflow axis is oriented at an intermediate angle between edge-on and face-on, the red-shifted component is expected to be more attenuated than the blue-shifted component. 
The portions of the NLR with the highest line-of-sight velocity components are likely to be the most heavily obscured, while those with the smallest line-of-sight velocity components experience relatively less obscuration, unless there is asymmetrical obscuration. 

According to the top-middle panel of Figure \ref{fig12}, it appears that the mean value of line flux of the blue-shifted components is generally larger than that of the red-shifted components, consistent with the notion above.
However, the $V_{b}$ and $V_{r}$ should be similar in the symmetrical outflows, which is inconsistent with the negative correlation between $V_{b}/V_{r}$ and $F_{b}/F_{r}$.



\subsection{Dual AGN}
It is possible that AGNs may host two black holes following a galaxy merger, through tidal interactions, and galaxy merger is effective at enhancing nuclei activity by funneling dust and gas toward potential of each black hole \citep{Co09,Ba12}. In this scenario, each nucleus may possess its own NLR and the double peak may be from two distinct NLRs, which will move with its own active black hole.
Therefore, it will be expected that there are two sets of narrow emission lines from two distinct NLRs and each set of narrow emission lines has similar redshift. So we recheck the narrow H$\alpha$ emission line in our sample and find nine objects exhibiting obvious double peak. The spectra of the nine objects around H$\alpha$ emission line are refit and the fitting parameters are done similar as before, excepting two Gaussian functions applied to describe narrow H$\alpha$ emission lines.
The fitting results are shown in Figure \ref{fig06} and the corresponding model parameters of the emission lines are listed in Table \ref{tab3}, and all the nine objects show similar velocity offsets in blue-shifted (red-shifted) components of narrow H$\alpha$ and \oiii narrow emission lines. Here, the object with offset between \oiii and narrow H$\alpha$ emission line in velocity space less than 100 km/s is considered as having similar velocity offset.

Meanwhile, if some narrow emission lines in one set are obscured and only the other set of narrow emission lines can be fully observed, the single-peaked narrow emission line should have same velocity offset with one of the double peak of \oiii.
Therefore, we put single-peaked H$\alpha$ emission line and double-peaked \oiii~in velocity space and found 31 objects exhibit coincident velocity offsets in H$\alpha$ and either of blue-shifted and red-shifted components of \oiii emission lines, and 22 objects show different velocity offsets between H$\alpha$ and both blue-shifted and red-shifted components of \oiii.
Similarly, the object with offset between one set of \oiii and narrow H$\alpha$ emission line in velocity space less than 100 km/s is considered as having similar shift.
Here, we show an example of the 53 objects with single-peaked narrow H$\alpha$ emission line in Figure \ref{fig18}, and the best fitting results to all the 53 targets are shown in Figure \ref{figure18} in Appendix A.

Meanwhile, there are also two objects 
showing tidal tails, which might be caused by merger, four objects showing clear two cores and one object showing three cores in the images. However, For the objects (Plate-Mjd-Fiberid: 0307-51663-0219, 2785-54537-0499 and 7582-56960-0660), the two cores of the objects are so far that the scope of fiber position can hardly cover both NLRs of two cores.
Even for the objects (Plate-Mjd-Fiberid: 0979-52427-0072 and 1762-53415-0628) with NLRs of two nearest cores covered in the scope of fiber position, the double-peaked \oiii~also can be produced by other models related to a single AGN rather than a dual AGN, as discussed in \citet{ZZ23,Ma23}.

Although most sources do not show two cores or merger in the images, they cannot be completely ruled out as dual AGNs.
In our case, with a mean redshift of z$\sim$0.195, the 1-arcsec limit for unresolved objects corresponds to approximately 3.3 kpc. This suggests that spatial resolution constraints may lead to some dual cores being optically unresolved in images.

Moreover, the profiles of broad H$\alpha$ emission lines in our sample might be determined by two broad line regions in dual AGNs.
So the line widths and line fluxes of broad H$\alpha$ emission lines in dual AGNs tend to be broader and larger than those in normal type 1 AGNs, and then it is expected to see larger black hole masses of dual AGNs than normal AGNs. 
However, through Student's t-test statistical technique, we find no significant difference in black hole masses between type 1 AGNs with DPNELs in our sample and normal type 1 AGNs in \citet{sh112}, which is at odds with dual AGN systems.

\begin{figure*} \centering\includegraphics[width = 15cm,height=9cm]{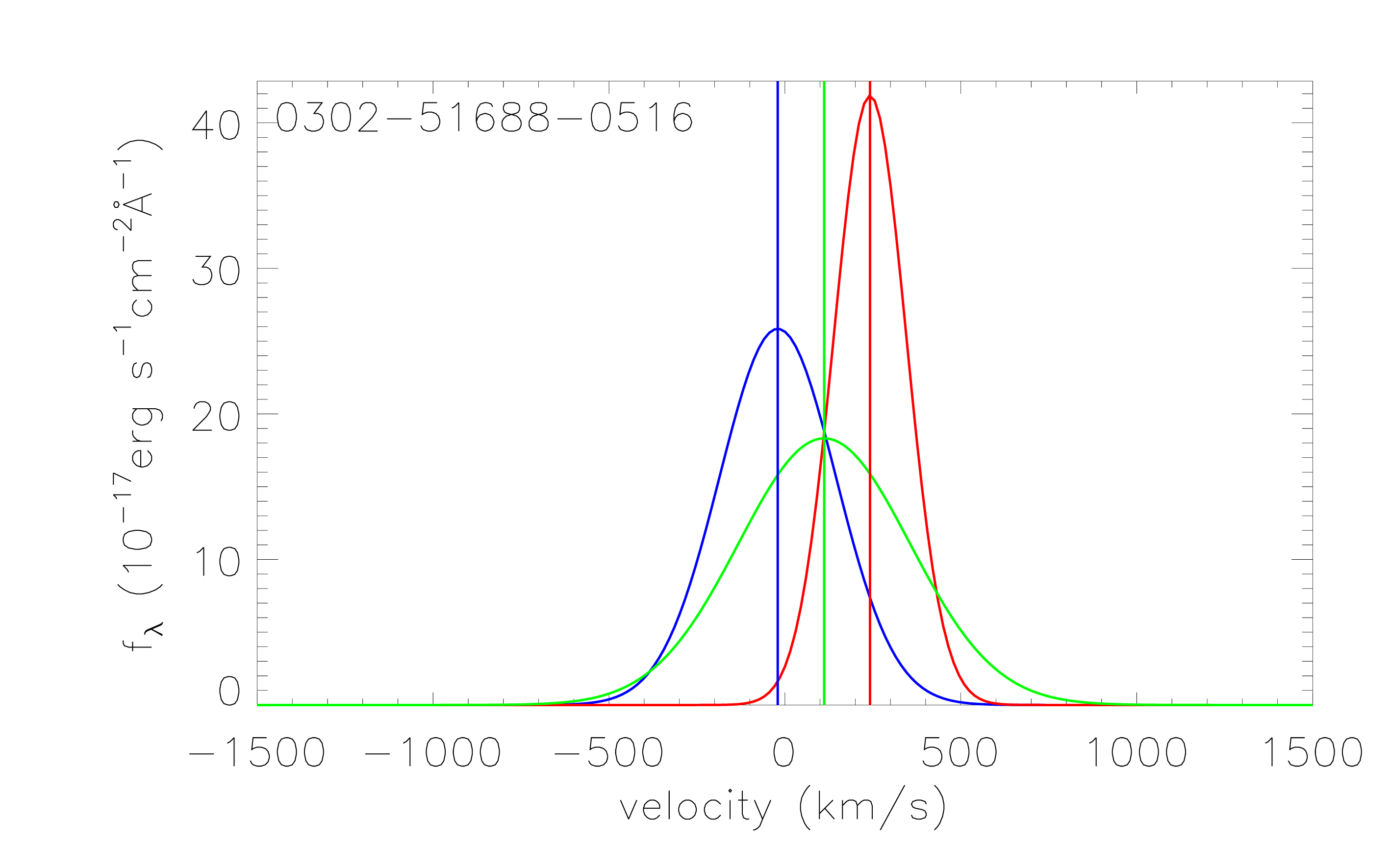} \caption{An example of the offsets of \oiii and narrow H$\alpha$ emission lines the object with single-peaked narrow H$\alpha$ emission lines in our sample in velocity space. The solid blue line represents the determined blue-shifted component of \oiii, and the vertical blue line marks the position of the central wavelength of the blue-shifted component of \oiii, the solid red line represents the determined red-shifted component of \oiii, and the vertical red line marks the position of the central wavelength of the red-shifted component of \oiii, the solid green line shows the determined narrow H$\alpha$ component, and the vertical green line marks the position of the central wavelength of the narrow H$\alpha$. 		
} \label{fig18} \end{figure*}


\subsection{Rotating disk}
In purely geometrical terms, the model is consisted of a flat extended disk component plus a spherical core component \citep{Xu09}.
The blue-shifted and red-shifted narrow emission lines would arise in the classical NLR, which follows a strict disk geometry \citep{gh05a} and the extended emission lines, if exist, are produced in spherical very inner part of the NLR (or outer broad line region).
The shape of the lines and relative positions of the peaks are dependent on the orientation of the disk plane and the disk outer radius \citep{Ba12}.
Based on axisymmetric models of rotating disk, \citet{Ma23} found that the double-peaked profiles primarily depend on the observation angle: as the inclination increases, the separation between the peaks becomes larger.
As discussed in \citet{Sm12}, two peaks of the \oiii~line with nearly equal intensities often represent rotating disks.
In addition, the great similarity of the line flux ratio (0.75$\leq$$F_{b}/F_{r}$$\leq$1.25, as shown in \citealp{Sm12}) in the blue-shifted and red-shifted narrow emission lines hints only a single ionizing continuum and similar physical condition (e.g., cloud densities, column densities and metal abundances) in the NLR.

However, the symmetric rotating disk model cannot account for the larger mean line flux of the blue-shifted components compared to the red-shifted components in our sample.
This case, however, can be explained by a lopsided disk, as found in \citet{Ga03}.
Our findings do not rule out the possibility of an asymmetric disk configuration.



\subsection{future applications}

Although the cause of the double-peaked \oiii~can be initially judged according to the current results, different models are not completely at odds with each other. Therefore, it is hard to give a unique explanation for double-peaked \oiii~of each object. But at least the 22 objects in our sample with different velocity offsets between double-peaked \oiii~and narrow H$\alpha$ emission lines could be excluded as dual AGN candidates.
If possible, more observational information, such as high-resolution imaging and long-slit spectroscopy should be obtained to provide further information on the NLRs in order to determine the cause of the double-peaked \oiii~properties in the future.

Future work will focus on investigating differences in long-term variability between the 62 type 1 AGNs with double-peaked \oiii in our sample and normal type 1 AGNs. These analyses aim to provide deeper insights into the physical mechanisms responsible for the observed double-peaked features and their connections to AGN variability and orientation effects.

The new sample presented in this work provides a significant update to the catalog of type 1 AGNs with double-peaked \oiii, introducing 47 previously unreported objects. This expanded sample could serve as an important complement to studies of type 2 AGNs with double-peaked \oiii (selected using similar criteria), enabling a more comprehensive investigation of AGN activity.

Additionally, a separate sample of 25 type 1.9 AGNs (AGNs with broad H$\alpha$ but no broad H$\beta$ emission line due to extinction) is excluded during the fifth step in Section 2. This sample holds potential for studying the effects of line-of-sight orientation on the observed double-peaked \oiii profiles.




\section{Summary and Conclusions}
We systematically select candidates of type 1 AGNs with double-peaked \oiii of spectra through strict criteria from 11557 QSOs at $z<0.3$ in SDSS DR16.
After visual check and Gaussian fitting of \oiii, fitting of broad H$\alpha$ emission lines, F-test for \oiii~emission lines and check of broad H$\beta$ and \oiii~emission lines, a sample of 62 type 1 AGNs with double-peaked \oiii is built.
The main summary and conclusions are shown as follows.
\begin{itemize}
\item The fraction of type 1 AGNs with double-peaked \oiii~in the final sample is about 0.5\% of the parent QSO sample. 
\item Among the 62 type 1 AGNs, there are 26 objects without host galaxy contributions and 36 objects showing host galaxy contributions.
\item The Fe~{\sc ii} emission lines around \oiii~emission lines may affect our sample selection, and the components of Fe~{\sc ii} emission lines should be determined to avoid the influence on double-peaked \oiii.
\item The mean value of the black hole masses for the 62 type 1 AGNs is about log$\rm M_{\rm BH}$/$\rm M_{\odot}$ $\sim$7.96$\pm$0.05, and the result is statistically similar to normal type 1 AGNs, which is inconsistent with dual AGN model.
\item There are 35\% of type 1 AGNs with double-peaked \oiii in the FIRST footprint detected with radio fluxes, while only 16\% of parent sample is detected with radio fluxes. In addition, 11 type 1 AGNs in our final sample are radio loud AGNs.
\item There are seven AGNs exhibiting sign of merger with four objects showing two cores, one object showing three cores and two objects showing tidal tails in the images.
\item The line widths and line fluxes of the blue-shifted components are larger than these of red-shifted components.
\item There is a trend in our sample that the observed peak separations increase with redshifts, and the Spearman Rank correlation coefficient is 0.36.
\item Considering the negligible probability of superposition model, it is hard to find the double-peaked \oiii~in our sample produced by chance superposition.
\item There is a very weak correlation between the line widths and velocity offsets in both blue-shifted and red-shifted components of \oiii, which disfavors AGN outflow model.
\item There is a negative relationship between $V_{b}/V_{r}$ and $F_{b}/F_{r}$ with the Spearman Rank correlation coefficient of -0.34, which is consistent with dual AGN model.
\item Among the 62 type 1 AGNs with double-peaked \oiii, there are nine type 1 AGNs with double-peaked profiles in both \oiii and H$\alpha$ emission lines, and all of the nine AGNs show similar velocity offsets between blue-shifted (red-shifted) components of \oiii and H$\alpha$ emission lines. In addition, in the remaining 53 type 1 AGNs with single-peaked narrow H$\alpha$ emission lines, 31 AGNs exhibit coincident velocity offsets in narrow H$\alpha$ and one shifted component of \oiii, and 22 AGNs show different velocity offsets between H$\alpha$ and double-peaked~\oiii, which can be excluded as dual AGN candidates.
\item The similar line flux ratio in the blue-shifted and red-shifted components of \oiii~may hints only a single ionizing continuum (symmetric rotating disk model), and the larger mean value of line flux of the blue-shifted component than that of red-shifted component in our sample would be possible in the asymmetric case.
\end{itemize}
The current findings do not strongly favor any single explanation, and additional observational data would be beneficial to better understand the details of NLRs.
This work offers a significant update to the sample of type 1 AGNs with double-peaked \oiii, introducing 47 newly identified type 1 AGNs.

\section*{Acknowledgments}
This work is supported by the National Natural Science Foundation of China (Nos. 12173020, 12373014, 12273013).
This manuscript has made use of the data from the Sloan Digital Sky Survey (SDSS).
The SDSS DR16 web site is (http://skyserver.sdss.org/dr16/en/home.aspx). 
The MPFIT website is (\url{http://cow.physics.wisc.edu/~craigm/idl/idl.html}).
The FIRST website is (\url{http://sundog.stsci.edu/index.html}).

\appendix
\section{Spectra of the 62 type 1 AGNs}
The SDSS spectra of the 62 type 1 AGNs in our final sample are shown in Figure \ref{figure11}, and the best fitting results of the emission lines around H$\alpha$ and \oiii are shown in Figure \ref{figure12} and \ref{figure13}, respectively.
The offsets of \oiii and narrow H$\alpha$ emission lines in velocity space are shown in Figure \ref{figure18}.

\begin{figure*} \centering\includegraphics[width = 18cm,height=21cm]{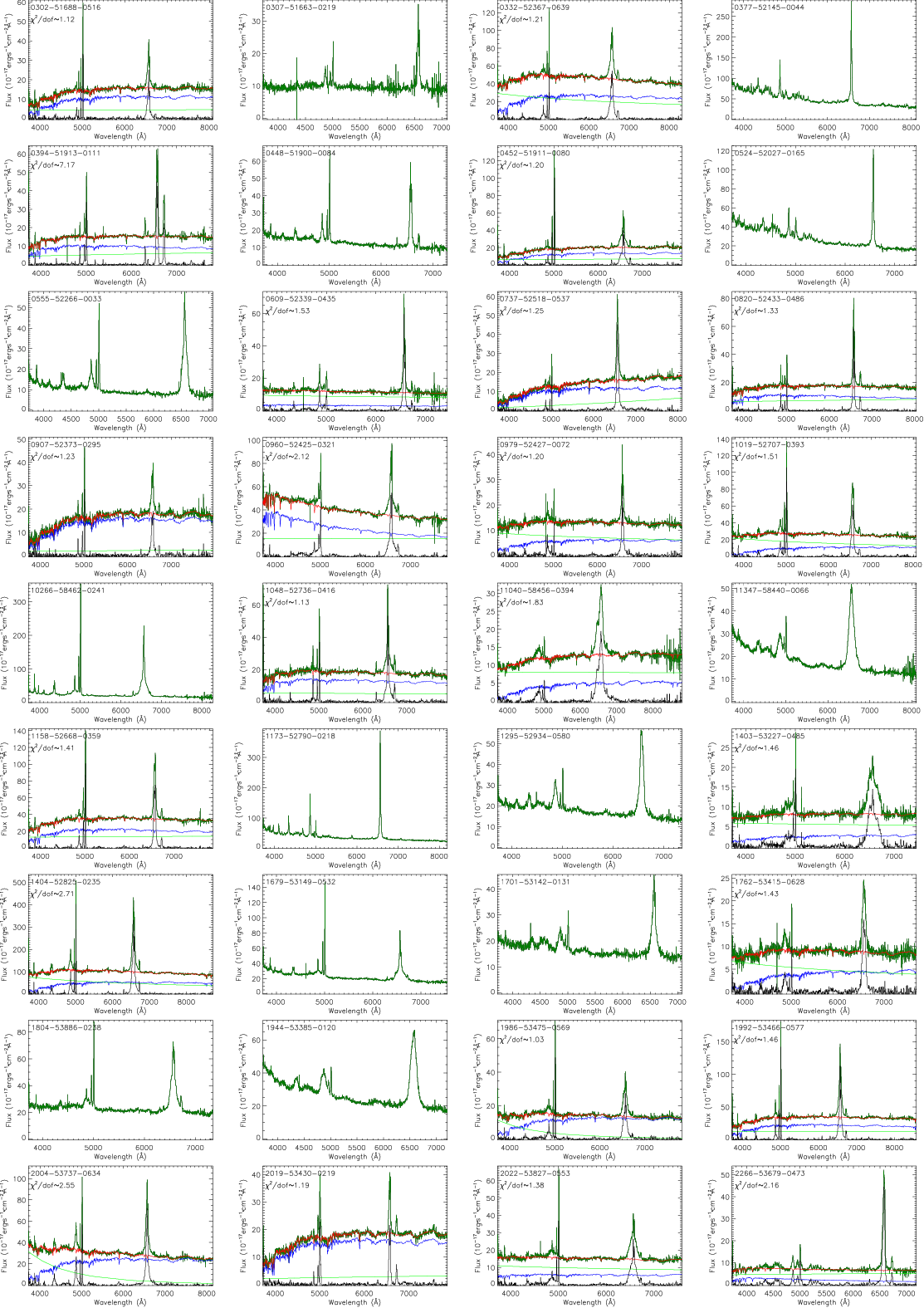} \caption{The SDSS spectra of the 62 type 1 AGNs in the final sample. 
In each panel, the solid dark green line represents the spectrum with Plate-Mjd-Fiberid shown in the top-left of corner, the solid red line shows the best fitting results with $\chi^2/\rm dof$ (if present) shown in the top-left of corner, the solid blue line shows starlight (if present), the solid green line represents the AGN continuum emission, and the solid black line represents the line spectrum calculated by the SDSS spectrum minus the best fitting results (if present).
} \label{figure11} \end{figure*} \setcounter{figure}{11} 
\begin{figure*} \centering\includegraphics[width = 18cm,height=16cm]{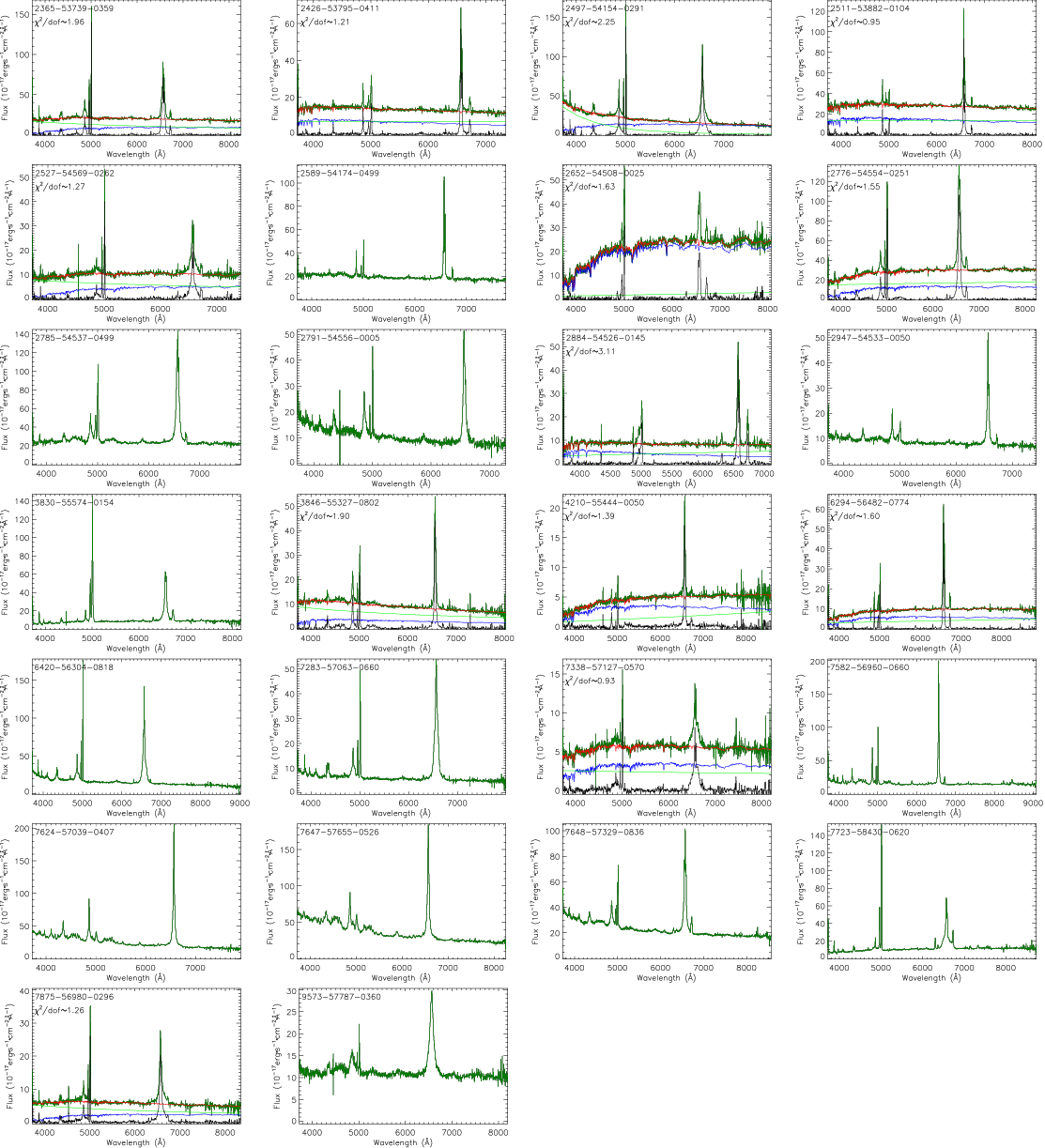} \caption{-- to be continued} \label{figure11} \end{figure*}

\begin{figure*} \centering\includegraphics[width = 18cm,height=21cm]{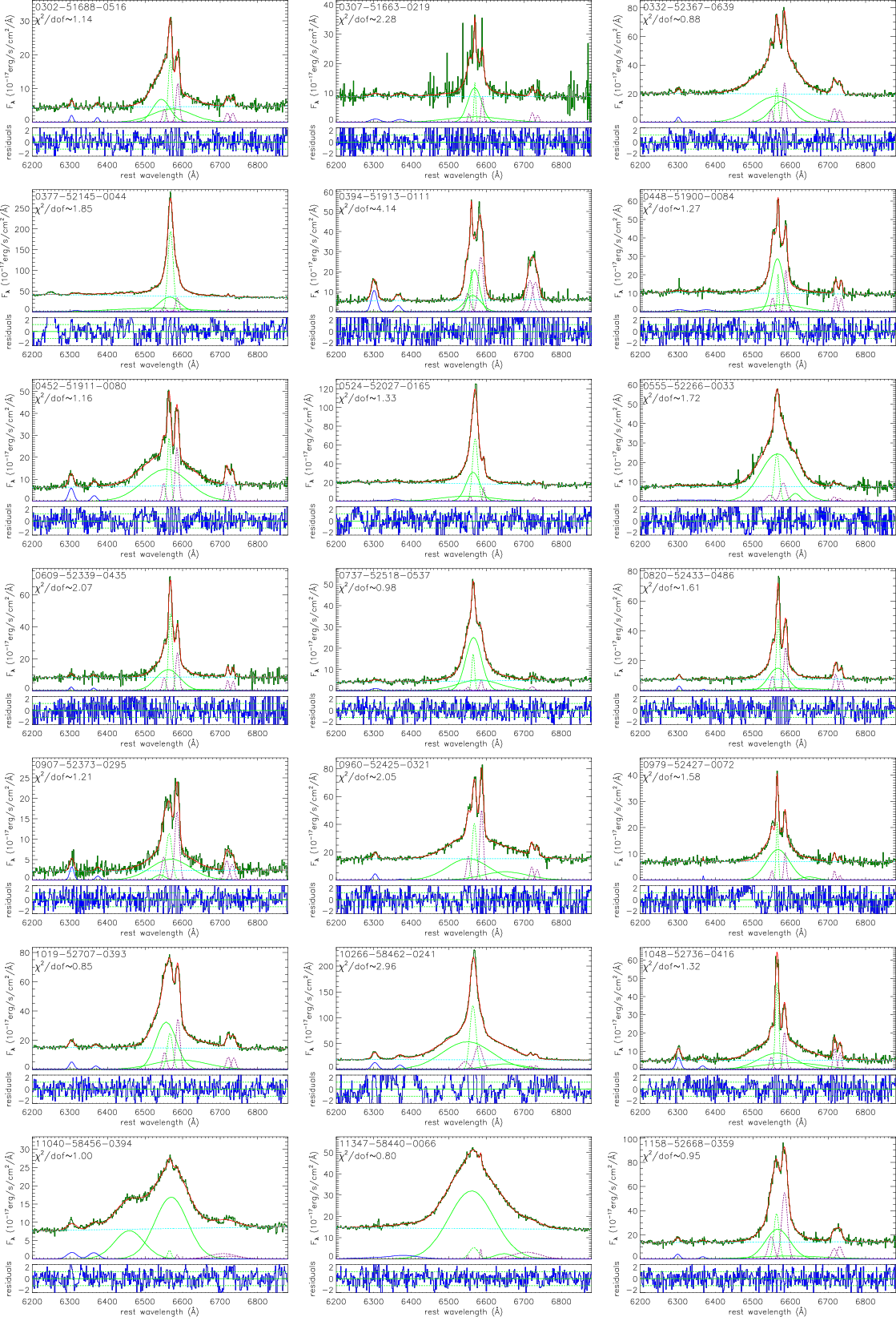} \caption{The best fitting results of the emission lines around H$\alpha$.
In the top of each panel, the solid dark green line shows the line spectrum after subtracting starlight (if present) from host galaxy with Plate-Mjd-Fiberid shown in the top-left of corner, the solid red line represents the best fitting results with $\chi^2/\rm dof$ shown in the top-left of corner, the dashed cyan line represents continuum emission,
the solid green lines represent the determined components of broad H$\alpha$ emission line, the dashed green line represents narrow H$\alpha$ emission line, the solid blue lines represent [O~{\sc i}] emission lines (if present), and the dashed purple lines represent [N~{\sc ii}] and [S~{\sc ii}] doublets. 
In the bottom of each panel, the solid blue line represents the residuals calculated by the line spectrum minus the best fitting results and then divided by uncertainties of SDSS spectrum, the horizontal solid and dashed green lines show residuals=0, ±1, respectively.} \label{figure12} \end{figure*} \setcounter{figure}{12}
\begin{figure*} \centering\includegraphics[width = 18cm,height=21cm]{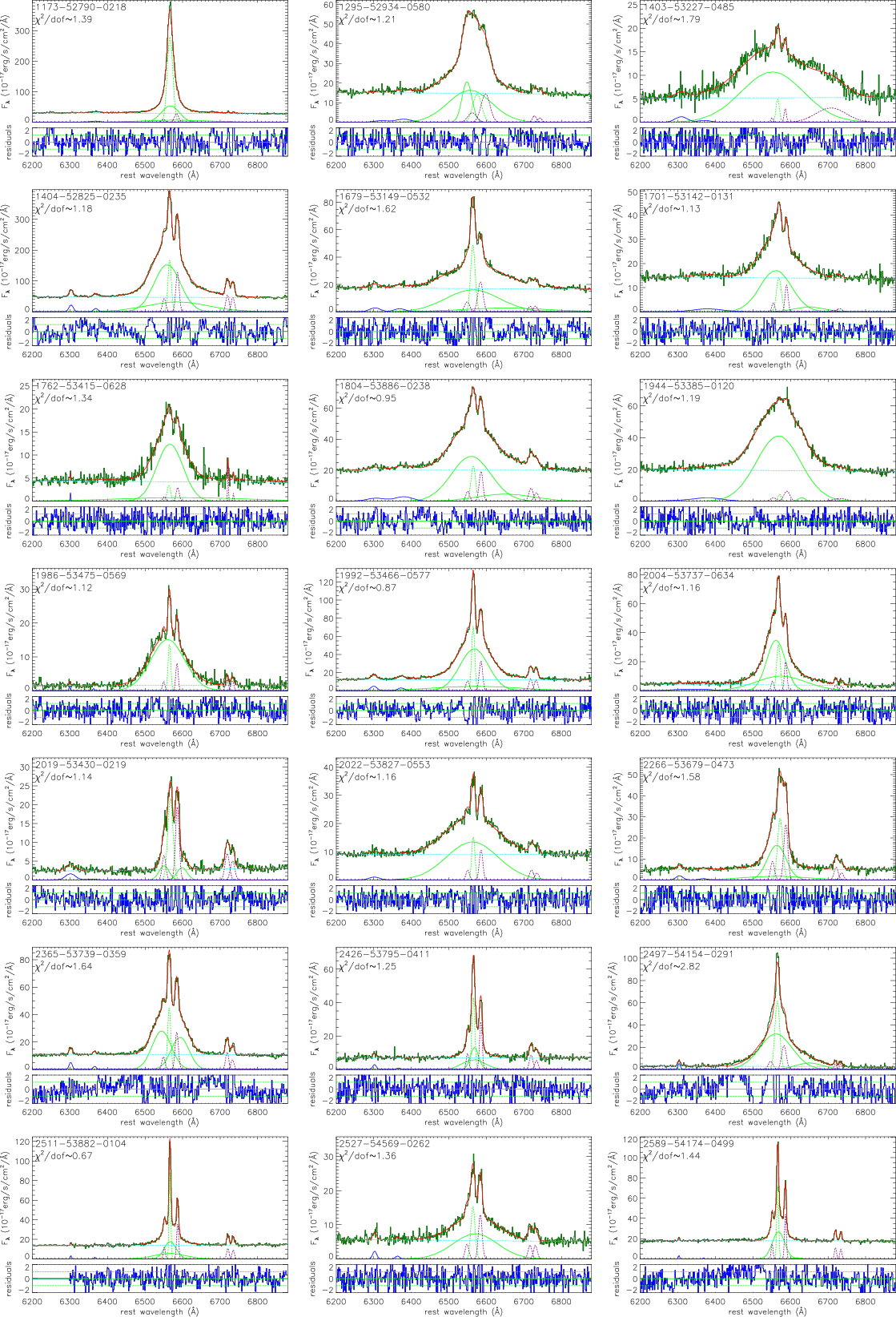} \caption{-- to be continued} \label{figure12} \end{figure*} \setcounter{figure}{12}
\begin{figure*} \centering\includegraphics[width = 18cm,height=21cm]{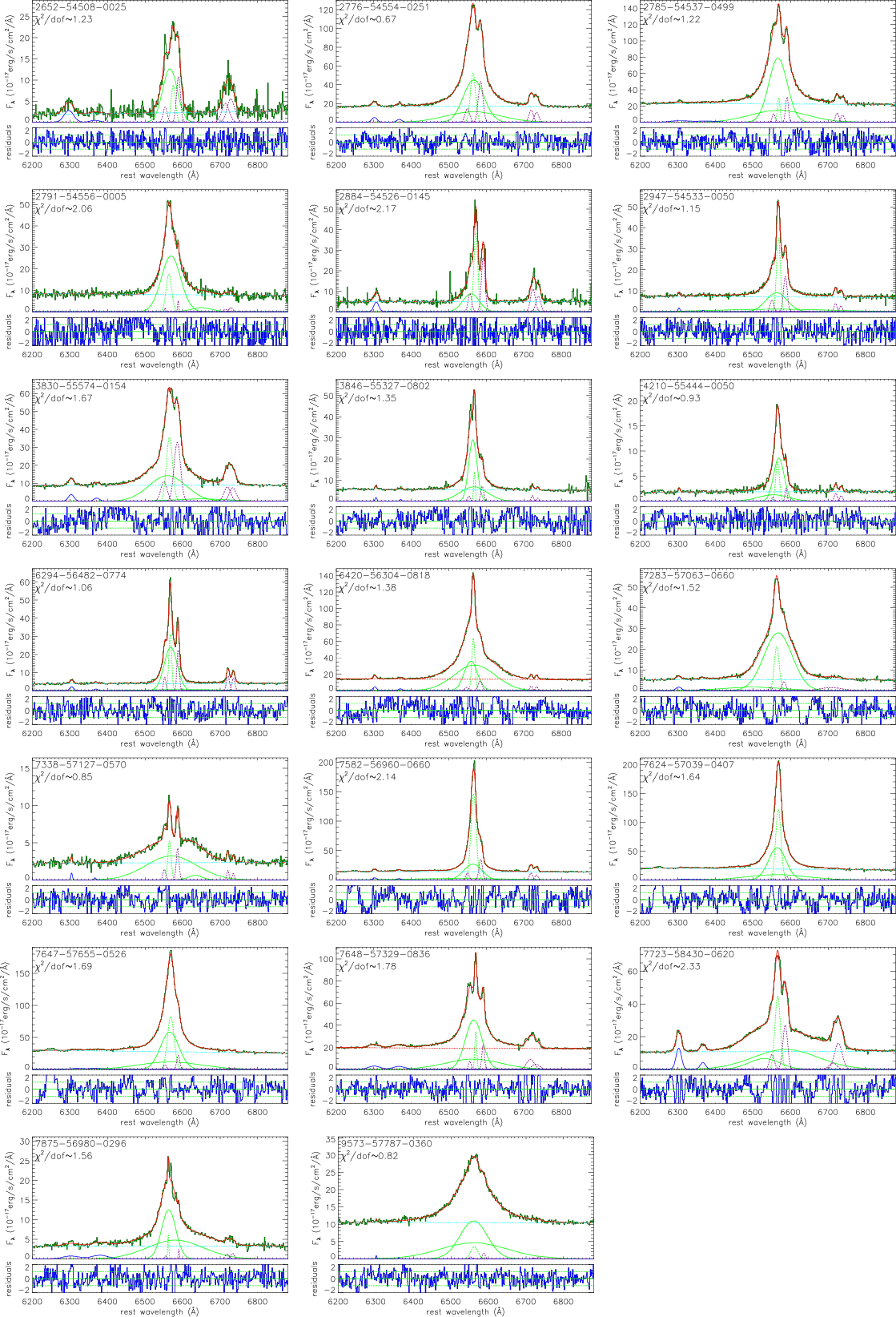} \caption{-- to be continued} \label{figure12} \end{figure*}

\begin{figure*} \centering\includegraphics[width = 18cm,height=21cm]{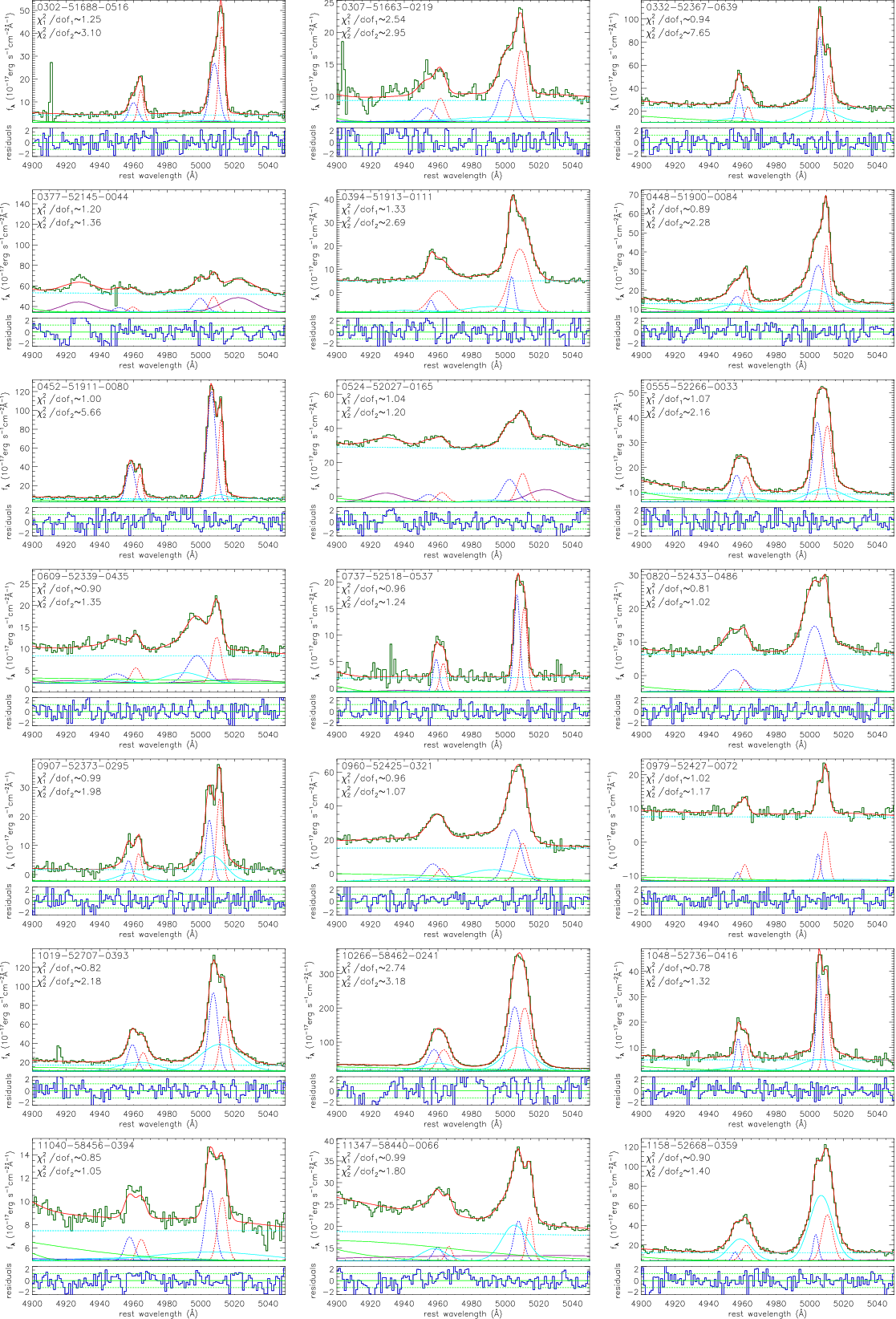} \caption{The best fitting results of the emission lines around \oiii.
In the top of each panel, the solid dark green line shows the line spectrum after subtracting starlight (if present) determined from host galaxy with Plate-Mjd-Fiberid shown in the top-left of corner, the solid red line represents the best fitting results with two narrow Gaussian functions describing each core component of \oiii~doublet, and the corresponding $\chi_1^2/\rm dof_1$ is shown in the top-left of corner, the $\chi_2^2/\rm dof_2$ determined by one narrow Gaussian function for each core component of \oiii~doublet is shown in the top-left of corner, the dashed cyan line represents continuum emission, the solid purple lines represent Fe~{\sc ii} emission lines, the solid green lines represent broad H$\beta$ emission line, the solid cyan lines represent extended component of \oiii, and the dashed blue and red lines represent the blue-shifted and the red-shifted components of \oiii, respectively. 
In the bottom of each panel, the solid blue line represents the residuals calculated by the line spectrum minus the best fitting results and then divided by the uncertainties of the SDSS spectrum, the horizontal solid and dashed green lines show residuals=0,±1, respectively.} \label{figure13} \end{figure*} \setcounter{figure}{13}
\begin{figure*} \centering\includegraphics[width = 18cm,height=21cm]{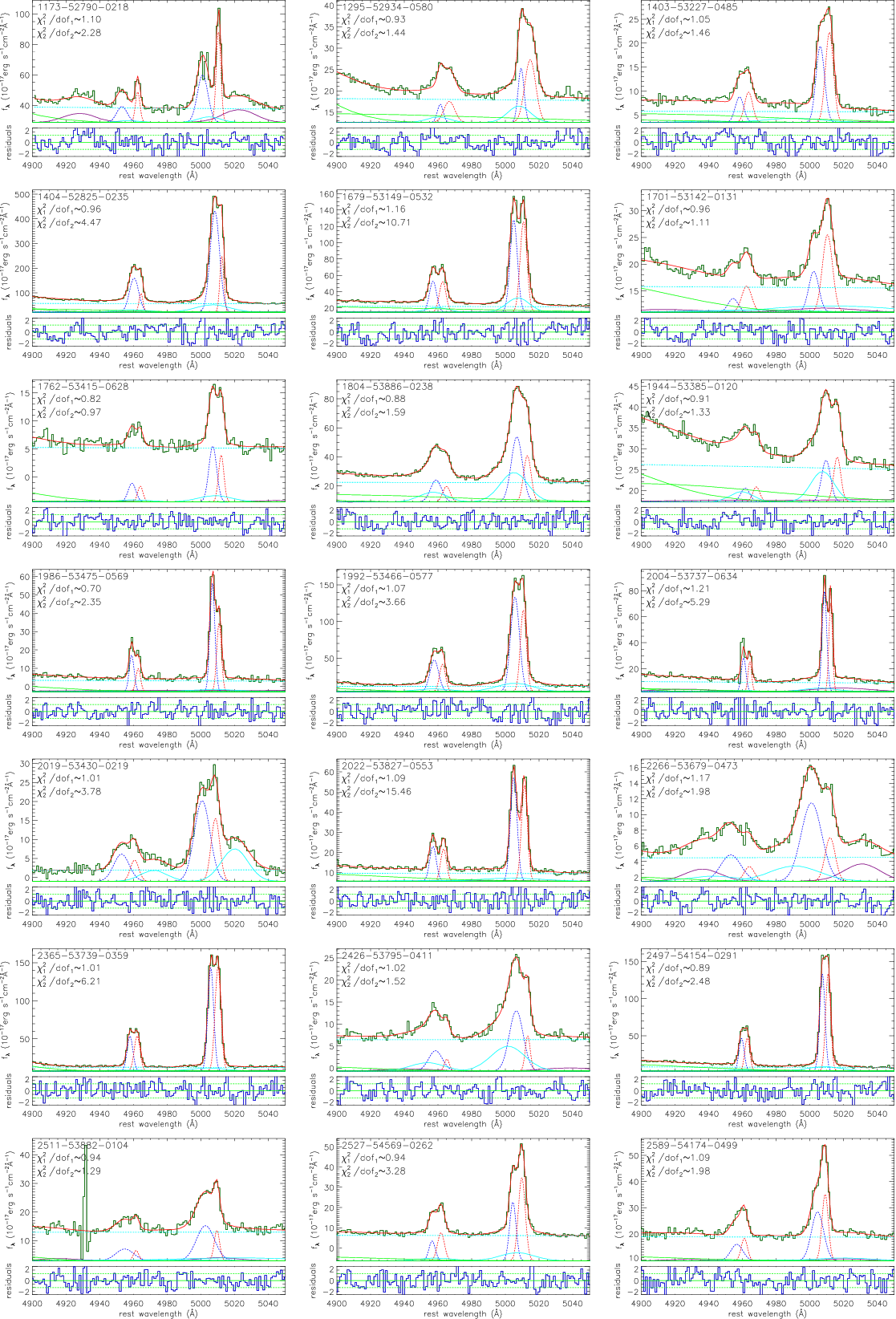} \caption{-- to be continued} \label{figure13} \end{figure*} \setcounter{figure}{13}
\begin{figure*} \centering\includegraphics[width = 18cm,height=21cm]{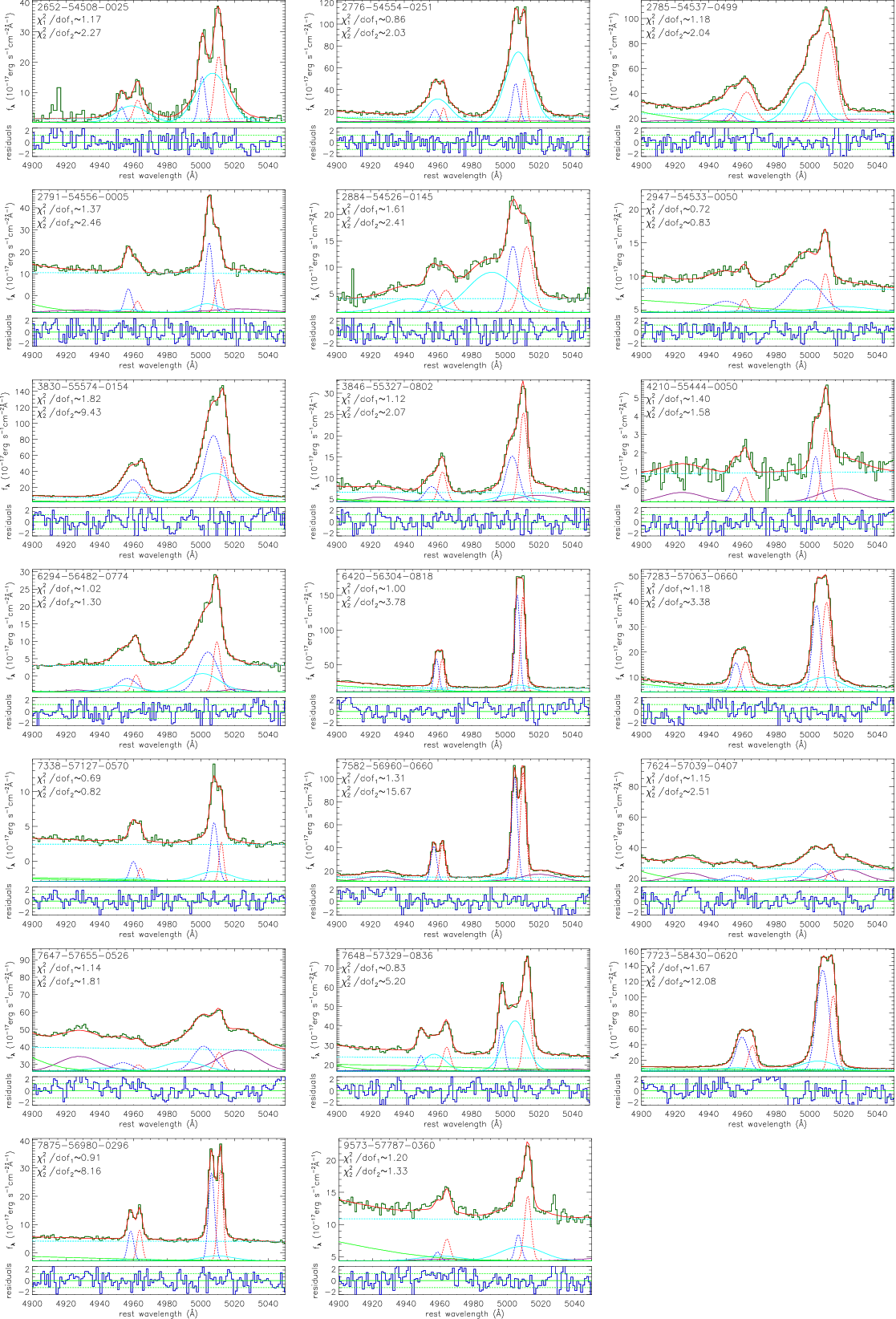} \caption{-- to be continued} \label{figure13} \end{figure*}

\begin{figure*} \centering\includegraphics[width = 18cm,height=21cm]{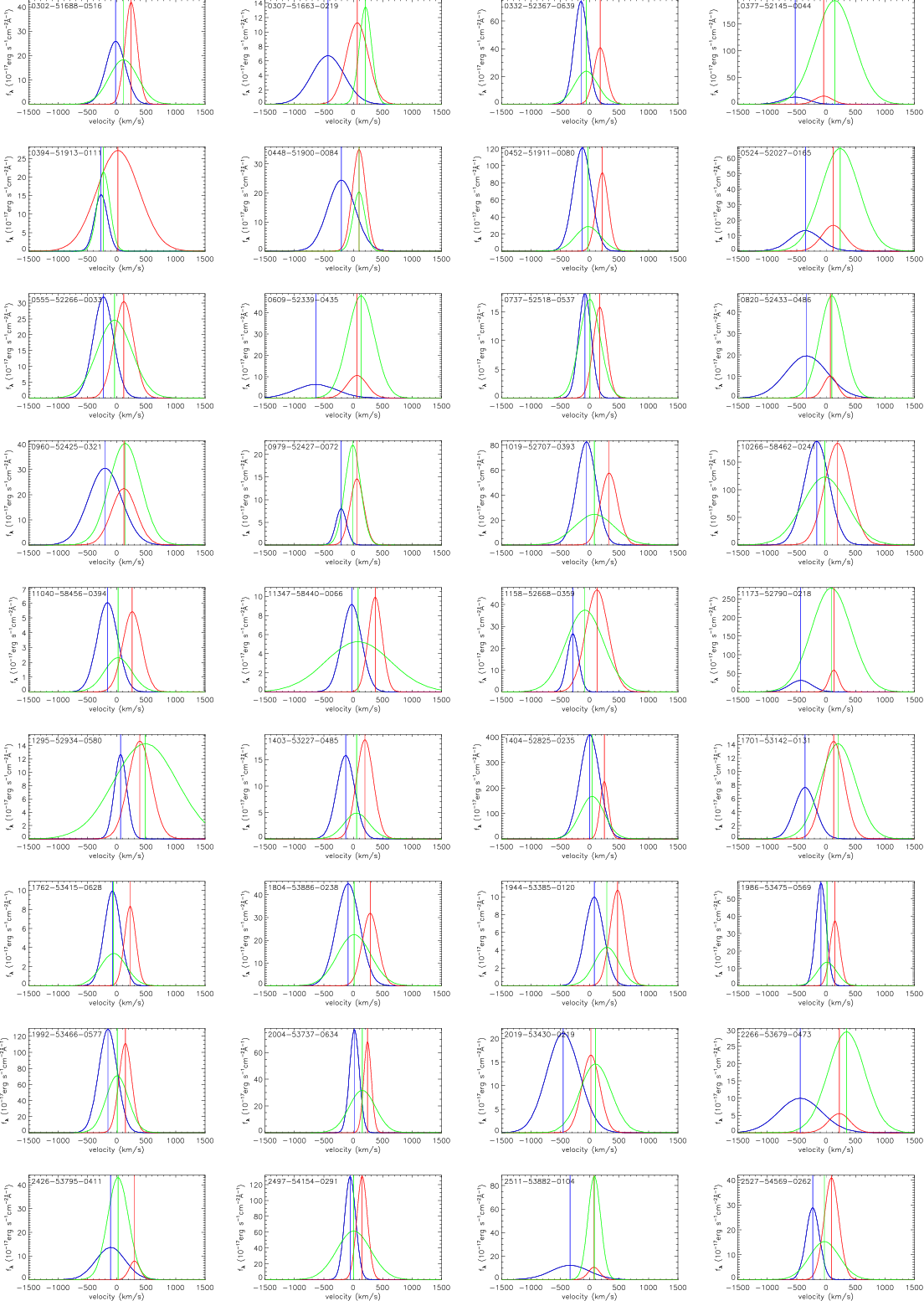} \caption{The offsets of \oiii and narrow H$\alpha$ emission lines for the 53 objects with single-peaked narrow H$\alpha$ emission lines in our sample in velocity space. In each panel, the solid blue line represents the determined blue-shifted component of \oiii, and the vertical blue line marks the position of the central wavelength of the blue-shifted component of \oiii, the solid red line represents the determined red-shifted component of \oiii, and the vertical red line marks the position of the central wavelength of the red-shifted component of \oiii, the solid green line shows the determined narrow H$\alpha$ component, and the vertical green line marks the position of the central wavelength of the narrow H$\alpha$. 		
	} \label{figure18} \end{figure*} \setcounter{figure}{14} \begin{figure*} \centering\includegraphics[width = 18cm,height=11cm]{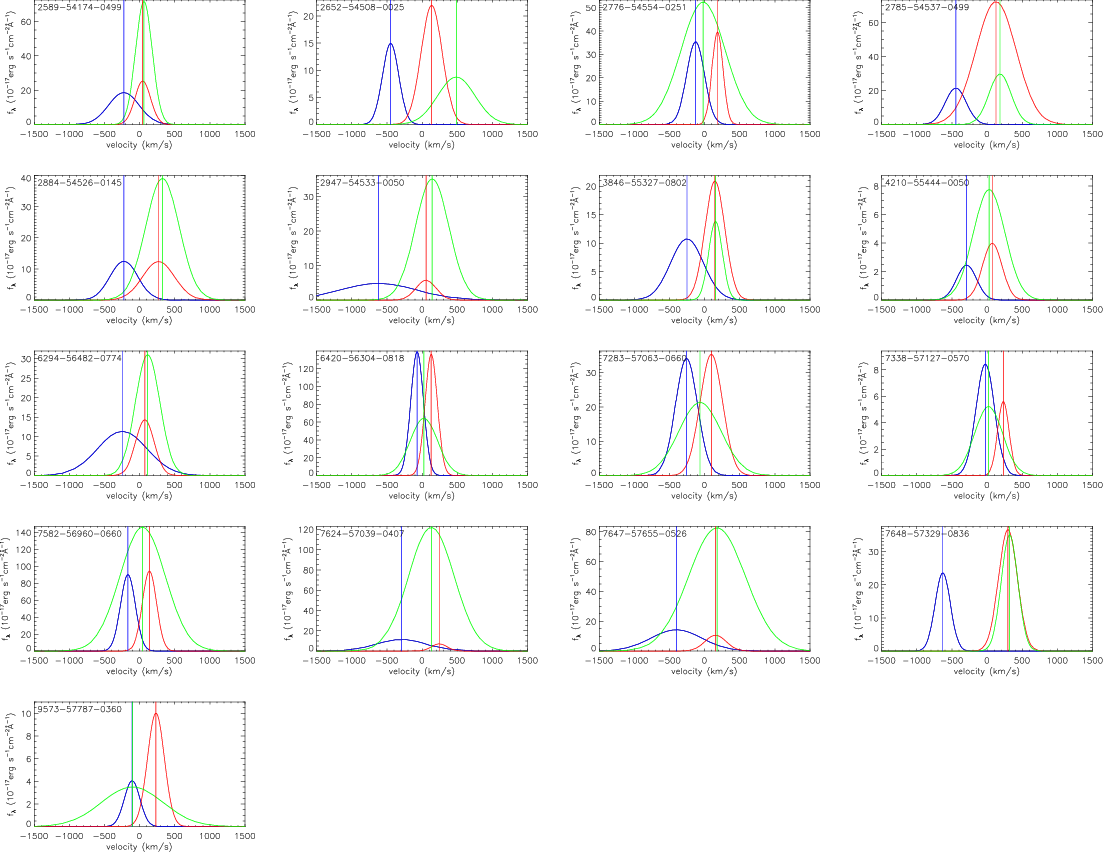} \caption{-- to be continued} \label{figure18} \end{figure*}


\section{Full data tables of the 62 type 1 AGNs}
In this appendix, the tables of the 62 type 1 AGNs are provide.
Information of the 62 AGNs in the final sample is listed in Table \ref{tabb1} including Plate-Mjd-Fiberid, RA, DEC, redshift, photometric magnitude in $r$-band, S/N.
The parameters of broad H$\alpha$ and H$\beta$ emission lines are listed in Table \ref{tabb5}. 
The features of [N~{\sc ii}] and narrow H$\alpha$ emission lines are listed in Table \ref{tabb4}.
Meanwhile, the modeled parameters of double-peaked \oiii and the results of F-test are shown in Table \ref{tabb2}.

\begin{longtable*}{c|c|c|c|c|c|c|c|c}
	\caption{sample information}\label{tabb1}\\
	\toprule
	Plate-Mjd-Fiberid & RA & DEC & z& $\rm mag_{r}$ & S/N & FIRST &$R$& Reference\\ 
	(1)&(2)&(3)&(4)&(5)&(6)&(7)&(8)&(9)\\ 
	\midrule
	\endfirsthead
	\multicolumn{9}{r}{Continued}\\
	\toprule
	Plate-Mjd-Fiberid & RA & DEC & z& $\rm mag_{r}$ & S/N & FIRST &$R$  & Reference\\ 
	(1)&(2)&(3)&(4)&(5)&(6)&(7)&(8)&(9)\\ 
	\endhead
	\multicolumn{9}{c}{Continued on next page}\\
	\endfoot
	\bottomrule
	\endlastfoot
	0302-51688-0516	&	212.51527	&	0.21395	&	0.14	&	18.4$\pm$0.01	&	15.19	&	0	&	0	&	-		\\ \hline
	0307-51663-0219	&	219.25504	&	-1.07168	&	0.29	&	19.03$\pm$0.08	&	11.15	&	0	&	0	&	-		\\ \hline
	0332-52367-0639	&	184.03066	&	-2.23828	&	0.1	&	17.28$\pm$0.01	&	29.36	&	0	&	0	&	 (1),(2)		\\ \hline
	0377-52145-0044	&	340.12026	&	-1.11381	&	0.13	&	17.25$\pm$0.02	&	27.98	&	0	&	0	&	-		\\ \hline
	0394-51913-0111	&	13.6107	&	-0.339	&	0.17	&	19.03$\pm$0.01	&		25.54	&	10.53	&	237.98	&	 -	\\ \hline
	0448-51900-0084	&	133.73845	&	54.80571	&	0.26	&	18.2$\pm$0.02	&	15.73	&	1.31	& 11.78	&	-		\\ \hline
	0452-51911-0080	&	145.43677	&	57.85658	&	0.16	&	17.97$\pm$0.01	&	15.12	&	0	&	0	&	(1),(2),(6),(7)		\\ \hline
	0524-52027-0165	&	196.07083	&	2.09365	&	0.23	&	17.33$\pm$0.02	&	25.71	&	0	&	0	&	-		\\ \hline
	0555-52266-0033	&	144.88269	&	54.81924	&	0.29	&	18.31$\pm$0.01	&	15.56	&	0	&	0	&	-		\\ \hline
	0609-52339-0435	&	221.95334	&	62.74577	&	0.23	&	18.49$\pm$0.02	&	11.75	&	0	&	0	&	(1),(2),(8)		\\ \hline
	0737-52518-0537	&	336.40999	&	14.11419	&	0.13	&	18.53$\pm$0.02	&	14.02	&	-	&	-	&	-		\\ \hline
	0820-52433-0486	&	254.17532	&	36.51368	&	0.14	&	18.29$\pm$0.01	&	18.07	&	1.46	&	9.92	&	-		\\ \hline
	0907-52373-0295	&	162.24439	&	54.8897	&	0.15	&	18.31$\pm$0.02	&	12.98	&	0	&	0	&			-	\\ \hline
	0960-52425-0321	&	203.1581	&	59.51494	&	0.17	&	17.29$\pm$0.02	&	38.4	&	0.92	&	2.51	&	-		\\ \hline
	0979-52427-0072	&	260.33538	&	26.61622	&	0.16	&	18.69$\pm$0.03	&	13.88	&	0	&	0	&	-	\\ \hline
	1019-52707-0393	&	183.67094	&	55.60951	&	0.13	&	17.84$\pm$0.01	&	19.79	&	0	&	0	&	-		\\ \hline
	10266-58462-0241	&	166.76885	&	32.10834	&	0.24	&	17.34$\pm$0.02	&	29.54	&	2.82	&	15.28	&	-		\\ \hline
	1048-52736-0416	&	222.79193	&	49.13712	&	0.16	&	18.22$\pm$0.02	&	14.91	&	0.89	&	5.59&	(2),(3),(4)		\\ \hline
	11040-58456-0394	&	10.42232	&	16.77411	&	0.17	&	18.9$\pm$0.02	&	19.87	&	-	&	-	&	-		\\ \hline
	11347-58440-0066	&	145.77147	&	28.09336	&	0.27	&	17.52$\pm$0.01	&	29.17	&	0	&	0	&	-		\\ \hline
	1158-52668-0359	&	206.56623	&	58.00226	&	0.16	&	17.45$\pm$0.03	&	23.48	&	1.71	&	5.95	&	-		\\ \hline
	1173-52790-0218	&	248.4095	&	37.22063	&	0.12	&	17.65$\pm$0.08	&	23.13	&	0	&	0	&	-		\\ \hline
	1295-52934-0580	&	123.92722	&	6.58972	&	0.24	&	17.98$\pm$0.02	&	21.7	&	0	&	0	&	(1),(2),(4),(7),(8),(9),(10)		\\ \hline
	1403-53227-0485	&	237.96583	&	33.76054	&	0.23	&	19.07$\pm$0.02	&	10.33	&	0	&	0	&	-		\\ \hline
	1404-52825-0235	&	238.57263	&	32.64384	&	0.05	&	16.37$\pm$0.01	&	37.92	&	2.52	&	2.86	&	-		\\ \hline
	1679-53149-0532	&	233.13253	&	42.06188	&	0.21	&	18.17$\pm$0.02	&	24.65	&	0	&	0	&	(2),(4),(6),(8),(10),(11),(12)		\\ \hline
	1701-53142-0131	&	206.62306	&	10.87407	&	0.29	&	18.25$\pm$0.01	&	15.91	&	0	&	0	&	-		\\ \hline
	1762-53415-0628	&	179.15898	&	15.08357	&	0.19	&	18.70$\pm$0.03	&	10.28	&	0	&	0	&	-	\\ \hline
	1804-53886-0238	&	205.86365	&	8.37615	&	0.24	&	18.07$\pm$0.01	&	19.15	&	1.73	&	8.90	&	-		\\ \hline
	1944-53385-0120	&	145.77147	&	28.09337	&	0.27	&	17.52$\pm$0.01	&	27.9	&	0	&	0	&	-		\\ \hline
	1986-53475-0569	&	187.3235	&	40.7632	&	0.17	&	18.31$\pm$0.02	&	14.14	&	0	&	0	&	-		\\ \hline
	1992-53466-0577	&	187.95448	&	39.09172	&	0.07	&	17.36$\pm$0.01	&	26.24	&	2.31	&	8.21	&	-		\\ \hline
	2004-53737-0634	&	184.24977	&	32.51835	&	0.13	&	17.69$\pm$0.02	&	28.34	&	0	&	0	&	(1),(2)		\\ \hline
	2019-53430-0219	&	159.46766	&	31.41677	&		0.15	&	18.26$\pm$0.01 	&	18.68	&	2.15	&	121.32	&	-		\\ \hline
	2022-53827-0553	&	192.05762	&	36.40656	&	0.21	&	18.34$\pm$0.02	&	16	&	2.93	&	22.04	&	(1),(2),(4),(7),(8),(9)		\\ \hline
	2266-53679-0473	&	120.00024	&	15.39059	&	0.27	&	19.09$\pm$0.02	&	11.55	&	6.73	&	112.92	&	-		\\ \hline
	2365-53739-0359	&	153.17171	&	21.93225	&	0.11	&	18.49$\pm$0.02	&	21.93	&	0	&	0	&	(1),(2),(4),(6),(10)		\\ \hline
	2426-53795-0411	&	128.66508	&	12.04501	&	0.25	&	18.53$\pm$0.01	&	15.42	&	2.18	&	18.66	&	-	\\ \hline
	2497-54154-0291	&	169.94838	&	23.59432	&	0.15	&	17.88$\pm$0.01	&	21.38	&	0	&	0	&	-		\\ \hline
	2511-53882-0104	&	177.51928	&	21.04227	&		0.13	&	17.79$\pm$0.02 	&	20.81	&		0	&	0	&	-		\\ \hline
	2527-54569-0262	&	242.61423	&	13.13523	&	0.23	&	18.79$\pm$0.45	&	11.36	&	0	&	0	&	(1),(2),(6),(7),(10),(11),(12)	\\ \hline
	2589-54174-0499	&	155.07937	&	17.61328	&	0.18	&	17.99$\pm$0.02	&	21.11	&	0.85	&	5.18	&	-	\\ \hline
	2652-54508-0025	&	202.23245	&	21.59238	&	0.13	&	17.88$\pm$0.03	&		16.57	&	5.92	&	459.61	&	(2)		\\ \hline
	2776-54554-0251	&	220.27352	&	18.08556	&	0.11	&	17.69$\pm$0.02	&	22.86	&	0	&	0	&	(1),(2),(4)		\\ \hline
	2785-54537-0499	&	212.67293	&	22.5603	&	0.17	&	17.5$\pm$0.02	&	22.67	&	3.68	&	17.61	&	(8)		\\ \hline
	2791-54556-0005	&	225.61673	&	19.04929	&	0.26	&	18.83$\pm$0.02	&	11.29	&	0	&	0	&	-		\\ \hline
	2884-54526-0145	&	244.96114	&	50.09315	&		0.28	&	19.24$\pm$0.01	&	12.1	&	2.34	&	65.56		&	-		\\ \hline
	2947-54533-0050	&	221.95334	&	62.74576	&		0.23	&	18.49$\pm$0.01 	&	15.94	&	0	&	0	&	-		\\ \hline
	3830-55574-0154	&	153.35855	&	-0.02679	&	0.26	&	18.51$\pm$0.02	&	16.62	&	1.69	&	24.49	&	-	\\ \hline
	3846-55327-0802	&	184.80754	&	0.6809	&	0.29	&	18.4$\pm$0.02	&	21.5	&	0	&	0	&	-		\\ \hline
	4210-55444-0050	&	351.07114	&	-0.94571	&	0.2	&	19.67$\pm$0.02	&	12.33	&	0	&	0	&	(2)	\\ \hline
	6294-56482-0774	&	342.15175	&	28.88209	&	0.17	&	18.59$\pm$0.01	&	20.24	&	-	&	-	&	-		\\ \hline
	6420-56304-0818	&	169.94837	&	23.59432	&	0.15	&	17.88$\pm$0.01	&	25.37	&	0	&	0	&	-		\\ \hline
	7283-57063-0660	&	144.88269	&	54.81922	&	0.29	&	18.31$\pm$0.01	&	17.47	&	0	&	0	&	-\\ \hline
	7338-57127-0570	&	212.3154	&	53.45606	&	0.26	&	19.24$\pm$0.03	&	13.65	&	0	&	0	&	-	\\ \hline
	7582-56960-0660	&	336.32782	&	21.03436	&	0.14	&	18.17$\pm$0.03	&	29.16	&	-	&	-	&	-		\\ \hline
	7624-57039-0407	&	15.00198	&	19.19814	&	0.3	&	17.2$\pm$0.02	&	36.62	&	-	&	-	&	-		\\ \hline
	7647-57655-0526	&	333.09655	&	29.51795	&	0.25	&	17.25$\pm$0.01	&	42.61	&	-	&	-	&	-		\\ \hline
	7648-57329-0836	&	336.11891	&	26.23979	&	0.21	&	17.95$\pm$0.02	&	33.06	&	-	&	-	&	(5)		\\ \hline
	7723-58430-0620-	&	16.67971	&	34.04106	&	0.18	&	18.5$\pm$0.01	&	21.95	&	-	&	-	&	-		\\ \hline
	7875-56980-0296	&	17.96095	&	-2.54367	&	0.23	&	19.45$\pm$0.03	&	13.54	&	-	&	-	&	-		\\ \hline
	9573-57787-0360	&	155.7179	&	17.99481	&	0.25	&	18.39$\pm$0.02	&	19.27	&	0	&	0	&	-		\\ 
\end{longtable*}
\begin{tablenotes}
	\item[*]Notes. Column 1: Plate, MJD, Fiberid of spectroscopic observation; Column 2: RA; Column 3: DEC; Column 4:  Redshift of spectroscopic observation; Column 5: Apparent psf magnitude in r band; Column 6: Median signal-to-noise over all good pixels; Column 7: Integrated FIRST radio flux(mJy); Column 8: Radio-loudness; Column 9: Reference:(1) \citet{Zh16}; (2) \citet{Sm10}; (3) \citet{Ge12}; (4)\citet{Sm12}; (5) \citet{Zh24}; (6) \citet{Fu12}; (7) \citet{Co18}; (8) \citet{Ki202}; (9) \citet{Co12}; (10) \citet{Mc15}; (11) \citet{Ro11}; (12) \citet{Fu11}.
\end{tablenotes}

\begin{longtable*}{c|c|c|c|c|c|c|c}
	\caption{Line parameters of broad H$\alpha$ and broad H$\beta$}\label{tabb5}\\
	\toprule
	Plate-Mjd-Fiberid & $\lambda_{H\alpha}$  & $\rm FWHM_{H\alpha}$  & $F_{H\alpha}$ & $\lambda_{H\beta}$  & $\rm FWHM_{H\beta}$  & $F_{H\beta}$ & mass\\ 
	(1)&(2)&(3)&(4)&(5)&(6)&(7)&(8)\\ 
	\midrule
	\endfirsthead
	\multicolumn{7}{r}{Continued}\\
	\toprule
	Plate-Mjd-Fiberid & $\lambda_{H\alpha}$  & $\rm FWHM_{H\alpha}$  & $F_{H\alpha}$ & $\lambda_{H\beta}$  & $\rm FWHM_{H\beta}$  & $F_{H\beta}$ & mass\\ 
	(1)&(2)&(3)&(4)&(5)&(6)&(7)&(8)\\ 
	\endhead
	\multicolumn{7}{c}{Continued on next page}\\
	\endfoot
	\bottomrule
	\endlastfoot
	0302-51688-0516	&	6559.7 	$\pm$	5.0 	&	72.3 	$\pm$	5.7 	&	918.7 	$\pm$	116.2 	&	4850.9 	$\pm$	3.1 	&	35.9 	$\pm$	5.4 	&	96.4 	$\pm$	17.7 	&	7.81	$\pm$	0.10 	\\ \hline
	0307-51663-0219	&	6569.2 	$\pm$	0.6 	&	47.7 	$\pm$	5.0 	&	881.3 	$\pm$	55.5 	&	4864.9 	$\pm$	0.8 	&	43.7 	$\pm$	2.6 	&	188.6 	$\pm$	9.3 	&	7.50	$\pm$	0.11 	\\ \hline
	0332-52367-0639	&	6565.3 	$\pm$	1.6 	&	99.4 	$\pm$	3.9 	&	3770.8 	$\pm$	226.9 	&	4870.1 	$\pm$	2.1 	&	102.5 	$\pm$	5.9 	&	768.8 	$\pm$	50.6 	&	7.98	$\pm$	0.05 	\\ \hline
	0377-52145-0044	&	6546.1 	$\pm$	0.5 	&	73.3 	$\pm$	5.4 	&	4751.0 	$\pm$	153.0 	&	4859.9 	$\pm$	1.6 	&	50.8 	$\pm$	2.8 	&	1154.8 	$\pm$	66.7 	&	7.80	$\pm$	0.07 	\\ \hline
	0394-51913-0111	&	6566.3 	$\pm$	1.7 	&	35.2 	$\pm$	4.0 	&	1141.8 	$\pm$	94.9 	&	4863.3 	$\pm$	1.9 	&	23.2 	$\pm$	4.6 	&	46.0 	$\pm$	10.5 	&	7.77	$\pm$	0.12 	\\ \hline
	0448-51900-0084	&	6567.9 	$\pm$	0.3 	&	40.8 	$\pm$	4.5 	&	1700.1 	$\pm$	58.8 	&	4865.3 	$\pm$	1.7 	&	35.2 	$\pm$	5.9 	&	544.7 	$\pm$	28.4 	&	7.28	$\pm$	0.11 	\\ \hline
	0452-51911-0080	&	6555.0 	$\pm$	0.8 	&	48.8 	$\pm$	2.1 	&	2213.3 	$\pm$	33.9 	&	4857.4 	$\pm$	4.3 	&	130.1 	$\pm$	12.4 	&	369.4 	$\pm$	31.6 	&	7.59	$\pm$	0.04 	\\ \hline
	0524-52027-0165	&	6557.9 	$\pm$	0.6 	&	48.8 	$\pm$	3.0 	&	2387.4 	$\pm$	110.1 	&	4857.4 	$\pm$	4.3 	&	29.3 	$\pm$	12.4 	&	369.4 	$\pm$	31.6 	&	7.79	$\pm$	0.07 	\\ \hline
	0555-52266-0033	&	6566.7 	$\pm$	0.7 	&	113.3 	$\pm$	1.3 	&	2854.9 	$\pm$	66.8 	&	4869.4 	$\pm$	16.5 	&	87.2 	$\pm$	33.1 	&	824.1 	$\pm$	64.6 	&	8.44	$\pm$	0.02 	\\ \hline
	0609-52339-0435	&	6561.2 	$\pm$	0.8 	&	76.5 	$\pm$	1.7 	&	1051.7 	$\pm$	30.6 	&	4878.4 	$\pm$	15.8 	&	51.3 	$\pm$	35.6 	&	541.9 	$\pm$	72.0 	&	7.69	$\pm$	0.03 	\\ \hline
	0737-52518-0537	&	6570.6 	$\pm$	0.4 	&	54.7 	$\pm$	2.2 	&	2010.9 	$\pm$	86.9 	&	4867.0 	$\pm$	1.9 	&	38.9 	$\pm$	5.6 	&	140.4 	$\pm$	14.8 	&	7.91	$\pm$	0.05 	\\ \hline
	0820-52433-0486	&	6572.7 	$\pm$	0.6 	&	63.3 	$\pm$	4.1 	&	1329.8 	$\pm$	68.0 	&	4868.8 	$\pm$	4.0 	&	49.7 	$\pm$	8.6 	&	230.1 	$\pm$	24.8 	&	7.57	$\pm$	0.07 	\\ \hline
	0907-52373-0295	&	6566.7 	$\pm$	2.4 	&	114.1 	$\pm$	7.2 	&	711.2 	$\pm$	41.2 	&	4863.9 	$\pm$	6.1 	&	58.1 	$\pm$	18.0 	&	70.8 	$\pm$	20.6 	&	8.22	$\pm$	0.07 	\\ \hline
	0960-52425-0321	&	6579.6 	$\pm$	1.3 	&	165.0 	$\pm$	3.3 	&	2965.6 	$\pm$	61.9 	&	4891.4 	$\pm$	5.4 	&	152.1 	$\pm$	13.6 	&	935.0 	$\pm$	85.6 	&	8.45	$\pm$	0.02 	\\ \hline
	0979-52427-0072	&	6572.0 	$\pm$	0.6 	&	73.0 	$\pm$	1.8 	&	951.6 	$\pm$	21.8 	&	4881.0 	$\pm$	20.8 	&	56.6 	$\pm$	9.0 	&	316.5 	$\pm$	51.1 	&	7.39	$\pm$	0.03 	\\ \hline
	1019-52707-0393	&	6569.2 	$\pm$	0.9 	&	70.9 	$\pm$	2.7 	&	3344.8 	$\pm$	162.9 	&	4873.4 	$\pm$	1.7 	&	65.5 	$\pm$	14.5 	&	998.0 	$\pm$	54.6 	&	7.63	$\pm$	0.05 	\\ \hline
	10266-58462-0241	&	6562.1 	$\pm$	0.7 	&	177.7 	$\pm$	1.3 	&	10224.7 	$\pm$	103.2 	&	4866.9 	$\pm$	1.0 	&	131.5 	$\pm$	3.1 	&	2926.8 	$\pm$	37.4 	&	9.05	$\pm$	0.01 	\\ \hline
	1048-52736-0416	&	6578.3 	$\pm$	3.9 	&	134.4 	$\pm$	15.5 	&	1845.5 	$\pm$	601.7 	&	4877.2 	$\pm$	6.5 	&	85.2 	$\pm$	11.9 	&	193.7 	$\pm$	32.5 	&	8.60	$\pm$	0.18 	\\ \hline
	11040-58456-0394	&	6537.1 	$\pm$	0.9 	&	145.5 	$\pm$	4.3 	&	2581.4 	$\pm$	56.0 	&	4842.7 	$\pm$	2.7 	&	147.1 	$\pm$	11.2 	&	416.7 	$\pm$	33.6 	&	8.60	$\pm$	0.03 	\\ \hline
	11347-58440-0066	&	6562.6 	$\pm$	1.6 	&	149.4 	$\pm$	2.0 	&	4960.8 	$\pm$	128.5 	&	4879.0 	$\pm$	2.2 	&	123.5 	$\pm$	11.4 	&	1363.5 	$\pm$	62.7 	&	8.79	$\pm$	0.02 	\\ \hline
	1158-52668-0359	&	6572.5 	$\pm$	0.7 	&	80.5 	$\pm$	6.1 	&	2545.9 	$\pm$	150.8 	&	4867.5 	$\pm$	2.2 	&	62.0 	$\pm$	4.5 	&	559.9 	$\pm$	35.1 	&	7.91	$\pm$	0.08 	\\ \hline
	1173-52790-0218	&	6571.8 	$\pm$	0.3 	&	60.5 	$\pm$	4.0 	&	4879.2 	$\pm$	209.3 	&	4865.2 	$\pm$	0.2 	&	26.1 	$\pm$	3.7 	&	2227.0 	$\pm$	91.9 	&	7.46	$\pm$	0.07 	\\ \hline
	1295-52934-0580	&	6553.9 	$\pm$	0.6 	&	50.2 	$\pm$	2.0 	&	3060.3 	$\pm$	41.2 	&	4873.3 	$\pm$	0.5 	&	73.5 	$\pm$	1.7 	&	1257.6 	$\pm$	58.4 	&	7.57	$\pm$	0.04 	\\ \hline
	1403-53227-0485	&	6551.6 	$\pm$	1.1 	&	196.0 	$\pm$	3.5 	&	2225.8 	$\pm$	35.6 	&	4882.9 	$\pm$	3.4 	&	181.6 	$\pm$	16.2 	&	458.1 	$\pm$	62.6 	&	8.89	$\pm$	0.02 	\\ \hline
	1404-52825-0235	&	6568.2 	$\pm$	0.3 	&	91.9 	$\pm$	1.7 	&	19953.0 	$\pm$	477.4 	&	4871.1 	$\pm$	1.2 	&	78.2 	$\pm$	10.3 	&	4664.7 	$\pm$	117.1 	&	7.88	$\pm$	0.02 	\\ \hline
	1679-53149-0532	&	6580.2 	$\pm$	3.1 	&	180.5 	$\pm$	3.4 	&	3620.9 	$\pm$	103.1 	&	4884.3 	$\pm$	5.8 	&	198.9 	$\pm$	7.9 	&	1382.9 	$\pm$	114.3 	&	8.68	$\pm$	0.02 	\\ \hline
	1701-53142-0131	&	6569.3 	$\pm$	1.0 	&	104.9 	$\pm$	2.4 	&	2001.1 	$\pm$	47.6 	&	4868.8 	$\pm$	2.9 	&	92.0 	$\pm$	6.5 	&	720.8 	$\pm$	41.8 	&	8.23	$\pm$	0.03 	\\ \hline
	1762-53415-0628	&	6570.1 	$\pm$	0.7 	&	86.6 	$\pm$	4.7 	&	1410.4 	$\pm$	37.1 	&	4860.7 	$\pm$	6.7 	&	78.6 	$\pm$	8.4 	&	277.3 	$\pm$	49.8 	&	7.98	$\pm$	0.05 	\\ \hline
	1804-53886-0238	&	6575.7 	$\pm$	0.9 	&	132.9 	$\pm$	2.2 	&	4639.5 	$\pm$	72.1 	&	4886.0 	$\pm$	3.0 	&	134.5 	$\pm$	11.8 	&	1180.5 	$\pm$	71.9 	&	8.65	$\pm$	0.02 	\\ \hline
	1944-53385-0120	&	6568.4 	$\pm$	0.5 	&	141.3 	$\pm$	1.1 	&	6090.6 	$\pm$	62.3 	&	4878.6 	$\pm$	1.4 	&	114.5 	$\pm$	15.6 	&	1810.7 	$\pm$	92.6 	&	8.75	$\pm$	0.01 	\\ \hline
	1986-53475-0569	&	6560.3 	$\pm$	0.6 	&	105.9 	$\pm$	1.5 	&	1698.9 	$\pm$	24.4 	&	4860.6 	$\pm$	2.0 	&	99.9 	$\pm$	5.3 	&	477.1 	$\pm$	22.3 	&	7.97	$\pm$	0.02 	\\ \hline
	1992-53466-0577	&	6568.2 	$\pm$	0.3 	&	94.9 	$\pm$	2.5 	&	5756.7 	$\pm$	130.5 	&	4867.9 	$\pm$	1.7 	&	77.0 	$\pm$	3.9 	&	621.5 	$\pm$	33.5 	&	8.13	$\pm$	0.03 	\\ \hline
	2004-53737-0634	&	6567.0 	$\pm$	0.5 	&	68.0 	$\pm$	2.3 	&	3821.1 	$\pm$	141.8 	&	4868.2 	$\pm$	0.8 	&	55.8 	$\pm$	17.7 	&	1068.2 	$\pm$	35.8 	&	7.61	$\pm$	0.04 	\\ \hline
	2019-53430-0219	&	6571.4 	$\pm$	0.6 	&	50.4 	$\pm$	3.3 	&	490.7 	$\pm$	60.2 	&	4861.0 	$\pm$	2.1 	&	23.0 	$\pm$	0.8 	&	47.9 	$\pm$	15.1 	&	7.45	$\pm$	0.09 	\\ \hline
	2022-53827-0553	&	6562.2 	$\pm$	0.8 	&	161.7 	$\pm$	2.3 	&	2317.6 	$\pm$	32.2 	&	4865.2 	$\pm$	3.8 	&	144.1 	$\pm$	9.2 	&	631.4 	$\pm$	36.5 	&	8.54	$\pm$	0.02 	\\ \hline
	2266-53679-0473	&	6562.0 	$\pm$	0.7 	&	63.0 	$\pm$	3.0 	&	1405.6 	$\pm$	61.5 	&	4860.0 	$\pm$	1.1 	&	42.8 	$\pm$	2.3 	&	188.9 	$\pm$	11.0 	&	8.05	$\pm$	0.05 	\\ \hline
	2365-53739-0359	&	6565.7 	$\pm$	2.2 	&	98.3 	$\pm$	3.1 	&	3187.0 	$\pm$	269.9 	&	4865.1 	$\pm$	2.8 	&	77.3 	$\pm$	3.7 	&	842.2 	$\pm$	66.7 	&	7.86	$\pm$	0.05 	\\ \hline
	2426-53795-0411	&	6567.1 	$\pm$	1.3 	&	32.5 	$\pm$	4.6 	&	797.0 	$\pm$	103.1 	&	4868.2 	$\pm$	9.0 	&	33.5 	$\pm$	18.9 	&	197.1 	$\pm$	26.4 	&	6.99	$\pm$	0.16 	\\ \hline
	2497-54154-0291	&	6572.3 	$\pm$	0.4 	&	112.4 	$\pm$	1.3 	&	3998.7 	$\pm$	40.3 	&	4867.5 	$\pm$	1.2 	&	77.4 	$\pm$	5.1 	&	1634.4 	$\pm$	39.1 	&	8.09	$\pm$	0.01 	\\ \hline
	2511-53882-0104	&	6566.4 	$\pm$	0.5 	&	44.5 	$\pm$	2.6 	&	1283.4 	$\pm$	88.0 	&	4861.8 	$\pm$	3.4 	&	39.7 	$\pm$	6.3 	&	251.2 	$\pm$	47.9 	&	7.16	$\pm$	0.07 	\\ \hline
	2527-54569-0262	&	6570.3 	$\pm$	1.1 	&	139.2 	$\pm$	3.6 	&	1087.4 	$\pm$	25.7 	&	4880.2 	$\pm$	8.5 	&	109.9 	$\pm$	13.2 	&	230.5 	$\pm$	38.1 	&	8.40	$\pm$	0.03 	\\ \hline
	2589-54174-0499	&	6569.2 	$\pm$	0.2 	&	38.5 	$\pm$	0.6 	&	1121.0 	$\pm$	20.4 	&	4868.7 	$\pm$	5.0 	&	28.1 	$\pm$	14.3 	&	345.2 	$\pm$	27.4 	&	6.95	$\pm$	0.02 	\\ \hline
	2652-54508-0025	&	6566.2 	$\pm$	1.2 	&	45.9 	$\pm$	1.5 	&	616.5 	$\pm$	48.1 	&	4861.7 	$\pm$	1.7 	&	23.2 	$\pm$	0.2 	&	63.9 	$\pm$	9.7 	&	7.31	$\pm$	0.05 	\\ \hline
	2776-54554-0251	&	6565.7 	$\pm$	0.4 	&	74.2 	$\pm$	3.8 	&	5179.5 	$\pm$	180.4 	&	4867.3 	$\pm$	4.9 	&	61.0 	$\pm$	9.3 	&	929.5 	$\pm$	123.1 	&	7.88	$\pm$	0.05 	\\ \hline
	2785-54537-0499	&	6567.2 	$\pm$	0.2 	&	65.1 	$\pm$	1.3 	&	7659.4 	$\pm$	137.9 	&	4866.1 	$\pm$	0.8 	&	54.7 	$\pm$	2.0 	&	1339.9 	$\pm$	29.7 	&	8.12	$\pm$	0.02 	\\ \hline
	2791-54556-0005	&	6574.9 	$\pm$	0.4 	&	61.0 	$\pm$	1.3 	&	1802.8 	$\pm$	33.4 	&	4868.2 	$\pm$	0.7 	&	57.5 	$\pm$	1.8 	&	533.1 	$\pm$	16.3 	&	7.69	$\pm$	0.02 	\\ \hline
	2884-54526-0145	&	6558.9 	$\pm$	1.9 	&	50.8 	$\pm$	2.1 	&	477.0 	$\pm$	44.8 	&	4858.4 	$\pm$	4.9 	&	25.4 	$\pm$	0.5 	&	28.5 	$\pm$	8.3 	&	7.98	$\pm$	0.06 	\\ \hline
	2947-54533-0050	&	6560.0 	$\pm$	0.7 	&	72.1 	$\pm$	1.9 	&	1100.2 	$\pm$	117.9 	&	4867.7 	$\pm$	3.5 	&	46.0 	$\pm$	16.8 	&	467.7 	$\pm$	24.4 	&	7.63	$\pm$	0.05 	\\ \hline
	3830-55574-0154	&	6570.6 	$\pm$	0.8 	&	121.2 	$\pm$	1.9 	&	2005.1 	$\pm$	33.7 	&	4867.3 	$\pm$	1.1 	&	63.4 	$\pm$	4.2 	&	195.5 	$\pm$	13.1 	&	8.82	$\pm$	0.02 	\\ \hline
	3846-55327-0802	&	6564.4 	$\pm$	0.1 	&	30.1 	$\pm$	1.0 	&	1438.7 	$\pm$	31.3 	&	4862.1 	$\pm$	0.9 	&	46.9 	$\pm$	2.6 	&	249.3 	$\pm$	10.0 	&	7.30	$\pm$	0.04 	\\ \hline
	4210-55444-0050	&	6564.3 	$\pm$	0.4 	&	36.0 	$\pm$	2.6 	&	432.1 	$\pm$	23.9 	&	4861.6 	$\pm$	1.6 	&	31.5 	$\pm$	5.3 	&	34.6 	$\pm$	5.3 	&	7.32	$\pm$	0.08 	\\ \hline
	6294-56482-0774	&	6573.6 	$\pm$	0.2 	&	41.6 	$\pm$	1.1 	&	1208.1 	$\pm$	20.5 	&	4860.8 	$\pm$	0.9 	&	23.9 	$\pm$	0.7 	&	81.8 	$\pm$	7.7 	&	7.68	$\pm$	0.03 	\\ \hline
	6420-56304-0818	&	6565.6 	$\pm$	1.1 	&	64.2	$\pm$	1.3 	&	5869.2	$\pm$	129.5 	&	4865.4 	$\pm$	0.6 	&	82.2 	$\pm$	2.0 	&	1486.1 	$\pm$	22.1 	&	7.79	$\pm$	0.02 	\\ \hline
	7283-57063-0660	&	6559.2 	$\pm$	0.2 	&	91.5 	$\pm$	0.8 	&	2982.2 	$\pm$	19.2 	&	4863.9 	$\pm$	0.6 	&	89.9 	$\pm$	1.7 	&	476.4 	$\pm$	8.0 	&	8.52	$\pm$	0.01 	\\ \hline
	7338-57127-0570	&	6573.2 	$\pm$	2.5 	&	168.0 	$\pm$	4.0 	&	576.1 	$\pm$	22.8 	&	4864.0 	$\pm$	1.5 	&	146.6 	$\pm$	16.4 	&	147.0 	$\pm$	15.6 	&	8.39	$\pm$	0.03 	\\ \hline
	7582-56960-0660	&	6552.8 	$\pm$	0.2 	&	59.8 	$\pm$	2.7 	&	2439.2 	$\pm$	51.2 	&	4861.1 	$\pm$	0.8 	&	43.0 	$\pm$	1.7 	&	668.8 	$\pm$	22.8 	&	7.46	$\pm$	0.04 	\\ \hline
	7624-57039-0407	&	6564.6 	$\pm$	0.2 	&	60.6 	$\pm$	1.4 	&	5090.0 	$\pm$	73.0 	&	4862.9 	$\pm$	0.4 	&	35.2 	$\pm$	1.2 	&	1541.9 	$\pm$	42.7 	&	8.01	$\pm$	0.02 	\\ \hline
	7647-57655-0526	&	6565.8 	$\pm$	0.2 	&	65.6 	$\pm$	2.0 	&	6017.8 	$\pm$	106.8 	&	4863.1 	$\pm$	0.3 	&	58.3 	$\pm$	0.8 	&	1481.5 	$\pm$	17.8 	&	8.10	$\pm$	0.03 	\\ \hline
	7648-57329-0836	&	6565.6	$\pm$	0.2 	&	59.7	$\pm$	0.5 	&	4232.3	$\pm$	33.8 	&	4880.3 	$\pm$	0.8 	&	74.5 	$\pm$	11.6 	&	1189.5 	$\pm$	51.4 	&	7.78	$\pm$	0.01 	\\ \hline
	7723-58430-0620	&	6575.0 	$\pm$	0.8 	&	189.7 	$\pm$	1.9 	&	3324.5 	$\pm$	41.4 	&	4867.1 	$\pm$	2.3 	&	229.8 	$\pm$	6.3 	&	796.5 	$\pm$	27.1 	&	8.74	$\pm$	0.01 	\\ \hline
	7875-56980-0296	&	6573.4 	$\pm$	1.1 	&	56.3 	$\pm$	2.8 	&	1475.9 	$\pm$	26.9 	&	4873.6 	$\pm$	6.7 	&	59.5 	$\pm$	13.9 	&	285.9 	$\pm$	27.7 	&	7.71	$\pm$	0.05 	\\ \hline
	9573-57787-0360	&	6560.7 	$\pm$	2.0 	&	105.1 	$\pm$	8.3 	&	2049.2 	$\pm$	103.7 	&	4862.6 	$\pm$	1.3 	&	111.4 	$\pm$	3.4 	&	480.1 	$\pm$	14.1 	&	8.31	$\pm$	0.08 	\\ 
\end{longtable*}
\begin{tablenotes}
	\item[*]Notes. Column 1: Plate, MJD, Fiberid of spectroscopic observation; Column 2 and Column 3: the central wavelength (the first moment) of board H$\alpha$ in units of \AA~and corresponding FWHM of broad H$\alpha$ in units of \AA~ determined through the line profiles described by two Gaussian functions, respectively; 
	Column 4: corresponding flux of broad H$\alpha$ in units of $10^{-17}{\rm erg/s/cm^{2}}$ determined by the line flux described by two Gaussian functions;
	Column 5 and Column 6: the central wavelength (the first moment) of board H$\beta$ in units of \AA~and corresponding FWHM of broad H$\beta$ in units of \AA~determined through the line profiles described by two Gaussian functions, respectively;
	Column 7: corresponding flux of broad H$\beta$ in units of $10^{-17}{\rm erg/s/cm^{2}}$ determined by the line flux described by two Gaussian functions; Column 8: the virial black hole mass log($\rm M_{\rm BH}$/$\rm M_{\odot}$).
\end{tablenotes}

\begin{longtable*}{c|c|c|c|c|c|c}
	\caption{Features of narrow H$\alpha$ and \nii emission lines}\label{tabb4}\\
	\toprule
	Plate-Mjd-Fiberid & $\lambda_{H\alpha}$  & $\sigma_{H\alpha}$   & $F_{H\alpha}$  & $\lambda_{[N~{\textsc{ii}}]}$   & $\sigma_{[N~{\textsc{ii}}]}$  & $F_{[N~{\textsc{ii}}]}$   \\ 
	(1)&(2)&(3)&(4)&(5)&(6)&(7)\\ 
	\midrule
	\endfirsthead
	\multicolumn{7}{r}{Continued}\\
	\toprule
	Plate-Mjd-Fiberid & $\lambda_{H\alpha}$  & $\sigma_{H\alpha}$   & $F_{H\alpha}$  & $\lambda_{[N~{\textsc{ii}}]}$   & $\sigma_{[N~{\textsc{ii}}]}$  & $F_{[N~{\textsc{ii}}]}$   \\ 
	(1)&(2)&(3)&(4)&(5)&(6)&(7)\\ 
	\endhead
	\multicolumn{7}{c}{Continued on next page}\\
	\endfoot
	\bottomrule
	\endlastfoot
	0302-51688-0516&6567.1$\pm$0.2&5.3$\pm$0.3&244.6$\pm$23.8&6587.0$\pm$0.2&4.7$\pm$0.3&135.8$\pm$14.0\\ \hline
	0307-51663-0219&6569.1$\pm$0.2&2.3$\pm$0.2&77.4$\pm$7.8&6589.3$\pm$0.3&3.2$\pm$0.3&70.7$\pm$8.8\\ \hline
	0332-52367-0639&6563.3$\pm$0.2&4.3$\pm$0.3&254.6$\pm$25.8&6583.9$\pm$0.2&4.2$\pm$0.3&289.2$\pm$26.5\\ \hline
	0377-52145-0044&6567.8$\pm$0.1&7.2$\pm$0.1&3502.1$\pm$55.5&6586.1$\pm$0.3&4.9$\pm$0.3&409.6$\pm$38.0\\ \hline
	0394-51913-0111&6559.6$\pm$0.1&2.3$\pm$0.1 &125.7$\pm$5.2&	6585.2$\pm$0.1&	6.9$\pm$0.2&471.1$\pm$28.3\\ \hline
	0448-51900-0084&6566.7$\pm$0.1&2.2$\pm$0.2&113.0$\pm$8.9&6587.7$\pm$0.2&4.2$\pm$0.2&235.3$\pm$13.6\\ \hline
	0452-51911-0080&6564.0$\pm$0.2&4.7$\pm$0.2&339.4$\pm$13.2&6585.0$\pm$0.2&4.6$\pm$0.2&281.5$\pm$11.4\\ \hline
	0524-52027-0165&6569.8$\pm$0.1&7.1$\pm$0.2&1176.8$\pm$54.4&6592.7$\pm$0.2&2.9$\pm$0.2&107.9$\pm$11.4\\ \hline
	0555-52266-0033&6563.8$\pm$0.4&6.5$\pm$0.2&401.0$\pm$19.8&6580.6$\pm$1.0&8.0$\pm$0.9&190.7$\pm$35.3\\ \hline
	0609-52339-0435&6567.5$\pm$0.1&5.0$\pm$0.1&600.3$\pm$13.8&6586.4$\pm$0.1&4.7$\pm$0.2&276.5$\pm$12.0\\ \hline
	0737-52518-0537&6564.7$\pm$0.2&3.9$\pm$0.3&166.7$\pm$17.3&6585.9$\pm$0.6&4.1$\pm$0.8&49.7$\pm$13.3\\ \hline
	0820-52433-0486&6566.7$\pm$0.1&4.5$\pm$0.1&534.4$\pm$19.5&6587.0$\pm$0.1&4.5$\pm$0.1&322.0$\pm$15.2\\ \hline
	0907-52373-0295&6564.1$\pm$0.6&7.9$\pm$0.6&224.8$\pm$21.2&6585.3$\pm$0.3&6.0$\pm$0.3&249.2$\pm$15.3\\ \hline
	0960-52425-0321&6567.7$\pm$0.1&5.9$\pm$0.1&595.4$\pm$10.9&6587.1$\pm$0.1&4.7$\pm$0.1&591.4$\pm$11.5\\ \hline
	0979-52427-0072&6564.4$\pm$0.1&2.9$\pm$0.1&161.2$\pm$6.7&6585.7$\pm$0.2&3.3$\pm$0.3&86.1$\pm$8.1\\ \hline
	1019-52707-0393&6566.3$\pm$0.4&7.3$\pm$0.4&451.7$\pm$34.0&6587.5$\pm$0.2&5.5$\pm$0.2&472.1$\pm$26.5\\ \hline
	10266-58462-0241&6564.1$\pm$0.1&8.3$\pm$0.1&2546.3$\pm$46.3&6578.7$\pm$0.4&10.8$\pm$0.3&1215.2$\pm$50.2\\ \hline
	1048-52736-0416&6564.2$\pm$0.1&4.5$\pm$0.1&540.1$\pm$12.4&6584.0$\pm$0.2&4.5$\pm$0.3&226.7$\pm$14.7\\ \hline
	11040-58456-0394&6565.0$\pm$1.0&5.0$\pm$1.2&29.2$\pm$8.6&6585.6$\pm$1.5&3.0$\pm$1.8&8.0$\pm$5.6\\ \hline
	11347-58440-0066&6566.4$\pm$1.2&12.2$\pm$1.9&160.2$\pm$34.4&6584.9$\pm$0.4&2.1$\pm$0.5&23.4$\pm$6.7\\ \hline
	1158-52668-0359&6562.8$\pm$0.4&7.0$\pm$0.3&659.3$\pm$42.7&6584.0$\pm$0.3&7.7$\pm$0.3&1052.7$\pm$58.6\\ \hline
	1173-52790-0218&6566.5$\pm$0.1&7.4$\pm$0.1&5152.0$\pm$89.1&6583.9$\pm$0.6&5.6$\pm$0.5&397.6$\pm$70.1\\ \hline
	1295-52934-0580&6575.2$\pm$0.9&11.8$\pm$0.6&422.0$\pm$18.2&6597.6$\pm$0.6&11.8$\pm$0.4&427.5$\pm$18.4\\ \hline
	1403-53227-0485&6565.9$\pm$0.5&4.6$\pm$0.5&56.8$\pm$5.9&6585.9$\pm$0.6&3.0$\pm$0.6&22.0$\pm$4.4\\ \hline
	1404-52825-0235&6565.5$\pm$0.1&4.5$\pm$0.1&1888.1$\pm$59.4&6586.4$\pm$0.1&4.5$\pm$0.1&1420.8$\pm$52.6\\ \hline
	1679-53149-0532&6563.9$\pm$0.1&6.0$\pm$0.1&731.9$\pm$14.9&6584.0$\pm$0.3&6.8$\pm$0.3&373.0$\pm$16.4\\ \hline
	1701-53142-0131&6569.0$\pm$0.4&6.5$\pm$0.4&230.1$\pm$16.4&6589.5$\pm$0.3&4.0$\pm$0.4&110.7$\pm$11.0\\ \hline
	1762-53415-0628&6563.4$\pm$0.9&4.8$\pm$1.1&40.7$\pm$11.3&6587.0$\pm$0.8&4.1$\pm$1.0&30.4$\pm$9.2\\ \hline
	1804-53886-0238&6564.9$\pm$0.3&6.2$\pm$0.3&350.7$\pm$16.5&6585.6$\pm$0.3&5.1$\pm$0.3&243.4$\pm$15.3\\ \hline
	1944-53385-0120&6570.9$\pm$1.3&4.5$\pm$1.2&49.5$\pm$15.1&6590.2$\pm$1.2&8.0$\pm$1.5&127.2$\pm$25.8\\ \hline
	1986-53475-0569&6564.9$\pm$0.2&3.6$\pm$0.2&121.9$\pm$7.5&6585.1$\pm$0.3&3.3$\pm$0.3&65.4$\pm$6.7\\ \hline
	1992-53466-0577&6564.7$\pm$0.1&4.1$\pm$0.1&733.1$\pm$20.6&6585.3$\pm$0.2&4.6$\pm$0.2&372.0$\pm$20.5\\ \hline
	2004-53737-0634&6567.9$\pm$0.2&4.9$\pm$0.2&390.3$\pm$23.8&6587.2$\pm$0.2&3.5$\pm$0.2&173.8$\pm$13.9\\ \hline
	2019-53430-0219&6566.8$\pm$	0.3&5.8$\pm$0.5&212.1$\pm$	36.7&6586.1$\pm$0.2&4.5$\pm$0.3&169.7$\pm$21.1\\ \hline
	2022-53827-0553&6565.1$\pm$0.2&4.4$\pm$0.2&166.1$\pm$8.9&6585.6$\pm$0.3&4.7$\pm$0.3&122.8$\pm$8.2\\ \hline
	2266-53679-0473&6572.2$\pm$0.3&6.8$\pm$0.2&500.7$\pm$19.6&6587.4$\pm$0.2&5.3$\pm$0.2&347.9$\pm$21.8\\ \hline
	2365-53739-0359&6564.4$\pm$0.1&4.3$\pm$0.2&475.9$\pm$30.0&6584.5$\pm$0.2&4.4$\pm$0.2&297.1$\pm$26.4\\ \hline
	2426-53795-0411&6565.1$\pm$0.1&3.9$\pm$0.2&421.6$\pm$32.9&6585.1$\pm$0.1&3.5$\pm$0.2&252.1$\pm$25.1\\ \hline
	2497-54154-0291&6564.8$\pm$0.2&6.4$\pm$0.1&981.6$\pm$20.2&6582.5$\pm$0.4&6.3$\pm$0.3&337.5$\pm$23.2\\ \hline
	2511-53882-0104&	6566.3$\pm$0.1&	2.4$\pm$0.1&	520.7$\pm$12.7&	6586.7$\pm$0.1&	2.7$\pm$0.1&241.0$\pm$	10.3\\ \hline
	2527-54569-0262&6563.9$\pm$0.2&5.6$\pm$0.2&213.9$\pm$8.2&6583.6$\pm$0.2&4.9$\pm$0.2&158.9$\pm$8.0\\ \hline
	2589-54174-0499&6565.9$\pm$0.1&2.6$\pm$0.1&465.6$\pm$11.1&6586.3$\pm$0.1&2.5$\pm$0.1&274.1$\pm$7.8\\ \hline
	2652-54508-0025&6575.3$\pm$	0.9&5.8$\pm$0.8&	127.9$\pm$29.5&6589.0$\pm$0.6&5.3$\pm$0.5&142.7$\pm$	23.2\\ \hline	
	2776-54554-0251&6564.1$\pm$0.2&6.9$\pm$0.2&906.5$\pm$53.7&6584.5$\pm$0.2&6.2$\pm$0.3&669.3$\pm$46.4\\ \hline
	2785-54537-0499&6568.6$\pm$0.2&3.5$\pm$0.3&258.0$\pm$24.8&6590.6$\pm$0.2&3.9$\pm$0.2&300.8$\pm$20.6\\ \hline
	2791-54556-0005&6563.5$\pm$0.3&7.3$\pm$0.4&323.8$\pm$26.0&6588.3$\pm$0.3&1.5$\pm$0.4&19.6$\pm$4.9\\ \hline
	2884-54526-0145&6571.7$\pm$0.1&5.2$\pm$0.1&	510.7$\pm$	19.6&6591.4$\pm$0.1&4.9$\pm$0.2&324.7$\pm$17.8\\ \hline
	2947-54533-0050&6567.7$\pm$0.1&	5.2$\pm$0.1&	458.3$\pm$10.8&6586.1$\pm$0.1&5.0$\pm$0.2&	209.5$\pm$	8.8	\\ \hline
	3830-55574-0154&6565.1$\pm$0.2&8.1$\pm$0.2&722.4$\pm$17.6&6585.9$\pm$0.2&8.5$\pm$0.2&698.4$\pm$21.7\\ \hline
	3846-55327-0802&6568.0$\pm$0.1&2.1$\pm$0.1&74.1$\pm$6.0&6588.8$\pm$0.3&3.9$\pm$0.2&69.9$\pm$5.4\\ \hline
	4210-55444-0050&6565.2$\pm$0.2&4.9$\pm$0.3&95.9$\pm$11.2&6588.1$\pm$0.4&2.7$\pm$0.4&17.9$\pm$3.2\\ \hline
	6294-56482-0774&6567.1$\pm$0.1&3.8$\pm$0.1&296.9$\pm$10.5&6587.9$\pm$0.1&3.2$\pm$0.1&180.5$\pm$5.6\\ \hline
	6420-56304-0818&6565.1$\pm$0.1&4.3$\pm$0.1&687.4$\pm$32.0&6583.1$\pm$0.3&5.4$\pm$0.5&160.3$\pm$32.1\\ \hline
	7283-57063-0660&6563.2$\pm$0.2&6.6$\pm$0.2&352.3$\pm$9.8&6582.5$\pm$0.6&5.1$\pm$0.6&57.2$\pm$8.8\\ \hline
	7338-57127-0570&6565.1$\pm$0.2&4.4$\pm$0.2&57.4$\pm$3.1&6586.9$\pm$0.3&3.7$\pm$0.3&39.4$\pm$3.6\\ \hline
	7582-56960-0660&6565.5$\pm$0.1&7.1$\pm$0.1&2581.0$\pm$27.1&6584.7$\pm$0.1&4.6$\pm$0.1&402.9$\pm$14.7\\ \hline
	7624-57039-0407&6567.5$\pm$0.1&7.1$\pm$0.1&2164.5$\pm$38.1&6589.7$\pm$0.4&3.1$\pm$0.5&47.7$\pm$9.1\\ \hline
	7647-57655-0526&6568.5$\pm$0.1&8.9$\pm$0.2&1834.1$\pm$59.6&6588.5$\pm$0.3&4.1$\pm$0.3&222.0$\pm$20.4\\ \hline
	7648-57329-0836&6571.6$\pm$0.1&2.6$\pm$0.1&226.3$\pm$11.4&6591.2$\pm$0.1&3.3$\pm$0.2&189.1$\pm$10.1\\ \hline
	7723-58430-0620-&6565.6$\pm$0.1&7.2$\pm$0.1&809.6$\pm$11.8&6585.8$\pm$0.2&6.4$\pm$0.1&437.2$\pm$11.8\\ \hline
	7875-56980-0296&6561.9$\pm$0.1&1.5$\pm$0.1&22.9$\pm$2.2&6590.1$\pm$0.3&1.6$\pm$0.3&9.4$\pm$1.7\\ \hline
	9573-57787-0360&6562.3$\pm$1.0&9.6$\pm$0.8&84.6$\pm$10.0&6587.9$\pm$1.1&3.7$\pm$1.2&14.8$\pm$5.2\\ 
\end{longtable*}
\begin{tablenotes}
	\item[*]Notes. Column 1: Plate, MJD, Fiberid of spectroscopic observation; Column 2: the central wavelength of narrow H$\alpha$ in units of~\AA; Column 3: corresponding second moment of narrow H$\alpha$ in units of \AA; Column 4: corresponding flux of narrow H$\alpha$ in units of $10^{-17}{\rm erg/s/cm^{2}}$; Column 5: the central wavelength of narrow \nii~in units of~\AA; Column 6: corresponding second moment of~\nii~in units of~\AA; Column 7:  corresponding flux of \nii~in units of $10^{-17}{\rm erg/s/cm^{2}}$.
\end{tablenotes}

\begin{longtable*}{c|c|c|c|c|c|c|c|c|c}
	\caption{Features of \oiii}\label{tabb2}\\
	\toprule
	Plate-Mjd-Fiberid & $\lambda_b$ & $\sigma_b$  & $F_b$& $\lambda_r$ & $\sigma_r$ & $F_r$&F-test&$V_b$&$V_r$\\
	(1)&(2)&(3)&(4)&(5)&(6)&(7)&(8)&(9)&(10)\\ 
	\midrule
	\endfirsthead
	\multicolumn{7}{r}{Continued}\\
	\toprule
	Plate-Mjd-Fiberid & $\lambda_b$ & $\sigma_b$  & $F_b$& $\lambda_r$ & $\sigma_r$ & $F_r$&F-test&$V_b$&$V_r$\\
	(1)&(2)&(3)&(4)&(5)&(6)&(7)&(8)&(9)&(10)\\ 
	\endhead
	\multicolumn{7}{c}{Continued on next page}\\
	\endfoot
	\bottomrule
	\endlastfoot
	0302-51688-0516	&	5007.9 	$\pm$	0.5 	&	2.8 	$\pm$	0.3 	&	178.9 	$\pm$	29.6 	&	5012.3 	$\pm$	0.1 	&	1.7 	$\pm$	0.1 	&	180.2 	$\pm$	28.4 	&	5 	&	173.2 	$\pm$	41.9 	&	435.6 	$\pm$	19.4 	\\ \hline
	0307-51663-0219	&	5001.0 	$\pm$	0.9 	&	4.6 	$\pm$	0.6 	&	78.2 	$\pm$	13.8 	&	5009.4 	$\pm$	0.3 	&	3.1 	$\pm$	0.2 	&	87.3 	$\pm$	13.0 	&	3 	&	-			&	-			\\ \hline
	0332-52367-0639	&	5005.8 	$\pm$	0.1 	&	1.9 	$\pm$	0.1 	&	360.5 	$\pm$	16.0 	&	5011.2 	$\pm$	0.2 	&	1.8 	$\pm$	0.1 	&	181.2 	$\pm$	15.6 	&	5 	&	112.5 	$\pm$	16.2 	&	214.0 	$\pm$	19.8 	\\ \hline
	0377-52145-0044	&	4999.5 	$\pm$	0.7 	&	3.6 	$\pm$	0.7 	&	121.2 	$\pm$	24.1 	&	5007.5 	$\pm$	0.5 	&	2.5 	$\pm$	0.4 	&	94.5 	$\pm$	21.8 	&	3 	&	-			&	-			\\ \hline
	0394-51913-0111	&	5003.7 	$\pm$	0.1 	&	1.8 	$\pm$	0.1 	&	69.5 	$\pm$	7.0 	&	5008.5 	$\pm$	0.2 	&	6.0 	$\pm$	0.2 	&	409.3 	$\pm$	32.4 	&	5 	&	203.3 	$\pm$	15.6 	&	83.8 	$\pm$	22.1 	\\ \hline
	0448-51900-0084	&	5004.9 	$\pm$	0.6 	&	3.9 	$\pm$	0.5 	&	239.1 	$\pm$	47.1 	&	5009.9 	$\pm$	0.1 	&	1.9 	$\pm$	0.1 	&	164.6 	$\pm$	30.4 	&	5 	&	-		&	-			\\ \hline
	0452-51911-0080	&	5006.1 	$\pm$	0.1 	&	2.6 	$\pm$	0.1 	&	776.8 	$\pm$	19.7 	&	5011.8 	$\pm$	0.1 	&	1.7 	$\pm$	0.1 	&	390.0 	$\pm$	17.9 	&	5 	&	21.2 	$\pm$	20.4 	&	319.3 	$\pm$	20.0 	\\ \hline
	0524-52027-0165	&	5002.4 	$\pm$	1.2 	&	4.3 	$\pm$	0.7 	&	143.0 	$\pm$	36.4 	&	5010.3 	$\pm$	0.6 	&	3.3 	$\pm$	0.3 	&	136.9 	$\pm$	35.8 	&	3 	&	-			&	-			\\ \hline
	0555-52266-0033	&	5004.4 	$\pm$	0.3 	&	2.8 	$\pm$	0.2 	&	227.4 	$\pm$	25.4 	&	5010.2 	$\pm$	0.3 	&	2.8 	$\pm$	0.2 	&	210.8 	$\pm$	26.4 	&	5 	&	-			&	-			\\ \hline
	0609-52339-0435	&	4997.7 	$\pm$	0.7 	&	5.8 	$\pm$	1.0 	&	93.6 	$\pm$	18.8 	&	5009.3 	$\pm$	0.2 	&	2.8 	$\pm$	0.2 	&	74.7 	$\pm$	9.7 	&	5 	&	387.7 	$\pm$	69.4 	&	311.2 	$\pm$	40.5 	\\ \hline
	0737-52518-0537	&	5006.9 	$\pm$	0.4 	&	1.8 	$\pm$	0.2 	&	82.2 	$\pm$	16.1 	&	5011.1 	$\pm$	0.4 	&	1.9 	$\pm$	0.3 	&	74.4 	$\pm$	16.2 	&	5 	&	39.8 	$\pm$	29.9 	&	292.5 	$\pm$	34.3 	\\ \hline
	0820-52433-0486	&	5002.7 	$\pm$	0.5 	&	5.9 	$\pm$	0.4 	&	290.6 	$\pm$	35.6 	&	5009.4 	$\pm$	0.2 	&	2.0 	$\pm$	0.3 	&	50.3 	$\pm$	14.1 	&	5 	&	144.4 	$\pm$	42.7 	&	260.2 	$\pm$	25.2 	\\ \hline
	0907-52373-0295	&	5005.0 	$\pm$	0.2 	&	2.2 	$\pm$	0.3 	&	114.4 	$\pm$	21.5 	&	5011.0 	$\pm$	0.1 	&	1.8 	$\pm$	0.1 	&	129.2 	$\pm$	14.9 	&	5 	&	124.3 	$\pm$	24.4 	&	237.4 	$\pm$	20.7 	\\ \hline
	0960-52425-0321	&	5004.9 	$\pm$	0.2 	&	4.7 	$\pm$	0.3 	&	362.2 	$\pm$	21.0 	&	5010.2 	$\pm$	0.3 	&	3.5 	$\pm$	0.2 	&	198.3 	$\pm$	15.4 	&	3 	&	52.2 	$\pm$	21.7 	&	373.0 	$\pm$	29.3 	\\ \hline
	0979-52427-0072	&	5004.8 	$\pm$	0.4 	&	1.5 	$\pm$	0.3 	&	30.3 	$\pm$	9.2 	&	5009.3 	$\pm$	0.3 	&	1.9 	$\pm$	0.2 	&	71.3 	$\pm$	9.8 	&	3 	&	83.9 	$\pm$	43.1 	&	184.1 	$\pm$	35.3 	\\ \hline
	1019-52707-0393	&	5007.3 	$\pm$	0.2 	&	2.8 	$\pm$	0.1 	&	572.3 	$\pm$	37.0 	&	5013.7 	$\pm$	0.2 	&	2.4 	$\pm$	0.2 	&	347.6 	$\pm$	35.3 	&	5 	&	172.2 	$\pm$	22.5 	&	556.4 	$\pm$	25.2 	\\ \hline
	10266-58462-0241	&	5005.6 	$\pm$	0.5 	&	3.5 	$\pm$	0.2 	&	1671.7 	$\pm$	299.4 	&	5011.5 	$\pm$	0.5 	&	3.6 	$\pm$	0.2 	&	1677.9 	$\pm$	295.7 	&	3 	&	-			&	-			\\ \hline
	1048-52736-0416	&	5005.4 	$\pm$	0.2 	&	1.7 	$\pm$	0.1 	&	160.9 	$\pm$	18.7 	&	5009.8 	$\pm$	0.3 	&	1.8 	$\pm$	0.2 	&	137.9 	$\pm$	18.8 	&	5 	&	99.9 	$\pm$	32.8 	&	166.2 	$\pm$	36.6 	\\ \hline
	11040-58456-0394	&	5005.6 	$\pm$	0.6 	&	2.9 	$\pm$	0.5 	&	44.4 	$\pm$	9.1 	&	5012.6 	$\pm$	0.7 	&	2.7 	$\pm$	0.5 	&	36.4 	$\pm$	8.9 	&	4 	&	45.4 	$\pm$	52.5 	&	375.2 	$\pm$	53.2 	\\ \hline
	11347-58440-0066	&	5007.8 	$\pm$	0.3 	&	2.8 	$\pm$	0.4 	&	63.9 	$\pm$	16.4 	&	5014.5 	$\pm$	0.2 	&	1.9 	$\pm$	0.2 	&	48.2 	$\pm$	8.7 	&	5 	&	-		&	-	\\ \hline
	1158-52668-0359	&	5003.4 	$\pm$	0.2 	&	1.6 	$\pm$	0.2 	&	106.6 	$\pm$	27.2 	&	5010.3 	$\pm$	0.2 	&	3.2 	$\pm$	0.3 	&	374.9 	$\pm$	74.0 	&	5 	&	269.9 	$\pm$	24.0 	&	141.4 	$\pm$	24.6 	\\ \hline
	1173-52790-0218	&	5001.0 	$\pm$	0.2 	&	3.2 	$\pm$	0.3 	&	245.8 	$\pm$	55.9 	&	5010.4 	$\pm$	0.1 	&	1.4 	$\pm$	0.1 	&	204.1 	$\pm$	14.1 	&	5 	&	-			&	-			\\ \hline
	1295-52934-0580	&	5009.3 	$\pm$	0.2 	&	1.8 	$\pm$	0.2 	&	55.8 	$\pm$	13.6 	&	5014.7 	$\pm$	0.4 	&	3.3 	$\pm$	0.4 	&	119.9 	$\pm$	23.4 	&	5 	&	-			&	-			\\ \hline
	1403-53227-0485	&	5006.1 	$\pm$	0.5 	&	2.5 	$\pm$	0.3 	&	101.3 	$\pm$	17.5 	&	5011.5 	$\pm$	0.4 	&	2.4 	$\pm$	0.2 	&	112.5 	$\pm$	17.5 	&	5 	&	119.3 	$\pm$	62.8 	&	445.8 	$\pm$	56.9 	\\ \hline
	1404-52825-0235	&	5008.2 	$\pm$	0.1 	&	2.8 	$\pm$	0.1 	&	2859.1 	$\pm$	87.9 	&	5012.4 	$\pm$	0.1 	&	1.2 	$\pm$	0.1 	&	669.4 	$\pm$	69.7 	&	5 	&	150.0 	$\pm$	12.1 	&	397.2 	$\pm$	9.6 	\\ \hline
	1679-53149-0532	&	5005.0 	$\pm$	0.1 	&	2.1 	$\pm$	0.1 	&	608.2 	$\pm$	17.6 	&	5010.8 	$\pm$	0.1 	&	2.0 	$\pm$	0.0 	&	563.2 	$\pm$	17.2 	&	5 	&	-			&	-			\\ \hline
	1701-53142-0131	&	5002.2 	$\pm$	0.5 	&	3.0 	$\pm$	0.4 	&	57.8 	$\pm$	9.1 	&	5010.4 	$\pm$	0.3 	&	3.0 	$\pm$	0.2 	&	108.9 	$\pm$	9.2 	&	3 	&	-			&	-			\\ \hline
	1762-53415-0628	&	5007.0 	$\pm$	0.4 	&	2.3 	$\pm$	0.4 	&	57.8 	$\pm$	10.1 	&	5012.0 	$\pm$	0.3 	&	1.6 	$\pm$	0.3 	&	33.3 	$\pm$	8.6 	&	4 	&	10.8 	$\pm$	45.5 	&	311.3 	$\pm$	42.0 	\\ \hline
	1804-53886-0238	&	5006.8 	$\pm$	0.3 	&	3.3 	$\pm$	0.3 	&	372.5 	$\pm$	53.3 	&	5013.0 	$\pm$	0.3 	&	2.2 	$\pm$	0.2 	&	179.1 	$\pm$	32.1 	&	5 	&	-			&	-			\\ \hline
	1944-53385-0120	&	5009.5 	$\pm$	0.4 	&	2.7 	$\pm$	0.5 	&	66.9 	$\pm$	22.2 	&	5016.2 	$\pm$	0.3 	&	2.1 	$\pm$	0.3 	&	57.9 	$\pm$	15.2 	&	5 	&	-			&	-			\\ \hline
	1986-53475-0569	&	5006.8 	$\pm$	0.1 	&	1.5 	$\pm$	0.1 	&	216.3 	$\pm$	11.3 	&	5010.7 	$\pm$	0.1 	&	1.4 	$\pm$	0.1 	&	130.2 	$\pm$	10.9 	&	5 	&	38.1 	$\pm$	22.5 	&	274.9 	$\pm$	25.0 	\\ \hline
	1992-53466-0577	&	5005.7 	$\pm$	0.1 	&	2.6 	$\pm$	0.1 	&	849.6 	$\pm$	43.2 	&	5010.7 	$\pm$	0.1 	&	1.8 	$\pm$	0.1 	&	501.1 	$\pm$	38.4 	&	5 	&	98.8 	$\pm$	17.4 	&	197.0 	$\pm$	15.0 	\\ \hline
	2004-53737-0634	&	5008.6 	$\pm$	0.0 	&	1.4 	$\pm$	0.0 	&	265.2 	$\pm$	8.1 	&	5012.3 	$\pm$	0.0 	&	1.1 	$\pm$	0.0 	&	184.8 	$\pm$	7.4 	&	5 	&	248.8 	$\pm$	11.6 	&	471.4 	$\pm$	11.5 	\\ \hline
	2019-53430-0219	&	5000.7 	$\pm$	0.5 	&	4.7 	$\pm$	0.3 	&	247.4 	$\pm$	20.5 	&	5008.6 	$\pm$	0.2 	&	2.6 	$\pm$	0.2 	&	105.6 	$\pm$	19.2 	&	5 	&	303.5 	$\pm$	38.6 	&	168.0 	$\pm$	26.2 	\\ \hline
	2022-53827-0553	&	5005.0 	$\pm$	0.0 	&	2.0 	$\pm$	0.0 	&	253.8 	$\pm$	5.9 	&	5011.3 	$\pm$	0.0 	&	1.8 	$\pm$	0.0 	&	211.1 	$\pm$	5.3 	&	5 	&	143.2 	$\pm$	25.9 	&	235.8 	$\pm$	25.9 	\\ \hline
	2266-53679-0473	&	5000.9 	$\pm$	0.4 	&	6.1 	$\pm$	0.4 	&	153.1 	$\pm$	17.0 	&	5012.0 	$\pm$	0.3 	&	3.1 	$\pm$	0.3 	&	42.5 	$\pm$	8.0 	&	5 	&	48.2 	$\pm$	48.9 	&	616.2 	$\pm$	46.3 	\\ \hline
	2365-53739-0359	&	5005.8 	$\pm$	0.1 	&	1.7 	$\pm$	0.0 	&	585.6 	$\pm$	22.6 	&	5010.1 	$\pm$	0.1 	&	1.8 	$\pm$	0.0 	&	600.0 	$\pm$	23.1 	&	5 	&	97.3 	$\pm$	17.4 	&	158.9 	$\pm$	17.4 	\\ \hline
	2426-53795-0411	&	5006.5 	$\pm$	0.3 	&	4.1 	$\pm$	0.4 	&	141.5 	$\pm$	18.6 	&	5013.3 	$\pm$	0.2 	&	1.5 	$\pm$	0.2 	&	29.7 	$\pm$	7.0 	&	5 	&	43.9 	$\pm$	37.1 	&	451.7 	$\pm$	30.3 	\\ \hline
	2497-54154-0291	&	5007.3 	$\pm$	0.1 	&	1.6 	$\pm$	0.1 	&	510.7 	$\pm$	39.0 	&	5010.7 	$\pm$	0.1 	&	1.5 	$\pm$	0.1 	&	490.3 	$\pm$	38.9 	&	5 	&	87.0 	$\pm$	22.8 	&	290.9 	$\pm$	22.4 	\\ \hline
	2511-53882-0104	&	5002.7 	$\pm$	0.7 	&	5.4 	$\pm$	0.6 	&	164.9 	$\pm$	21.1 	&	5009.5 	$\pm$	0.2 	&	1.6 	$\pm$	0.3 	&	42.9 	$\pm$	13.1 	&	5 	&	152.5 	$\pm$	55.0 	&	254.6 	$\pm$	26.5 	\\ \hline
	2527-54569-0262	&	5004.4 	$\pm$	0.1 	&	1.9 	$\pm$	0.1 	&	138.9 	$\pm$	9.4 	&	5009.7 	$\pm$	0.1 	&	2.1 	$\pm$	0.1 	&	214.0 	$\pm$	9.8 	&	5 	&	202.4 	$\pm$	32.9 	&	116.2 	$\pm$	31.0 	\\ \hline
	2589-54174-0499	&	5004.5 	$\pm$	0.9 	&	3.6 	$\pm$	0.5 	&	169.6 	$\pm$	37.6 	&	5009.0 	$\pm$	0.1 	&	1.9 	$\pm$	0.2 	&	119.0 	$\pm$	34.9 	&	5 	&	-			&	-			\\ \hline
	2652-54508-0025	&	5000.8 	$\pm$	0.2 	&	1.9 	$\pm$	0.3 	&	72.5 	$\pm$	15.0 	&	5010.5 	$\pm$	0.2 	&	2.5 	$\pm$	0.2 	&	138.0 	$\pm$	23.3 	&	5 	&	547.3 	$\pm$	24.4 	&	36.3 	$\pm$	23.4 	\\ \hline
	2776-54554-0251	&	5006.1 	$\pm$	0.2 	&	2.1 	$\pm$	0.2 	&	188.0 	$\pm$	24.0 	&	5011.3 	$\pm$	0.1 	&	1.3 	$\pm$	0.1 	&	128.9 	$\pm$	13.8 	&	5 	&	29.3 	$\pm$	21.0 	&	282.7 	$\pm$	18.4 	\\ \hline
	2785-54537-0499	&	5000.8 	$\pm$	0.2 	&	2.5 	$\pm$	0.3 	&	134.4 	$\pm$	36.5 	&	5010.4 	$\pm$	0.2 	&	4.8 	$\pm$	0.2 	&	866.5 	$\pm$	95.9 	&	5 	&	-			&	-			\\ \hline
	2791-54556-0005	&	5004.9 	$\pm$	0.1 	&	2.2 	$\pm$	0.1 	&	171.3 	$\pm$	13.6 	&	5010.3 	$\pm$	0.2 	&	1.8 	$\pm$	0.2 	&	69.3 	$\pm$	10.4 	&	5 	&	-			&	-			\\ \hline
	2884-54526-0145	&	5004.5 	$\pm$	0.4 	&	3.3 	$\pm$	0.2 	&	103.2 	$\pm$	12.9 	&	5012.7 	$\pm$	0.4 	&	3.8 	$\pm$	0.3 	&	117.8 	$\pm$	14.1 	&	5 	&	292.6 	$\pm$	38.7 	&	784.9 	$\pm$	42.3 	\\ \hline
	2947-54533-0050	&	4997.8 	$\pm$	0.7 	&	9.1 	$\pm$	0.7 	&	110.1 	$\pm$	9.8 	&	5009.2 	$\pm$	0.2 	&	2.4 	$\pm$	0.2 	&	34.2 	$\pm$	4.6 	&	3 	&	-			&	-			\\ \hline
	3830-55574-0154	&	5007.5 	$\pm$	0.2 	&	5.4 	$\pm$	0.2 	&	1118.8 	$\pm$	70.9 	&	5013.7 	$\pm$	0.1 	&	2.1 	$\pm$	0.1 	&	305.2 	$\pm$	27.7 	&	5 	&	-			&	-			\\ \hline
	3846-55327-0802	&	5004.1 	$\pm$	0.5 	&	3.9 	$\pm$	0.4 	&	105.3 	$\pm$	13.5 	&	5010.7 	$\pm$	0.1 	&	2.4 	$\pm$	0.1 	&	123.6 	$\pm$	11.2 	&	5 	&	41.4 	$\pm$	41.2 	&	438.5 	$\pm$	20.3 	\\ \hline
	4210-55444-0050	&	5003.4 	$\pm$	0.7 	&	2.4 	$\pm$	0.5 	&	14.5 	$\pm$	4.1 	&	5009.5 	$\pm$	0.4 	&	2.3 	$\pm$	0.3 	&	23.1 	$\pm$	4.1 	&	3 	&	50.9 	$\pm$	58.2 	&	314.2 	$\pm$	41.9 	\\ \hline
	6294-56482-0774	&	5004.1 	$\pm$	0.6 	&	5.9 	$\pm$	0.6 	&	167.6 	$\pm$	51.6 	&	5009.5 	$\pm$	0.1 	&	2.2 	$\pm$	0.1 	&	79.0 	$\pm$	9.9 	&	5 	&	6.7 	$\pm$	49.8 	&	314.9 	$\pm$	19.6 	\\ \hline
	6420-56304-0818	&	5007.1 	$\pm$	0.1 	&	1.5 	$\pm$	0.0 	&	523.5 	$\pm$	28.4 	&	5010.4 	$\pm$	0.1 	&	1.5 	$\pm$	0.0 	&	496.0 	$\pm$	28.9 	&	5 	&	-			&	-			\\ \hline
	7283-57063-0660	&	5003.9 	$\pm$	0.2 	&	2.6 	$\pm$	0.1 	&	223.4 	$\pm$	15.9 	&	5009.8 	$\pm$	0.2 	&	2.7 	$\pm$	0.1 	&	237.2 	$\pm$	16.5 	&	5 	&	-			&	-			\\ \hline
	7338-57127-0570	&	5007.8 	$\pm$	0.4 	&	2.2 	$\pm$	0.3 	&	46.4 	$\pm$	6.9 	&	5012.1 	$\pm$	0.3 	&	1.4 	$\pm$	0.2 	&	19.1 	$\pm$	5.8 	&	4 	&	137.4 	$\pm$	39.1 	&	394.1 	$\pm$	34.8 	\\ \hline
	7582-56960-0660	&	5005.5 	$\pm$	0.0 	&	1.8 	$\pm$	0.0 	&	397.8 	$\pm$	9.2 	&	5010.6 	$\pm$	0.0 	&	1.7 	$\pm$	0.0 	&	397.8 	$\pm$	8.2 	&	5 	&	-			&	-			\\ \hline
	7624-57039-0407	&	5003.3 	$\pm$	0.4 	&	5.9 	$\pm$	0.4 	&	170.5 	$\pm$	10.9 	&	5012.3 	$\pm$	0.2 	&	1.8 	$\pm$	0.3 	&	32.9 	$\pm$	6.7 	&	5 	&	-			&	-			\\ \hline
	7647-57655-0526	&	5001.5 	$\pm$	0.6 	&	6.1 	$\pm$	0.6 	&	217.5 	$\pm$	23.1 	&	5010.8 	$\pm$	0.2 	&	2.5 	$\pm$	0.3 	&	67.0 	$\pm$	16.5 	&	5 	&	-			&	-			\\ \hline
	7648-57329-0836	&	4997.7 	$\pm$	0.1 	&	1.8 	$\pm$	0.1 	&	106.8 	$\pm$	7.1 	&	5013.2 	$\pm$	0.1 	&	2.2 	$\pm$	0.1 	&	205.5 	$\pm$	10.2 	&	5 	&	-			&	-			\\ \hline
	7723-58430-0620	&	5007.6 	$\pm$	0.1 	&	3.7 	$\pm$	0.1 	&	1182.4 	$\pm$	38.9 	&	5013.8 	$\pm$	0.1 	&	2.2 	$\pm$	0.0 	&	533.0 	$\pm$	31.2 	&	5 	&	-			&	-			\\ \hline
	7875-56980-0296	&	5006.2 	$\pm$	0.1 	&	1.8 	$\pm$	0.0 	&	140.8 	$\pm$	3.8 	&	5011.5 	$\pm$	0.0 	&	1.6 	$\pm$	0.0 	&	130.1 	$\pm$	3.6 	&	5 	&	14.6 	$\pm$	20.6 	&	333.0 	$\pm$	20.5 	\\ \hline
	9573-57787-0360	&	5006.5 	$\pm$	0.4 	&	1.9 	$\pm$	0.4 	&	18.8 	$\pm$	4.5 	&	5012.1 	$\pm$	0.2 	&	2.0 	$\pm$	0.2 	&	50.4 	$\pm$	4.8 	&	3 	&	-			&	-			\\ 
\end{longtable*}
\begin{tablenotes}
	\item[*]Notes. Column 1: Plate, MJD, Fiberid of spectroscopic observation; Column 2: the central wavelength of blue-shifted component of \oiii in units of \AA; Column 3: corresponding second moment of blue-shifted component of \oiii in units of \AA; Column 4: corresponding flux of blue-shifted component of \oiii~in units of $10^{-17}{\rm erg/s/cm^{2}}$; Column 5: the central wavelength of red-shifted component of \oiii in units of \AA; Column 6: corresponding second moment of red-shifted component of \oiii in units of \AA; Column 7: corresponding flux of red-shifted component of \oiii~in units of $10^{-17}{\rm erg/s/cm^{2}}$; Column 8: the  confidence level ($\sigma$) of F-test; Column 9 and Column 10: the velocity offset of blue-shifted component and red-shifted component of \oiii determined by shifted velocity from SSP method in units of km/s, respectively, and `-' represents an object with no obvious absorption line.
\end{tablenotes}

\end{document}